\RequirePackage{fix-cm}
\documentclass[smallextended]{svjour3}       
\smartqed  
\usepackage{graphicx}
\usepackage{cite}
\usepackage{natbib}
\usepackage{amsmath,amssymb,amsfonts}
\usepackage{algorithmic}

\usepackage{graphicx}
\usepackage{textcomp}
\usepackage[dvipsnames]{xcolor}
\usepackage{booktabs}
\usepackage{multirow}
\usepackage{xspace}
\usepackage{caption}
\usepackage{subcaption}
\usepackage{footnote}
\usepackage{hyperref}
\usepackage{tcolorbox}
\usepackage{csquotes}
\usepackage[ruled,vlined]{algorithm2e}
\usepackage{orcidlink}

\usepackage{tikz}
\usepackage{tcolorbox}

\usepackage{mdframed}
\usepackage{ntheorem}
\theoremstyle{nonumberplain}
\newmdtheoremenv[%
  linecolor=blue,
  linewidth=2pt,
  rightline=false,
  leftline=false]{figrev}{}

\begin{document}

\title{Towards effective assessment of steady state performance in Java software: Are we there yet?}
\titlerunning{Towards effective assessment of steady state performance in Java software}

\author{Luca Traini  \orcidlink{0000-0003-3676-0645}     \and
        Vittorio Cortellessa  \orcidlink{0000-0002-4507-464X} \and
	Daniele {Di Pompeo}  \orcidlink{0000-0003-2041-7375} \and
        Michele Tucci  \orcidlink{0000-0002-0329-1101} } 

\authorrunning{Luca Traini et al.} 

\institute{L. Traini \at University of L'Aquila, Italy\\ 
              \email{luca.traini@univaq.it}           
\and
           V. Cortellessa  \at  University of L'Aquila, Italy\\ 
              \email{vittorio.cortellessa@univaq.it}
\and
	   D. {Di Pompeo}  \at  University of L'Aquila, Italy\\ 
              \email{daniele.dipompeo@univaq.it} 
\and
           M. Tucci \at  Charles University, Czech Republic\\ 
              \email{tucci@d3s.mff.cuni.cz} 
}

\date{Accepted: 30 September 2022}

\maketitle

\newcommand{\ie}{\emph{i.e.,}\xspace}
\newcommand{\eg}{\emph{e.g.,}\xspace}
\newcommand{\etc}{etc.\xspace}
\newcommand{\etal}{\emph{et~al.}\xspace}
\newcommand{\secref}[1]{Section~\ref{#1}\xspace}
\newcommand{\figref}[1]{Fig.~\ref{#1}\xspace}
\newcommand{\listref}[1]{Listing~\ref{#1}\xspace}
\newcommand{\tabref}[1]{Table~\ref{#1}\xspace}
\newcommand{\algoref}[1]{Algorithm~\ref{#1}\xspace}
\newcommand{\eqqref}[1]{Equation~(\ref{#1})\xspace}
\newcommand{\tool}[1]{{\sc #1}\xspace}
\newcommand{\vda}{$\hat{A}_{12}$\xspace}
\newcommand*\circled[1]{\tikz[baseline=(char.base)]{
  \node[shape=circle,draw,inner sep=1pt] (char) {#1};}}

\newcommand{\unknown}{\textcolor{red}{XXX}\xspace}
\newcommand{\systems}{30\xspace}
\newcommand{\benchmarks}{586\xspace}
\newcommand{\experimentationTime}{$\sim$93~days\xspace}
\newboolean{showcomments}

\setboolean{showcomments}{true}

\ifthenelse{\boolean{showcomments}}
  {\newcommand{\nb}[2]{
    \fbox{\bfseries\sffamily\scriptsize#1}
    {\sf\small$\blacktriangleright$\textit{#2}$\blacktriangleleft$}
   }
  }
  {\newcommand{\nb}[2]{}
  }

\newcommand\DDP[1]{\textcolor{Purple}{\nb{DANIELE}{#1}}}
\newcommand\MT[1]{\textcolor{teal}{\nb{MICHELE}{#1}}}
\newcommand\VIC[1]{\textcolor{red}{\nb{VITTORIO}{#1}}}
\newcommand\LUCA[1]{\textcolor{NavyBlue}{\nb{LUCA}{#1}}}

\newcommand\DDPg[1]{\textcolor{Gray}{\nb{DANIELE}{#1}}}
\newcommand\MTg[1]{\textcolor{Gray}{\nb{MICHELE}{#1}}}
\newcommand\VICg[1]{\textcolor{Gray}{\nb{VITTORIO}{#1}}}
\newcommand\LUCAg[1]{\textcolor{Gray}{\nb{LUCA}{#1}}}

\begin{abstract}
Microbenchmarking is a widely used form of performance testing in Java software. A microbenchmark repeatedly executes a small chunk of code while collecting measurements related to its performance.
Due to Java Virtual Machine optimizations, microbenchmarks are usually subject to severe performance fluctuations in the first phase of their execution (also known as warmup).
For this reason, software developers typically discard measurements of this phase and focus their analysis when benchmarks reach a steady state of performance.
Developers estimate the end of the warmup phase based on their expertise, and configure their benchmarks accordingly.
Unfortunately, this approach is based on two strong assumptions: (i) benchmarks always reach a steady state of performance and (ii) developers accurately estimate warmup.
In this paper, we show that Java microbenchmarks do not always reach a steady state, and often developers fail to accurately estimate the end of the warmup phase.
We found that a considerable portion of studied benchmarks do not hit the steady state, and warmup estimates provided by software developers are often inaccurate (with a large error).
This has significant implications both in terms of results quality and time-effort.
Furthermore, we found that dynamic reconfiguration significantly improves warmup estimation accuracy, but still it induces suboptimal warmup estimates and relevant side-effects. 
We envision this paper as a starting point for supporting the introduction of more sophisticated automated techniques that can ensure results quality in a timely fashion.
\end{abstract}

\section{Introduction}
Microbenchmarking is a form of lightweight performance testing, widely used to assess the execution time of Java software \citep{Leitner2017}.
Although less demanding than other performance testing techniques (\eg load tests \citep{Jiang2015}), Java microbenchmarking requires careful design \citep{Costa2019, Kalibera2013, Georges2007} to enable a reliable performance assessment.
A key challenge is the inherent non-linearity of Java performance: The Java Virtual Machine (JVM) uses just-in-time compilation to translate ``hot'' parts of the Java code into efficient machine code at run-time \citep{Barrett2017}, leading to (often severe) performance fluctuations and potentially unstable results.

To tackle this problem, practitioners rely on the \emph{assumption} that microbenchmarking is characterized by \emph{two distinct phases}. 
During an initial \emph{warmup phase}, the JVM determines which parts of the software under test would most benefit from dynamic compilation, then, in a subsequent phase, the benchmark reaches a \emph{steady state} of performance.
Based on that, benchmarks are typically designed to discard measurements of the \emph{warmup phase} and focus on \emph{steady state} performance~\citep{Georges2007, Kalibera2013, Barrett2017}.
Java Microbenchmark Harness (JMH), \ie the most popular Java microbenchmarking framework \citep{Leitner2017}, leverages this concept and enables developers to manually \emph{configure} the expected \emph{warmup time} of a benchmark.
Once launched, a JMH benchmark continuously executes the software under test for the configured warmup time, and, only after that, starts to collect steady state performance measurements. 

Although considered the cornerstone of most of the current Java microbenchmarking practice, the \emph{two-phase assumption} is not yet confirmed by empirical studies.
Quite the opposite, there are indications that such an assumption oversimplifies the actual microbenchmarking behavior.
In a recent study, \cite{Barrett2017} studied a set of small and deterministic benchmarks \citep{Bagley2004, Bolz2015} across different types of VMs (including the JVM), and found that a relevant portion of benchmarks never hit the steady state.
It is worth to notice that these benchmarks significantly differ from ``software testing oriented'' benchmarks (\eg JMH). Indeed, they are not aimed at assessing specific software, rather, they are typically used as optimization targets by VM authors. Even more, they are generally more effectively optimized by VMs than average software \citep{Ratanaworabhan2009}.
Despite the peculiarity of these benchmarks, the finding of \cite{Barrett2017} raises concerns on the current Java microbenchmarking practice, and calls for further empirical investigation.

Nevertheless, even when benchmarks consistently reach a steady state, performance assessment remains far from trivial.
A key challenge is to effectively estimate \emph{warmup time}. 
An overestimated warmup time may waste too much time, thereby potentially hampering the adoption of benchmarks in the Continuous Integration (CI) pipeline \citep{Laaber2020, traini2021}.
On the other hand, an underestimated warmup time may easily mislead steady state performance assessment \citep{Georges2007, Kalibera2013}.

The current state-of-practice mostly relies on software developers' knowledge to determine warmup time.
Software developers estimate warmup time based on their expertise, and statically \emph{configure} benchmark execution according to this estimation.
Unfortunately, no empirical studies so far have directly investigated the effectiveness of this practice for steady state performance assessment.

Recently, an alternative approach to static benchmark configuration has been proposed by \cite{Laaber2020}.
This approach, called \emph{dynamic reconfiguration}, leverages stability criteria \citep{Kalibera2013, He2019} to automatically determine the end of the warmup phase at run-time.
According to their results, when compared to JMH default configurations, dynamic reconfiguration can significantly reduce execution time with low impact on results quality.
Despite these promising results, there is still little knowledge on the effectiveness of dynamic reconfiguration for steady state performance assessment.
And even more, it is yet unclear whether such techniques can improve the effectiveness of the current practice, \ie developer static configurations.

In the past years, Java microbenchmarks have been widely studied in the literature.
\cite{Leitner2017} found that JMH was one of the predominant microbenchmarking framework in the Java community. \cite{Costa2019} empirically studied five JMH bad practices. \cite{Laaber2019} performed an exploratory study on software microbenchmarking in the cloud. \cite{Samoaa2021} studied the impact of parameterization in JMH microbenchmarks.

Despite these efforts, there is still a lack of knowledge on the effectiveness of modern Java microbenchmarking for steady state performance assessment.
In this paper, we aim to fill this gap by presenting the first comprehensive study that investigates steady state performance assessment in Java microbenchmarking.
After an extensive experimentation of \benchmarks JMH benchmarks from \systems Java systems, involving $\sim$9.056 billion benchmark invocations for an overall execution time of \experimentationTime, we determined whether and when each benchmark reaches a steady state using an automated statistical approach by \cite{Barrett2017} based on changepoint analysis \citep{Killick2012}.
At the time of writing, \citeauthor{Barrett2017}'s approach represents one of the most advanced automated technique to determine steady state execution in Java benchmarks.
Besides investigating whether benchmarks ever reach a steady state or not, we also comprehensively evaluated the effectiveness of the current state-of-practice and state-of-the-art.
In particular, we investigated to what extent statically-defined developer configurations (\ie state-of-practice), and dynamic reconfiguration techniques (\ie state-of-the-art) are effective in ensuring a reliable and time-efficient assessment of steady state performance in JMH benchmarks. Even more, we quantified the potential side effects due to inaccurate warmup estimation both in terms of execution time waste and misleading performance measurements.

\cite{Laaber2020} have already investigated the effectiveness of dynamic reconfiguration. However, their study was mainly concerned with a particular aspect of Java microbenchmarking, \ie reducing execution time.
In this paper, instead, we aim to provide a comprehensive investigation on the effectiveness of dynamic reconfiguration (and developer static configurations) for steady state performance assessment.
Due to our goal, we do not use JMH defaults as baselines, as done by Laaber~\etal, since there is no guarantee that they can effectively capture steady state performance.
Indeed, although JMH defaults are a reasonable baseline for comparison\footnote{JMH defaults are configurations defined by JMH developers, which are undoubtedly experts in the field of Java microbenchmarking.}, their use may have downsides when studying steady state performance.
Several studies have shown that benchmarks often reach their steady states in different numbers of iterations \citep{Georges2007, Kalibera2013}, thus it may be misleading to use a unique (though reasonable) configuration as baseline, \ie JMH defaults may be effective for some benchmarks and suboptimal for other ones. 
In order to avoid this problem, we leverage a different (and more rigorous) approach to assess dynamic reconfiguration effectiveness.
We first use a state-of-the-art steady state detection technique to determine if/when a benchmark reaches a steady state of performance, then we base on its outcome to assess dynamic reconfiguration effectiveness.
Besides this evaluation of dynamic reconfiguration techniques, this paper presents the following novel investigations.
We present the first study that investigates if/when JMH benchmarks reach a steady state of performance. Second, we perform the first evaluation on the effectiveness of statically defined developer configurations. Third, we introduce the first comparison between the effectiveness of developer configurations (\ie the current state-of-practice) and dynamic reconfiguration techniques (\ie the current state-of-the-art).

Our results show that JMH benchmarks do not always reach a steady state of performance, thereby demystifying the current cornerstone of Java microbenchmarking, \ie the \emph{two-phase assumption}. This finding implies that practitioners may rely on measurements that are not representative of ``actual'' steady state performance.
In addition, our results suggest that developer static configurations are often ineffective for warmup estimation, and may cause either improperly long execution times or misleading performance assessment.
On the other hand, dynamic reconfiguration techniques show significant improvement over the current state of practice, but they still produce inaccurate estimates of the warmup time, hence causing time-consuming benchmark executions and distorted results.
This finding highlights room for improvement for dynamic reconfiguration, and it calls for further research on this topic.\\

The main contributions of this paper are:
\begin{itemize}
	\item a statistically rigorous investigation of steady state performance in JMH microbenchmarks.
	\item an empirical evaluation of developer static configurations in JMH microbenchmarks.
	\item a comprehensive comparison among the effectiveness of developer static configurations and state-of-the-art dynamic reconfiguration techniques.
	\item a large dataset of labeled benchmark executions to facilitate future research on Java steady state performance assessment, and foster further innovations in dynamic reconfiguration.\\
\end{itemize}

The remainder of this paper is organized as follows. \secref{sec:background} introduces steady state performance assessment and JMH microbenchmarks. \secref{sec:rqs} describes our research questions and \secref{sec:design} explains the experimental design. \secref{sec:results} reports the results. \secref{sec:discussion} discusses some implications of our findings. \secref{sec:threats} describes threats to validity. \secref{sec:related} presents related work, and \secref{sec:conclusion} concludes this paper.

\section{Background}\label{sec:background}
\subsection{Steady State Performance} \label{sec:steadystate}
\begin{figure*}[p]
     \centering
     \begin{subfigure}[b]{10cm}
         \centering
		\includegraphics[width=\linewidth]{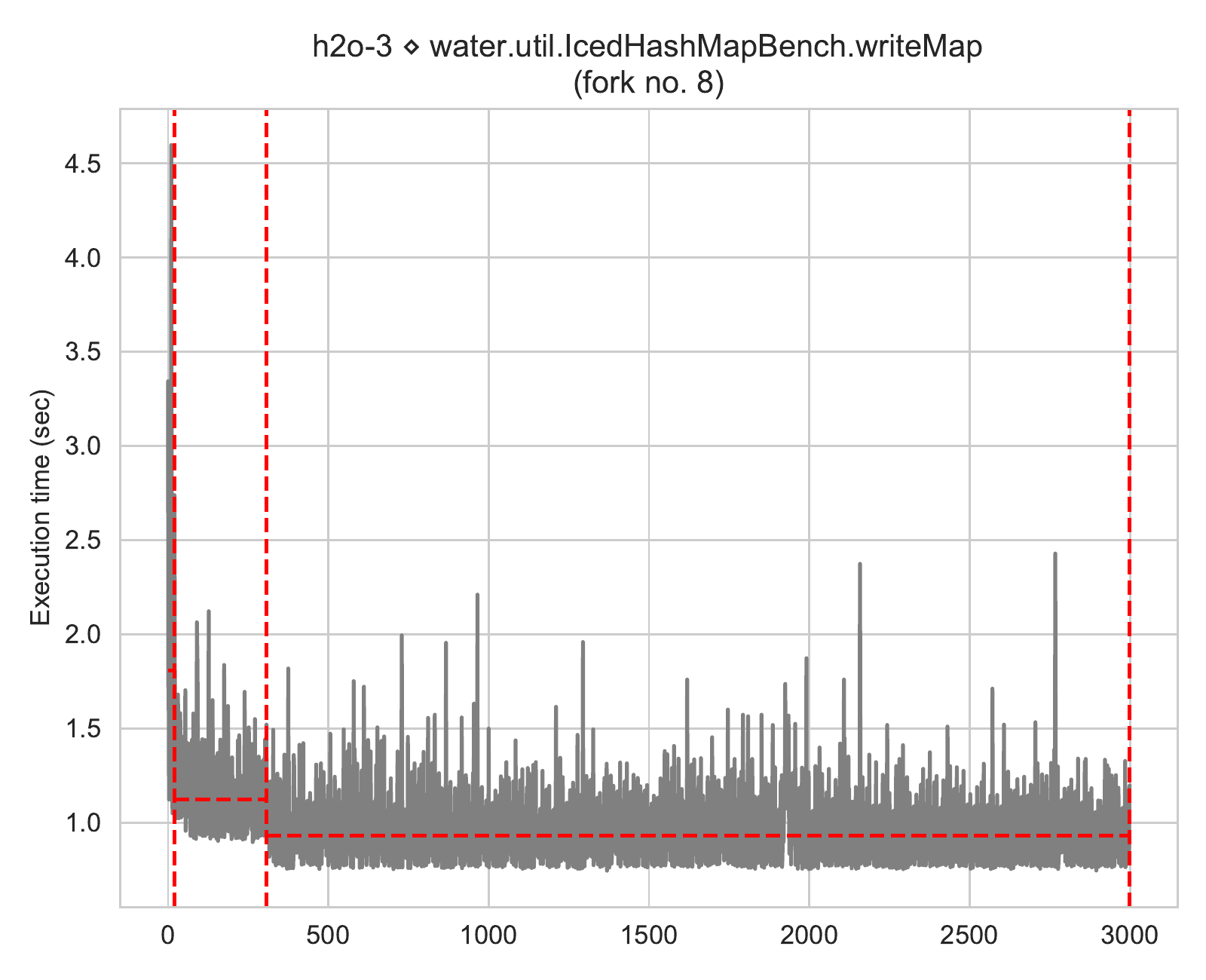}
  		\caption{Example of benchmark execution that consistently reaches a steady state of performance.
  		}
  		\label{fig:steady_example}
     \end{subfigure}
     \hfill
     \begin{subfigure}[b]{10cm}
         \centering
		\includegraphics[width=\linewidth]{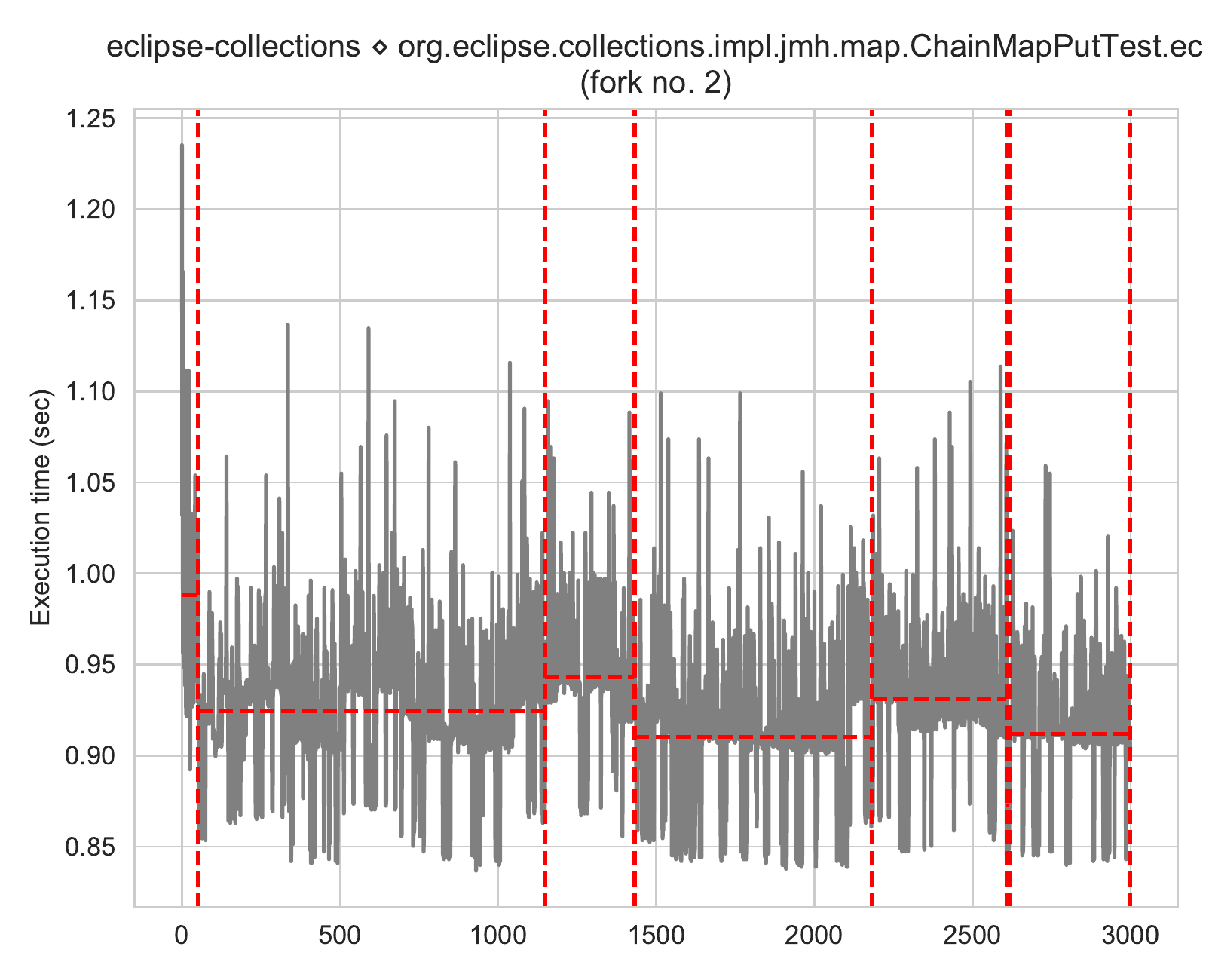}
  		\caption{Example of benchmark execution that doesn't reach a steady state of performance.}
  		\label{fig:non_steady_example}
     \end{subfigure}
     \caption{Two examples of benchmarks executions from our results. The grey line represents the execution times of each benchmark iteration visualized as a time-series. The x-axis represents the benchmark iteration number, while the y-axis represents the mean execution time within the iteration. Shifts in performance behaviour (\ie changepoints) are indicated by dashed vertical red lines. The height of each dashed horizontal red line denotes the mean execution time within the changepoint segment.  The plot titles show: the system that the benchmark belongs to (\eg \texttt{h2o-3}), the benchmark name (\eg~\texttt{water.util.IcedHashMapBench.writeMap}), and the JMH fork number.}
\end{figure*}

At the beginning of its execution, typically, a Java microbenchmark is slowly executed by the JVM.
In a subsequent phase, the JVM detects ``hots'' (\ie frequently executed) loops or methods, and it dynamically compiles them into optimized machine code.
As a consequence, subsequent executions of those loops or methods (usually) become faster.
Once dynamic compilation is completed, the JVM is said to have finished warming up, and the benchmark is said to be executing at a \emph{steady state of performance}.

Java microbenchmarking aims at assessing steady state performance (\eg execution time) of Java software. 
The typical approach to collect steady state measurements is straightforward: A microbenchmark is executed for a certain number of times, and the first $n$ benchmark executions are discarded (\ie those related to the \emph{warmup phase}) to prevent potentially misleading results.
Unfortunately, the fixed number of $n$ executions does not guarantee that warmup has ended.
In order to investigate this issue, researchers started to develop data-driven methodologies to identify the end of the warmup phase.
Prominent works in this regard are the methodologies proposed by \cite{Georges2007}, and \cite{Kalibera2013}.
The former uses preset thresholds on the coefficient of variation to determine whether the warmup phase is finished or not, while the latter leverages data visualization techniques (\ie auto-correlation function plots, lag plots and run-sequence plots).
Unfortunately, each one of these methodologies has its own drawback.
Kalibera and Jones showed that the Georges \etal's heuristic~(\citeyear{Georges2007}) often fails to accurately determine the end of the warmup phase \citep{Kalibera2013}.
On the other hand, the methodology proposed by \cite{Kalibera2013} is mostly based on a manual process, which typically implies some major limitations: (i) humans are prone to error/disagreement, (ii) manual analysis doesn't enable automation, and therefore it is not scalable.

To overcome these limitations, \cite{Barrett2017} recently proposed a novel automated technique based on change point detection~\citep{Eckley2011}.
The main advantage of this technique is that it provides a more rigorous approach compared to the Georges~\etal's simple heuristic, while still enabling a fully automated process (unlike Kalibera and Jones' approach).
The Barrett~\etal's technique leverages a standard change point detection algorithm, namely PELT~\citep{Killick2012}, to determine shifts in benchmark execution time. The identified shifts (\ie change points) are then post-processed (\eg by removing negligible performance shifts) to determine \emph{if}/\emph{when} a benchmark reaches a steady state of performance.
To the best of our knowledge, this technique currently represents the state-of-the-art for steady state detection.

\figref{fig:steady_example} and \figref{fig:non_steady_example} show two examples of benchmark executions along with the performance shifts identified by the PELT algorithm. The former consistently reaches a steady state of performance, while the latter doesn't.

\subsection{Java Microbenchmark Harness (JMH)}\label{sec:jmh}

JMH is the de-facto standard framework for writing and executing microbenchmarks for Java software.
It enables software developers to easily develop and execute microbenchmarks that measure fine-grained performance of specific units of Java code (\eg methods).
JMH supports steady state performance assessment by providing facilities that enable developers to statically \emph{configure} the number of times each benchmark execution will be repeated (without compromising the reliability of results\footnote{JMH code samples. Pitfalls of using loops in Java microbenchmarking. \url{https://bit.ly/3lCqCZ4}}).

\begin{figure}
    \centering
        \includegraphics[width=8cm]{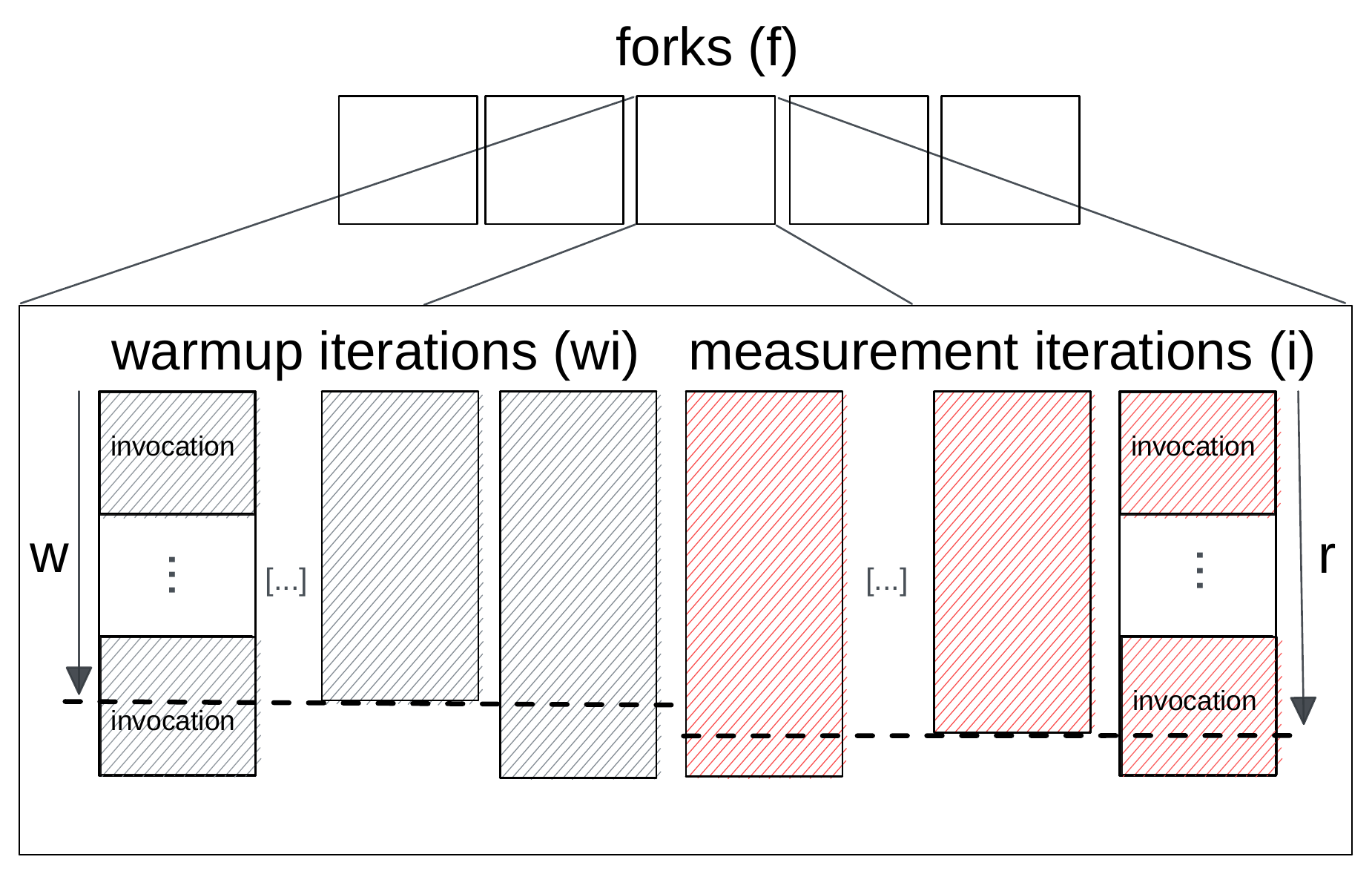}
    \caption{\label{fig:jmh} The JMH microbenchmark life cycle.}
\end{figure}

\figref{fig:jmh} depicts a typical JMH benchmark execution.
JMH supports three different levels of repetitions: \emph{forks}, \emph{iterations} and \emph{invocations}.
\emph{Invocations} (\ie the lower level of repetition) are nominal benchmark executions that are continuously performed within a predefined amount of time, namely an \emph{iteration}.
In turn, a \emph{fork} is constituted by a sequence of \emph{iterations} performed on a fully clear instantiation of new JVM.
Indeed, as suggested by best practices \citep{Georges2007, Kalibera2013, Barrett2017, Costa2019}, iterations should be repeated multiple times on fresh JVM instantiations (\ie forks) to mitigate the contextual effects of confounding factors.

Each fork is usually composed by two distinct types of iterations: \emph{warmup} and \emph{measurement iterations}.
\emph{Warmup iterations} are intended to bring the fork (\ie the fresh JVM) into a steady state of performance, while \emph{measurement iterations} are the ones where performance measurements are actually collected.
Each measurement iteration typically returns a set of performance measurements (\eg a sample of benchmark invocation execution times) or a performance statistic (\eg average execution time or throughput). 

JMH provides a set of configuration parameters to define the different levels of repetitions involved during microbenchmarking. These parameters include: warmup iteration time $w$, measurement iteration time $r$, warmup iterations $wi$, measurement iterations $i$, and forks $f$.
Iteration time parameters ($w$ and $r$)  define the minimum time spent within an iteration.
Given an iteration time $w$ (resp. $r$), a warmup (resp. measurement) iteration will continuously perform benchmark \emph{invocations} until the iteration time will expire.
Warmup and measurement iterations (\ie $wi$ and $i$), instead, define the number of iterations performed within each fork. Finally, the fork parameter $f$ defines the number of fresh JVM instantiation, \ie the higher level of benchmark repetition.

Typically, Java developers directly set JMH configuration parameters on benchmark code through Java annotations.
Nonetheless, when launching the benchmark, JMH allows to override developer configurations via Command Line Interface (CLI) arguments.

\subsection{Dynamic Reconfiguration} \label{sec:dyn-reconfig}

JMH allows to statically define the expected length of the warmup phase using configuration parameters, such as warmup iteration time $w$ and warmup iterations $wi$.
Such estimation is typically performed on the basis of developer expertise and/or benchmark nature.
Previous studies have shown that a static definition of the warmup time can be quite detrimental for steady state performance assessment~\citep{Georges2007, Kalibera2013, Barrett2017}.
An alternative to JMH static configuration can be found in a recent approach called dynamic reconfiguration~\citep{Laaber2020}.
This approach is able to determine, during a JMH benchmark execution, whether the measurements appear to be stable, and more executions are unlikely to improve their accuracy.
The rationale behind dynamic reconfiguration is that, by using automated stability criteria to halt the benchmark execution, developers can save some of the time dedicated to performance testing while keeping an acceptable level of accuracy.
To achieve this, the dynamic reconfiguration approach uses a sliding window to compare the last iterations to a stability criterion.
\cite{Laaber2020} propose and evaluate three stability criteria to dynamically estimate the end of the warmup phase:
\begin{itemize}
    \item \emph{Coefficient of variation (CV)\footnote{It is worth to notice that, although both the approaches of \cite{Laaber2020} and \cite{Georges2007} are based on coefficient of variation (CV), they have some relevant differences. Indeed, Georges~\etal's heuristic uses a fixed threshold on CV, while Laaber~\etal's approach computes the difference between the minimum and the maximum CV in a sliding window of iterations, and it determines whether this difference exceeds a predefined threshold. Interestingly, \cite{Laaber2020} reported that the usage of the Georges~\etal's heuristic to dynamically estimate the end of the warmup is unrealistic for JMH microbenchmarks. (Indeed, in our experimental setup, it stops warmup iterations in only 46.7\% of forks, when using a measurements window of 30, and a CV threshold of 0.02).}}: CV is the ratio of the standard deviation to the mean, and it can be used to compare normally distributed data. Even when data is not normally distributed, as it is often the case for benchmark data, CV can still provide an estimate of measurement variability. A fork is considered stable when the difference between the largest and the smallest value of CV computed on the sliding window is within a fixed threshold.
    \item \emph{Relative confidence interval width (RCIW)}: in this case, the variability in measurement data is estimated using a technique by \cite{Kalibera2013,kalibera2020} that employs hierarchical bootstrapping to compute the RCIW for the mean. The hierarchical levels are invocations, iterations, and forks.
    \item \emph{Kullback-Leibler divergence (KLD)}: a technique described by \cite{He2019} to compute the probability that two distributions are similar based on the Kullback-Leibler divergence (KLD)~\citep{kullback1951}. In this case, the first distribution contains all the measurements in the sliding window excluding the last one, while the second distribution includes also the last measurement. As a consequence, stability is reached when the mean of the computed similarity probabilities is above some threshold.
\end{itemize}

In order to further reduce benchmark execution time, dynamic reconfiguration leverages the same stability criteria also to dynamically determine whether to execute the next fork or not.
\newpage
\section{Research Questions} \label{sec:rqs}
In this work we aim to answer the following Research Questions (RQs):
\begin{enumerate}
	\item[RQ$_1$] \emph{Do Java microbenchmarks reach a steady state of performance?}
	\item[RQ$_2$] \emph{How does steady state impact microbenchmark performance?}
	\item[RQ$_3$] \emph{How effective are developer configurations in assessing Java steady state performance?}
	\item[RQ$_4$] \emph{How effective is dynamic reconfiguration in assessing Java steady state performance?}
	\item[RQ$_5$] \emph{Does dynamic reconfiguration provide more effective warmup estimates than developers do?}
\end{enumerate}

In the following subsections, we discuss in detail the motivation for each of the RQs and the methodology used to gather the answers. In \secref{sec:design}, we describe the experimental setup along with the benchmarks used in our empirical study.

\subsection{RQ$_1$ - Steady state assessment}
With this research question, we aim to evaluate whether Java microbenchmarks reach a steady state of performance.
We first use a state-of-the-art steady state detection technique~\citep{Barrett2017} to determine whether and when each fork reaches a steady state of performance (see \secref{sec:exp_steadystate} for details).
Based on these results, we then classify each benchmark as either (i) \emph{steady state}, if all the forks reach a steady state of performance, (ii) \emph{no steady state}, if none of the forks reach a steady state or (iii) \emph{inconsistent}, if the execution involves both \emph{steady state} and \emph{no steady state} forks.

We report the classification shares for both benchmarks and forks.
Additionally, we report the percentages of \emph{no steady state} forks for each benchmark.

\subsection{RQ$_2$ - Steady state impact}\label{sec:rqnew}
With this research question, we want to investigate to what extent the attainment of a steady state impacts benchmark performance. To do so, we compare the measurements collected during steady state phases against those collected in non-steady phases of benchmark execution. We perform two different analyses: the first one investigates the difference between steady and non-steady performance within the same fork, the second one assesses the same aspect across different forks.
Besides these two analyses, we also investigate two potential countermeasures to mitigate performance deviations in non-steady phases of benchmark execution.

All the aforementioned analyses involve a comparison between a set of steady measurements $M^{stable}$ and a set of non-steady measurements $M^{unstable}$. In order to assess to what extent non-steady measurements differ from steady measurements,  we use \emph{relative performance deviation} ($RPD$).

In the following, we first explain the process we use to compute $RPD$. Then, we describe in the detail the four analyses we use to gather the answer.  
\subsubsection*{Relative Performance Deviation}
In order to quantify the relative performance deviation of $M^{unstable}$ compared to $M^{stable}$, we use the technique proposed by \cite{Kalibera2013}.
The main benefit of this technique is that it provides a clear and rigorous account of the relative performance change and the uncertainty involved. For example, it can indicate that a set of execution time measurements is higher than another by $X\% \pm  Y\%$ with 95\% confidence. Following the guidelines of \cite{Kalibera2013,kalibera2020}, we build each confidence interval using bootstrapping with random re-sampling and replacement \citep{davison1997}, with a confidence level of 95\%.
We run 10,000 bootstrap iterations. At each iteration, new realizations $\hat{M}^{unstable}$ and $\hat{M}^{stable}$ (respectively, of $M^{unstable}$ and $M^{stable}$ measurements) are simulated and the relative performance change is computed.
The simulation of the $\hat{M}^{unstable}$ new realizations randomly selects a subset of real data from $M^{unstable}$ with replacement. Similarly, $\hat{M}^{stable}$ is simulated by randomly sampling $M^{stable}$. The two means ($\mu^{unstable}$ and $\mu^{stable}$) and the relative performance change ($\rho$) for simulated measurements are computed as follows:

{
\begin{equation*}
	\mu^{unstable}=\frac{\sum_{i=1}^{n}{\hat{M}^{unstable}_{i}}}{n}
\end{equation*}

\begin{equation*}
		\mu^{stable}=\frac{\sum_{i=1}^{m}{\hat{M}^{stable}_{i}}}{m}
\end{equation*}

\begin{equation*}	
		\rho=\frac{\mu^{unstable}-\mu^{stable}}{\mu^{stable}}
\end{equation*}
where $n$ is the number of measurements in $\hat{M}^{unstable}$, $m$ the number of measurements in $\hat{M}^{stable}$.
After the termination of all iterations, we collect a set of simulated realizations of the relative performance change $P=\{\rho_i \mid 1\leq i \leq 10,000\}$ and estimate the 0.025 and 0.975 quantiles on it, for a 95\% confidence interval.
We consider a relative performance change as statistically significant if the confidence interval does not contain 0.
For example, given a confidence interval of $(0.05, 0.07)$, we can say that the mean execution time in $M^{unstable}$ is higher than the one in $M^{stable}$ with a relative performance change that ranges between 5\% and 7\% with 95\% confidence.

We leverage the confidence interval of the mean relative performance change $(lb, ub)$ to compute the \emph{relative performance deviation}:

\begin{equation*}
 RPD=\begin{cases}
			\text{~~~\small$0$} & \text{\small if \xspace $lb \leq 0 \leq ub$} \\
    		\text{\large$|\frac{lb + ub}{2}|$} & \text{\small otherwise}
  		\end{cases}
\end{equation*}

In other words, we define $RPD$ as the center of the 95\% confidence interval of the mean relative performance change, if the interval doesn't contain zero. On the other hand, $RPD$ evaluates to 0 if the confidence interval does not report a statistically significant performance change, \ie the interval does contain 0.
Higher values of $RPD$ indicate that $M^{unstable}$ strongly deviates from $M^{stable}$.

\subsubsection*{Analyses}
\begin{figure}
  \centering
 \begin{subfigure}{12cm}
    \includegraphics[width=12cm]{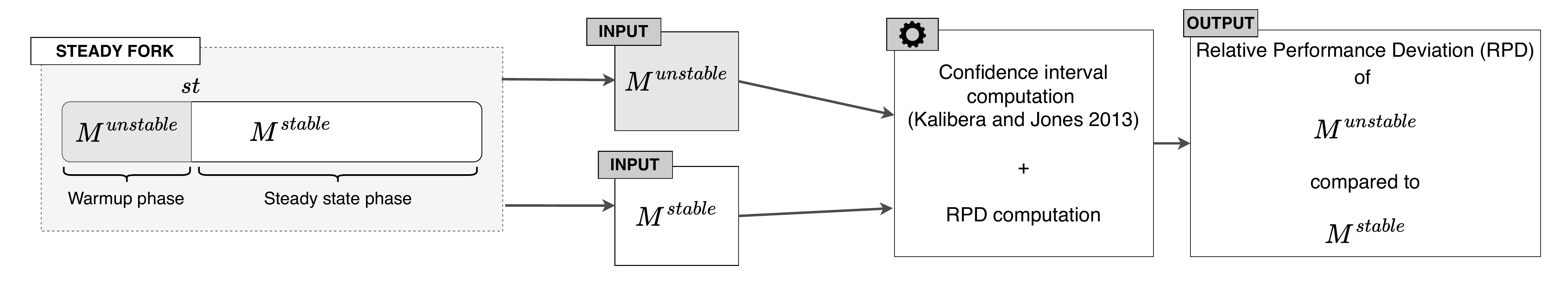}
    \caption{\label{fig:rq2_1} Performance deviation within fork.}
 \end{subfigure}
 \hfill
 \begin{subfigure}{12cm}
    \includegraphics[width=12cm]{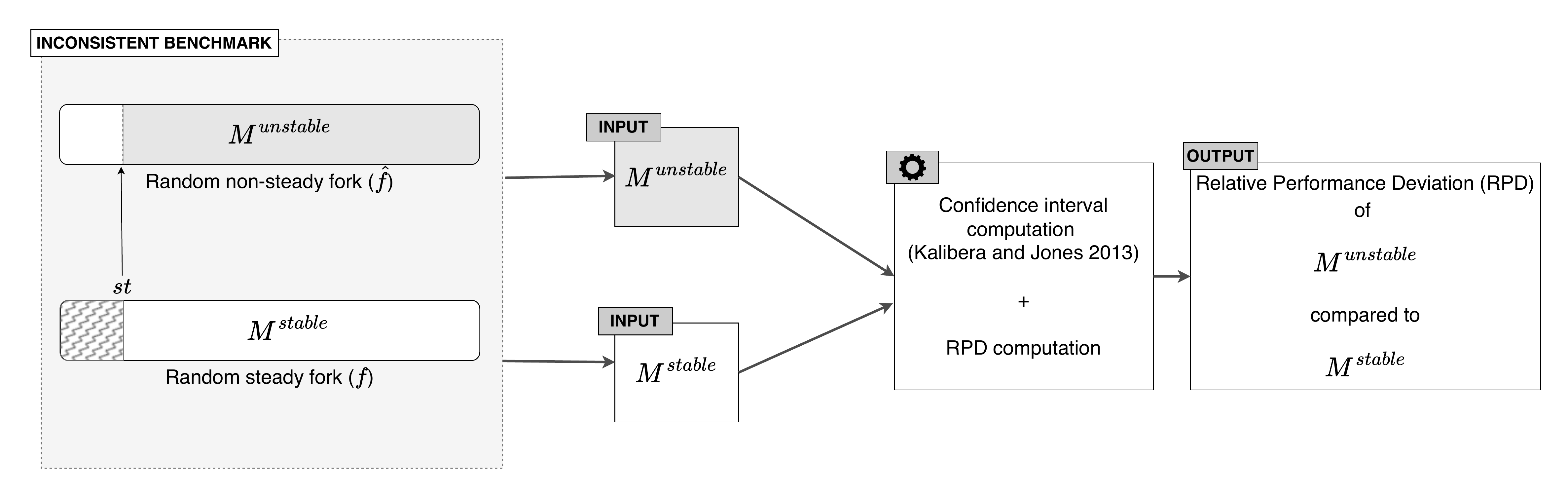}
    \caption{\label{fig:rq2_2} Performance deviation across forks.}
 \end{subfigure}
 \hfill
 \begin{subfigure}{12cm}
    \includegraphics[width=12cm]{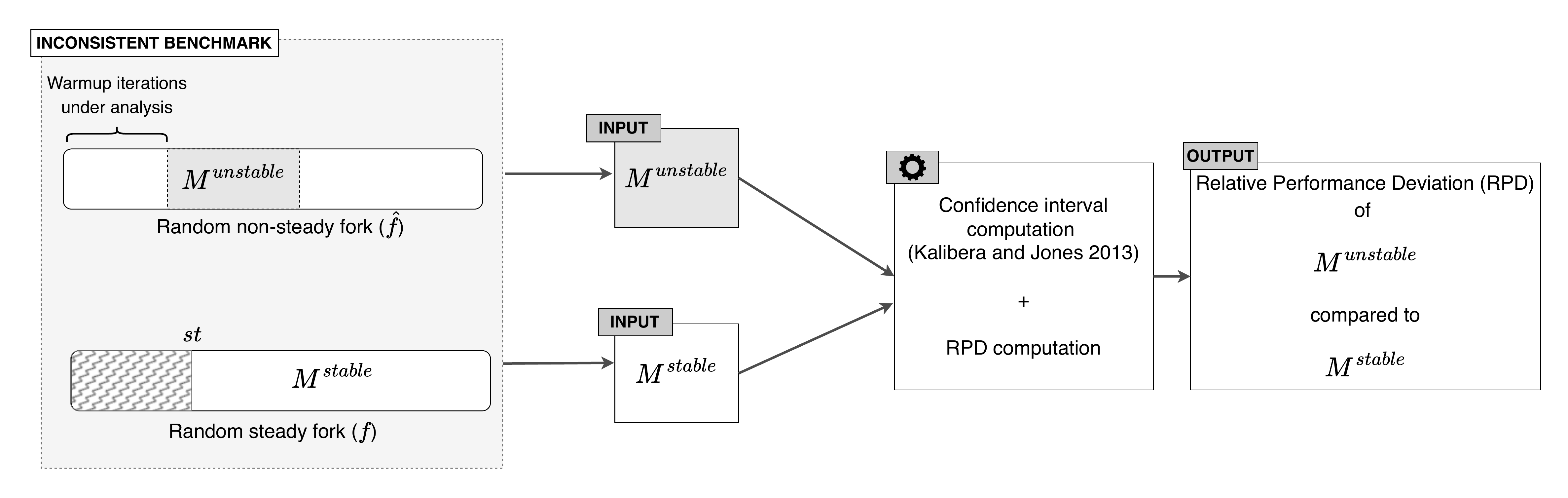}
    \caption{\label{fig:rq2_3} Influence of warmup iterations on performance deviation.}
 \end{subfigure}
 \hfill
 
 \begin{subfigure}{12cm}
    \includegraphics[width=12cm]{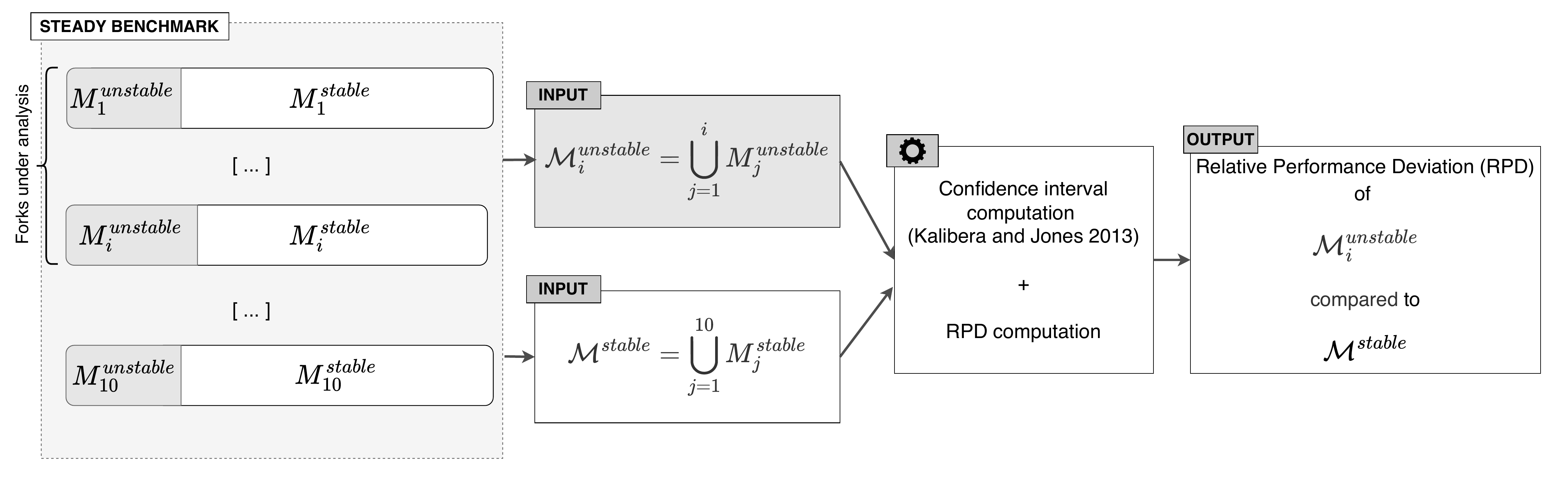}
    \caption{\label{fig:rq2_4} Influence of forks on performance deviation.}
 \end{subfigure}
 \caption{\label{fig:rq2_analyses} RQ$_2$ analyses.}
\end{figure}

In order to analyze performance deviation within forks, we consider only forks that have reached a steady state of performance, and we analyze how performance changes when the steady state is reached. In particular, we partition the set of measurements of each fork in two distinct sets, namely $M^{stable}$ and $M^{unstable}$, and we compare them to quantify the relative performance deviation ($RPD$). $M^{stable}$ contains the measurements collected during steady state execution, \ie those gathered after the \emph{steady state starting time} $st$, while $M^{unstable}$ contains measurements collected before $st$.
\figref{fig:rq2_1} provides a graphical representation of this process for a given steady fork.

To investigate performance deviation across forks, instead, we assess how performance differs between steady and non-steady forks. To do that, we  exclusively consider inconsistent benchmarks (\ie the ones that contain both steady and non-steady forks), and we randomly pick from each benchmark a pair of forks ($f$, $\hat{f}$): one that has reached a steady state of performance ($f$) and one that does not ($\hat{f}$). We then use each pair to compare the measurements collected in the steady fork ($M^{stable}$) against those collected in the non-steady fork ($M^{unstable}$), and we quantify $RPD$.
In order to enable a fair comparison, we use the same measurement window for both steady and non-steady forks. Namely, given a pair of forks ($f$, $\hat{f}$),  we define $M^{unstable}$ (resp., $M^{stable}$) as the set of measurements collected in $\hat{f}$ (resp., $f$) after $st$, where $st$ denotes the steady starting time of the steady fork $f$.
\figref{fig:rq2_2} depicts this process for one inconsistent benchmark.

Besides the aforementioned analyses, we investigate potential countermeasures to mitigate performance deviations of non-steady measurements.
In particular, we analyze two specific aspects: number of warmup iterations and number of forks.
With our analysis on warmup iterations, we aim to tackle performance deviations of non-steady forks, \ie forks that never reach a steady state of performance. In particular, we want to assess to what extent an increase in the number of warmup iterations can mitigate performance deviations of non-steady forks.
With our analysis on forks, instead, we aim to tackle deviations of non-steady measurements in forks that consistently reach a steady state of performance. That is, we investigate to what extent an increase in the number of forks mitigates performance deviations of measurements gathered during non-steady phases of benchmark execution.

In order to investigate the impact of warmup iterations, we first randomly sample one steady fork $f$ and one non-steady fork $\hat{f}$ from each inconsistent benchmark. Then, for each pair ($f$, $\hat{f}$), we use a sliding window of 50 consecutive measurements ($M^{unstable}$) in $\hat{f}$, and we assess the deviation from the set of steady state measurements collected in $f$, namely $M^{stable}$. In particular, we partition the sequence of measurements gathered from $\hat{f}$ in 60 segments of equal size (\ie 50 measurements), and we compute the relative performance deviation of each segment from $M^{stable}$. By doing so, we can assess whether increasing the number of warmup iterations can mitigate performance deviations of non-steady forks.
\figref{fig:rq2_3} shows the process used to compute the $RPD$ for one specific segment ($M^{unstable}$) of $\hat{f}$, \ie for a specific number of warmup iterations.

In order to investigate the influence of forks, instead, we assess how performance deviation of non-steady measurements changes when using different numbers of forks. Given the goal of this analysis, we consider only steady benchmarks, \ie benchmarks that exclusively involve steady forks. For each benchmark, we progressively increment the number of considered forks and, at each increment $i$, we form a set $\mathcal{M}^{unstable}_{i}$ composed by all the non-steady measurements gathered from these forks. Then, we compute, for each set $\mathcal{M}^{unstable}_{i}$, the deviation from the entire set of steady measurements gathered from all the forks ($\mathcal{M}^{stable}$), as showed in \figref{fig:rq2_4}.
In particular, we start by comparing the set of non-steady measurements collected from the first fork ($\mathcal{M}^{unstable}_{1}$) to the whole set of steady measurements $\mathcal{M}^{stable}$. Subsequently, we consider the set of non-steady measurements from the first two forks together ($\mathcal{M}^{unstable}_{2}$) and, again, we compare it to $\mathcal{M}^{stable}$. We proceed in this way for each set of non-steady measurements $\mathcal{M}^{unstable}_{i}$, with $i$ ranging from 1 to 10.
At the end of this process, we obtain one result (\ie relative performance deviation) for each pair ($b$, $i$), where $b$ denotes a steady benchmark and $i$ denotes the number of considered forks.
Through this analysis, we can assess if/how the increases in the number of forks mitigate the inherent deviations of non-steady measurements. It is worth to notice that we perform this analysis in an extreme situation, \ie by considering exclusively non-steady measurements.
We decided to do so because, if can we demonstrate that the increases in the number of forks can mitigate performance deviations in such an extreme case, then there are strong indications that they can effectively mitigate the impact of non-steady measurements.

\subsection{RQ$_3$ - Developer configuration assessment}\label{sec:rq2}
With this research question, we aim to evaluate the effectiveness of developer configurations for steady state performance assessment.
To do so, we assess how well software developers capture \emph{steady state starting time} (\emph{st}),  \ie the execution time required to reach a steady state of performance in a fork. 
Specifically, in each fork, we compare the \emph{estimated warmup time} (\emph{wt}) defined by software developers with the \emph{steady state starting time} \emph{st} detected by the technique of \cite{Barrett2017}.
We measure the error of $wt$ using \emph{warmup estimation error} ($WEE$), \ie how far $wt$ is to the steady state starting time $st$:
\begin{equation*}
	WEE = |wt - st|
\end{equation*}

Additionally, we report the proportion of underestimated and overestimated forks (\ie \emph{wt} is respectively smaller or larger than \emph{st} by at least 5 seconds), and we investigate potential side effects.

An overestimated warmup phase wastes execution time, since it leads to a surplus of warmup iterations, which, in turn, delay the beginning of measurement iterations.
This inevitably increases benchmark execution time and causes potentially harmful practical implications.
For example, the ``time effort'' of a performance test is often considered a critical factor when selecting the tests to execute before a software release \citep{traini2021,DBLP:conf/icsm/ChenS17}.
In that, an overestimated warmup phase can hamper the adoption of microbenchmarks for continuous performance assurance, especially when software releases happen frequently \citep{Rubin2016}, and tests are executed as part of a Continuous Integration (CI) pipeline \citep{Fowler2006}.
To investigate side effects of overestimated warmup time, we use \emph{time waste}, \ie the overestimation error induced by developer configurations.
We quantify \emph{time waste} through the difference (in terms of time) between \emph{wt} and \emph{st}.
In other words, \emph{time waste} represents the execution time that can be potentially saved in a fork.

On the other hand, an underestimated warmup phase may easily mislead steady state performance assessment.
Indeed, unstable measurements (\ie measurements gathered before $st$) may distort performance results, thereby leading to potentially wrong conclusions \citep{Georges2007, Kalibera2013}.
We assess the impact of underestimated warmup time using \emph{ relative performance deviation} ($RPD$), \ie the magnitude of performance deviation compared to steady state measurements.
That is, for each underestimated fork, we compare performance measurements in the steady state ($M^{stable}$) with those in the measurement time window defined by software developers ($M^{conf}$) using $RPD$.
A high $RPD$ indicates that performance measurements used by software developers strongly deviate from those collected in the steady state.

Besides investigating the effectiveness of developer configurations at fork level, we also analyze their effectiveness at benchmark level, \ie across multiple forks. Through this analysis, we aim to evaluate developer configurations not only in terms of warmup and measurement iterations, but also based on the number of configured forks.
To do that, we consider the entire set of measurements gathered through developer configuration (\ie across the configured forks), and we compare it to the whole set of steady measurements gathered across all the steady forks of the benchmark. 
In particular, for each benchmark, we assess the extent to which the set $\mathcal{M}^{conf}$ of developer measurements deviate from the set $\mathcal{M}^{stable}$ of steady measurements.
In addition, we report the time effort of benchmarks based on developer configurations by computing their overall execution time.

\subsection{RQ$_4$ - Dynamic reconfiguration assessment}
With this research question, we aim to evaluate the effectiveness of dynamic reconfiguration approaches~\citep{Laaber2020} for steady state performance assessment.
These approaches dynamically determine the warmup time during benchmark execution using stability criteria.
In our evaluation, we consider three stability criteria proposed by \cite{Laaber2020}: (i) Coefficient of variation (CV), (ii) Relative confidence interval width (RCIW), and (iii) Kullback-Leibler divergence (KLD).

In order to assess the effectiveness of each dynamic reconfiguration variant (\ie stability criteria), we use the same methodology adopted in RQ$_3$.
We first report the \emph{warmup estimation error} ($WEE$) of forks, along with the proportion of underestimated and overestimated warmup time.
Then, we consider the potential side effects of wrong warmup estimation using \emph{time waste} and \emph{relative performance deviation} ($RPD$).
Finally, we investigate the effectiveness of dynamic reconfiguration at benchmark level (\ie across forks) by assessing their RPDs and the overall execution time.

\subsection{RQ$_5$ - Dynamic vs Developer configurations}
To answer RQ$_5$, we compare dynamic reconfiguration techniques against developer configurations using three evaluation metrics: (i) \emph{warmup estimation error}, (ii) \emph{estimated warmup time}, and (iii) \emph{relative performance deviation}.


\emph{Warmup estimation error} ($WEE$) measures the accuracy of $wt$, \ie how close $wt$ is to the steady state starting time $st$.
Lower values of $WEE$ indicate better estimates of the warmup time.

\emph{Estimated warmup time} ($wt$) measures the time spent to warmup a fork with respect to a specific configuration.
Higher values of $wt$ increase the time effort devoted to performance testing, and, therefore, can potentially hamper the adoption of benchmarks for continuous performance assessment.

\emph{Relative performance deviation} ($RPD$), instead, measures in each fork the magnitude of performance deviation of $M^{conf}$ compared to steady state performance measurements ($M^{stable}$).
Higher values of $RPD$ indicate that $M^{conf}$ strongly deviates from $M^{stable}$, thus implying that performance measurements determined by the configuration may potentially mislead steady state performance assessment.
As a complementary analysis, we further use $RPD$ to measure performance deviations at benchmark level. That is, we quantify (for each benchmark) the magnitude of deviation of the entire set of measurements gathered through developers/dynamic configurations ($\mathcal{M}^{conf}$ ) when compared to the whole set of steady measurements collected through the entire benchmark execution ($\mathcal{M}^{stable}$).

The results of $WEE$, $wt$, and $RPD$ are compared using the Wilcoxon Rank-Sum test~\citep{cohen2013}, which is a non-parametric test that makes no assumption about underlying data distribution, hence, raises the bar for significance for both normally and non-normally distributed data.
Additionally, a standardized non-parametric effect size measure, namely the Vargha Delaney's \vda statistic~\citep{Vargha2000}, is used to assess the effect size.
Given a dynamic reconfiguration technique $D$, \vda measures the probability of $D$ performing better than developer configurations with reference to a specific evaluation metric. \vda is computed using \eqqref{eq:vda}, where $R_1$ is the rank sum of the first data group we are comparing, and $m$ and $n$ are the numbers of observations in the first and second data sample, respectively.

\begin{equation}\label{eq:vda}
	\hat{A}_{12} = \frac{(\frac{R_1}{m} - \frac{m+1}{2})}{n}
\end{equation}

We interpret \vda using the thresholds provided by \cite{Vargha2000}.
Based on \eqqref{eq:vda}, if dynamic configurations and developer configurations are equally good, \vda$=0.5$. Respectively, \vda higher than 0.5 means that dynamic reconfiguration is more likely to produce better results. The effect size is considered small for $0.56 \leq \hat{A}_{12} < 0.64$, medium for $0.64 \leq \hat{A}_{12} < 0.71$, and large for $\hat{A}_{12} \geq 0.71$.
On the other hand, \vda smaller than 0.5 means that developer configurations provide better results.
In this case, the effect size is considered small for $0.34 > \hat{A}_{12} \leq 0.44$, medium for $0.29 > \hat{A}_{12} \leq 0.34$, and large for $\hat{A}_{12} \leq 0.29$.

In order to avoid misleading interpretations, we perform transformation on the \vda effect size~\citep{Neumann2015}, since we consider values of $WEE$ smaller than 5 seconds and $RPD$ smaller than 5\%, as negligible~\citep{222561}, and, therefore, ``equally good''.
In particular, we apply \emph{Pre-Transforming Data}~\citep{Neumann2015} by replacing each value of $WEE<5sec$ and $RPD<0.05$ with zero.
When it comes to \emph{estimated warmup time} ($wt$), no transformation is performed on the \vda effect size, since we are interested in any improvement~\citep{Neumann2015, Sarro2016}.

\section{Experimental Design} \label{sec:design}
In order to answer our research questions, we first collect performance measurements from the execution of \benchmarks microbenchmarks across \systems systems.
Then, we analyze collected measurements to determine whether and when each fork reaches a steady state of performance.
Finally, we perform post-hoc analysis on the collected measurements to assess developers and dynamic configurations.

In this section, we first describe the microbenchmarking setup we use to collect performance measurements and the benchmark subjects. Then, we describe in detail the steady state detection technique used in our empirical study. Finally, we present the process we use to extract both developer and dynamic configurations.

\subsection{Microbenchmarking setup}\label{sec:microbench}
Following the methodology used in the study of \cite{Barrett2017}, we execute each benchmark for a substantially longer time than ``usual'' JMH configurations (on average 171~times longer than developer configurations).
We perform 10 JMH forks for each benchmark (as suggested by \cite{Barrett2017}), where each fork involves an overall execution time of at least 300 seconds and 3000 benchmark invocations. 
To do so, we configure the execution of each benchmark via JMH CLI arguments.
Specifically, we configure 3000 measurement iterations (\verb|-i 3000|) and 0 warmup iterations (\verb|-wi 0|) to collect all the measurements along the fork.
Each iteration continuously executes the benchmark method for 100ms (\verb|-r 100ms|)\footnote{We chose 100ms as iteration time because this value enables us to ``replicate'' every possible configuration considered in our study (see Section 4.4 for more details on how we obtain the set performance measurements for a particular JMH configuration).}.
The number of fork is configured to 10 (\verb|-f 10|). As benchmarking mode, we use sample (\verb|-bm sample|), which returns nominal execution times for a sample of benchmark invocations within the measurement iteration.

The execution environment and external events occurring during the benchmark runs have a remarkable influence on the accuracy of results.
This is especially true when executing microbenchmarks, as they tend to measure small portions of code that may last less than a microsecond and are, therefore, more prone to be affected by even small changes in the environment.
Hence, we tried to control as many sources of variability as possible in order to obtain more reliable measurements.

We disabled Intel Turbo Boost, \ie a feature that automatically raises the CPU operating frequency when demanding tasks are running~\citep{RHEL}.
We also disabled hyper-threading, \ie a feature in modern processors that executes two threads simultaneously on the same physical core~\citep{RHEL}.
This is achieved by replicating the architectural state but sharing execution resources such as ALUs and caches.
For this reason, hyper-threading may lead to contention patterns that continuously vary during the execution.

Another potential cause of variability among repeated runs is represented by Address Space Layout Randomization (ASLR), which is a security technique to randomly arrange the address space positions at each execution.
We disabled ASLR as it may cause variability in the measurements from one fork to another.

The amount of available memory can also affect execution times.
We fixed (through the \texttt{-Xmx} flag) the total amount of heap memory available to the JVM to 8GB, because this is the most important factor affecting garbage collection performance.
In fact, the throughput of garbage collections is inversely proportional to the amount of memory available, since collections occur when memory fills up~\citep{JavaPerformance}.

A large variety of operating system events may have a noticeable impact on execution times because they increase context switching in most cases.
For this reason, we tried to keep the events that are not related to the benchmarks to a minimum.
We disabled any Unix daemon that is not strictly necessary.
We also disabled SSH logins for the entire duration of the experiments.
To further reduce context switching, we used priority scheduling and increased the \emph{niceness} of the JMH process running the benchmarks and all its children.

Finally, we ensured that the state of the system was consistent at each run by monitoring the \emph{dmesg} log and the \emph{systemd} journal for anomalies, as well as the shell environment of the process for changes in size~\citep{MytkowiczDHS09}.

The benchmarks were executed on a bare metal server running Linux Ubuntu 18.04.2 LTS on a dual Intel Xeon E5-2650v3 CPU at 2.30GHz, with a total of 40 cores and 80GiB of RAM.

\subsection{Subject Benchmarks}
\begin{table}[ht]
\scriptsize
\center
\begin{tabular}{llrr}
\toprule
      Organization &                Name &  Stars &  No. Benchmarks \\
\midrule
            apache &               arrow &   8,065 &              46 \\
             raphw &          byte-buddy &   4,311 &              39 \\
            apache &               camel &   3,771 &              35 \\
cantaloupe-project &          cantaloupe &    195 &             103 \\
        prometheus &         client\_java &   1,545 &              33 \\
             crate &               crate &   3,109 &              39 \\
           eclipse & eclipse-collections &   1,711 &            2,415 \\
             h2oai &               h2o-3 &   5,400 &              73 \\
         hazelcast &           hazelcast &   4,406 &             144 \\
      HdrHistogram &        HdrHistogram &   1,871 &              75 \\
            apache &                hive &   3,781 &            1,402 \\
            imglib &             imglib2 &    240 &              25 \\
           JCTools &             JCTools &   2,697 &             172 \\
              jdbi &                jdbi &   1,533 &              76 \\
           eclipse &       jetty.project &   3,147 &             212 \\
           jgrapht &             jgrapht &   1,927 &              91 \\
            apache &               kafka &  19,224 &            3,578 \\
           zalando &             logbook &    900 &              20 \\
            apache &      logging-log4j2 &   1,207 &             572 \\
             netty &               netty &  26,984 &            1,746 \\
          prestodb &              presto &  12,153 &            1,534 \\
        protostuff &          protostuff &   1,649 &              31 \\
             r2dbc &            r2dbc-h2 &    128 &              20 \\
           eclipse &               rdf4j &    251 &             132 \\
     RoaringBitmap &       RoaringBitmap &   2,281 &            1,620 \\
         ReactiveX &              RxJava &  44,802 &            1,302 \\
  yellowstonegames &            SquidLib &    364 &             334 \\
            apache &           tinkerpop &   1,351 &              57 \\
     eclipse-vertx &              vert.x &  12,177 &              41 \\
        openzipkin &              zipkin &  14,468 &              63 \\
\bottomrule
\end{tabular}
\caption{This table reports the name of each subject system, the Github organization, the number of Github stars, and the number of benchmarks in the performance testing suite.}
\label{tab:systems}
\end{table}

\tabref{tab:systems} reports the list of the \systems Java open source systems we use in this study.
We selected such systems because they are relatively popular (\ie they have more than 100 Github stars), have non-trivial JMH suites (\ie have at least 20 benchmarks), and span different domains (\eg application servers, logging libraries, databases).
Given the large size of the benchmark suites, we randomly sample 20 benchmarks for each system.
14 out of the 600 sampled benchmarks failed in our experimental setup\footnote{The 14 failed benchmarks are distributed as follows: 3 \texttt{cantaloupe}, 1~\texttt{jetty.project}, 4~\texttt{vert.x}, 3~\texttt{hazelcast} and 5~\texttt{jbdi}}.

Overall, in our empirical study, we assess the behavior of \benchmarks randomly sampled benchmarks across \systems Java systems.

In order to investigate the correctness of the benchmarks selected for this study, we ran the SpotJMHBugs tool by \cite{Costa2019} on the systems.
The tool was able to detect only one potential bad practice, of type LOOP, in the \texttt{netty} project.
Therefore, we consider our selection of benchmarks suitable for the study.

\subsection{Steady State Detection}\label{sec:exp_steadystate}

Detecting the end of the warmup phase, and consequently the start of the steady state, is no trivial task, as the notion of ``\emph{steady}'' resides on how much stability of results one wants to achieve during performance testing or, to put it in another way, how much variability one is willing to tolerate.
Nonetheless, any analyst who wants to study steady state performance must establish the length of the warmup phase.
In our study, we are interested in automatically detecting the length of such phase to determine whether and when a benchmark reached a steady state.

For this task, we build on the steady state detection approach proposed by \cite{Barrett2017}, which we adapted to the purposes of our study.
The approach is fully automated, and it is based on changepoint analysis~\citep{Eckley2011}, which is a statistical technique to detect shifts in timeseries data. In our experimental setup, each datapoint of the timeseries represents the average execution time within a JMH iteration, and the whole timeseries represents a fork.

\begin{figure}[htbp]
    \centering
    \includegraphics[width=0.98\linewidth]{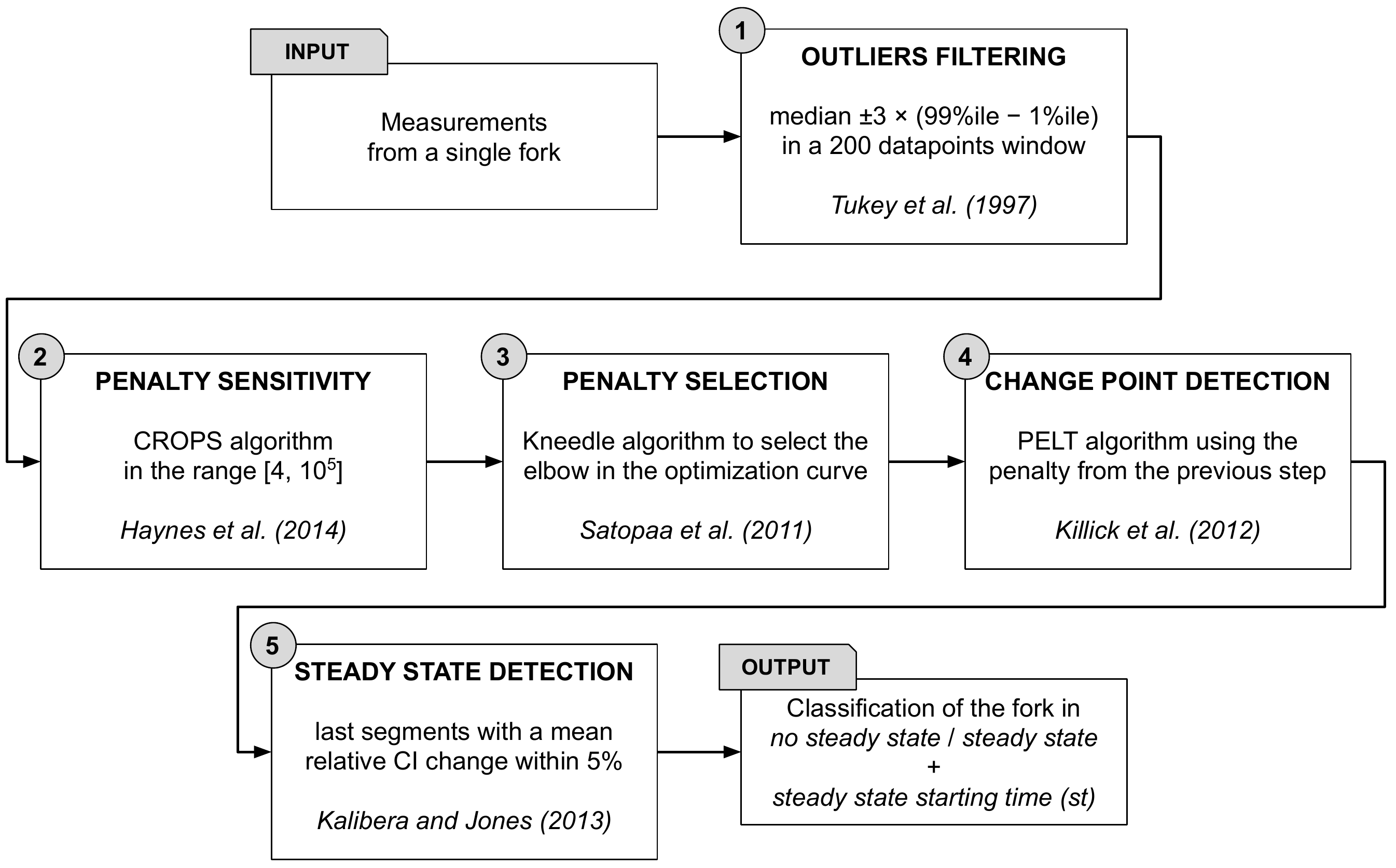}
    \caption{Steady state detection process.}%
    \label{fig:steady_state_detection}
\end{figure}

An overview of the approach can be found in \figref{fig:steady_state_detection}.
As a first step of the approach, we identify and remove potential outliers as we are only interested in detecting shifts in execution time that appear to stay consistent for a period of time. As done by \cite{Barrett2017}, we use the method by \cite{tukey1977} to identify as outliers the datapoints that lie outside the median $\pm3 \times (99\%ile - 1\%ile)$ in a 200 datapoints window \circled{1}.
Out of the $1.8 \times 10^7$ datapoints in the study, $0.27\%$ are classified as outliers, with the most of any fork being $1.3\%$.

After filtering outliers, we apply a changepoint algorithm to detect shifts in execution time.
Changepoint algorithms are designed to divide the entire timeseries into segments, within which the behavior of the timeseries is considered to remain unchanged.
On the basis of the segmentation of the timeseries we can detect if and when a benchmark execution reached a steady state.

The specific changepoint algorithm we used is called \emph{PELT}~\citep{Killick2012}.
We applied the algorithm to timeseries data gathered from individual forks in order to detect changes in both the mean and the variance of execution time.
An important parameter of the PELT algorithm is the \emph{penalty}, namely an argument designed to avoid under/over-fitting and, therefore, directly impacting the number of changepoints the algorithm will detect.
The higher the penalty value, the more difficult will be for the algorithm to detect changepoints.
Conversely, lower penalty values will result in more changepoints.
Barrett~\etal set this parameter to $15\log(n)$, where $n$ is the number of datapoints in the timeseries after discarding the outliers.
We concluded that a single penalty value for all the timeseries (\ie all the forks in the experiment) was not suitable for tuning the algorithm to the differences we found among benchmark execution data.
As a consequence, we decided to employ a method to derive an appropriate penalty value for each timeseries: we used the \emph{CROPS} algorithm~\citep{Haynes2014} to efficiently generate, for each timeseries, optimal changepoint segmentations for all penalty values in a continuous range ($[4, 10^5]$, in our case) \circled{2}.
The number of changepoints in the alternative segmentations and the corresponding penalty values can be used to derive an optimization curve (sometimes referred to as \emph{elbow diagram}).
\figref{fig:crops_example} shows an example of such a curve computed on the timeseries in \figref{fig:non_steady_example}.
As suggested by \cite{Lavielle05}, this diagram can be visually inspected to find suitably parsimonious penalty values in the area of the \emph{elbow}.
A penalty point in the elbow area can be automatically selected using the \emph{Kneedle} algorithm~\citep{SatopaaAIR11}, which is a method to find the point of maximum curvature in the continuous approximation of an optimization curve \circled{3}.
A red \emph{x} in \figref{fig:crops_example} marks the point in the curve that was chosen by the \emph{Kneedle} algorithm in that case.
Using this procedure, we were able to automatically derive a different penalty value to guide the segmentation of each fork.

\begin{figure}
    \centering
    \includegraphics[width=10cm]{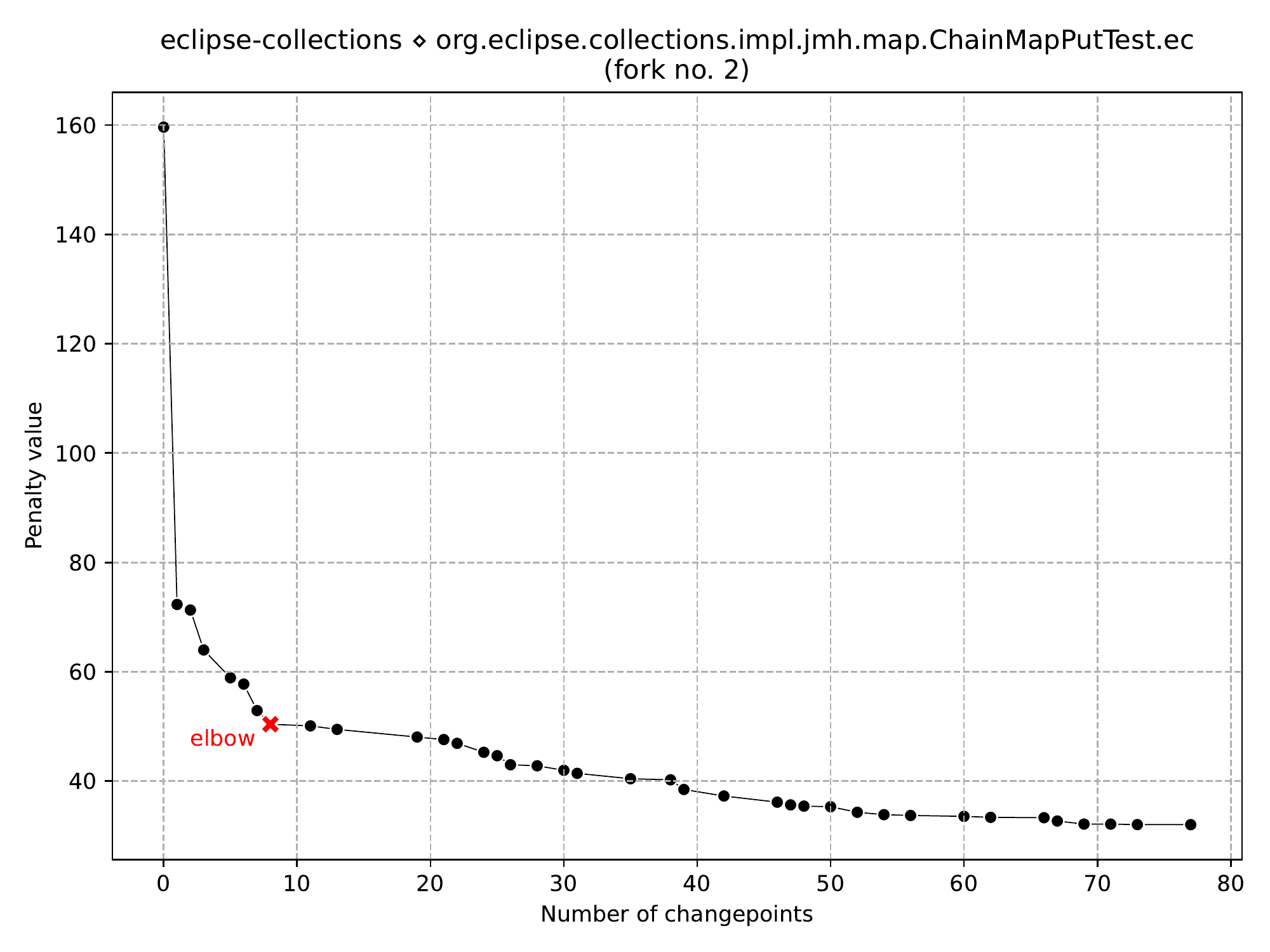}
    \caption{Example of the selection procedure of a penalty value from an elbow diagram.}
    \label{fig:crops_example}
\end{figure}

Once we obtain a segmentation of the timeseries \circled{4}, we can proceed to detect a possible steady state.
A steady state should be detected when the execution time is reasonably stable after the end of the warmup phase.
It is a matter of interpretation how much the execution time is allowed to vary but, following the approach from Barrett~\etal, we consider a fork to have reached a steady state if the last 500 measurements are contained in a single segment (\ie no changepoint was detected within the last 500 datapoints).
In general, to find when the steady state was first reached we could just take the start of the last segment.
However, this would not consider practical cases in which the smallest variation in mean or variance would generate different segments, even if, from a performance evaluation point of view, the benchmark has completed its warmup phase.
Therefore, we need to establish a suitable tolerance to allow the steady state period to span multiple segments whose variation is not meaningful to determine the execution time of the benchmark.
In Barrett~\etal the tolerance was provided by combining the maximum number of consecutive segments from the last one, such that a segment $s_i$ is equivalent to the final segment $s_f$ if mean($s_i$) is within (mean($s_f$) $\pm$ max(variance($s_f$), 0.001s)).
We did not intend to apply a fixed threshold in units of time (like 0.001 seconds) because our benchmarks vary from tens of nanoseconds to few seconds, nor we wanted to use the variance of the last segment as a threshold because it leads to an extremely low tolerance for most of the benchmarks\footnote{In 60\% of benchmarks forks, the ratio between the variance and mean of last segment is smaller than 0.0000002. Using variance as a threshold in these forks, it would imply that any negligible performance shift larger than 0.00002\%  would be considered as a meaningful performance change.}.
Hence, we preferred to compare the segments by applying a 5\% tolerance on the confidence interval for the mean relative performance change, by using the approach of \cite{Kalibera2013} \circled{5} (see \secref{sec:rq2} for details).

On the basis of this information, we can classify individual forks as \emph{steady state} if we were able to detect a steady state, or \emph{no steady state} when the opposite occurred.
Moreover, we classify a benchmark as \emph{steady state} if all its forks reached a steady state, \emph{no steady state} if all its forks did not reach a steady state, and \emph{inconsistent} if at least one fork was classified as \emph{steady state} while at least another one was classified as \emph{no steady state}.
The same information is used to derive the \emph{steady state starting time} (\emph{st}), which is the beginning of the first segment $s_i$ that is considered to be steady.
Consequently, for a given fork, the set of measurements in the range between \emph{st} and the end of the timeseries is the set of the steady state performance measurements ($M^{stable}$) of our experiment.
Based on that, given a specific benchmark, we can further define the entire set of steady measurements $\mathcal{M}^{stable}$ as the union of the steady measurements sets $M^{stable}$ collected across all the steady forks of the benchmark.

\subsection{Benchmark configurations}\label{sec:bench_conf}
Each benchmark configuration determines for each fork two relevant pieces of information: the \emph{estimated warmup time} (\emph{wt}) and a set of performance measurements $M^{conf}$.

We use this information to (i) compute warmup estimation error ($WEE$), (ii) determine whether a configuration underestimates or overestimates the \emph{steady state starting time} (\emph{st}), and (iii) assess potential side effects due to wrong estimation (\ie \emph{time waste} or \emph{relative performance deviation}).

In the following, we describe the process we use to derive \emph{wt} and $M^{conf}$ for each benchmark fork, for both developer and dynamic configurations.

\begin{algorithm}
\caption{Estimate overall warmup time}
\footnotesize
\SetKwComment{Comment}{{$\triangleright$}\ }{}
\SetCommentSty{algcommentfont}
\label{alg:estimate_wt}
 \KwData{warmup iteration time $w$, no. warmup iterations $wi$,  fork measurements $M$
}
 \KwResult{Overall warmup time $wt$}
 $wt \leftarrow 0$ \; 
 $sit\leftarrow 0$ \Comment*[r]{initialize simulated iteration time}
 $sic\leftarrow 0$ \Comment*[r]{initialize simulated iterations count}

 \For {$e$ in $M$}{
  $ni\leftarrow \lceil e/100ms\rceil$ \Comment*[r]{ compute no. benchmark invocations}
  $it\leftarrow ni\cdot e$ \Comment*[r]{ estimate time spent in the iteration }
  add $it$ to $sit$ \;
  \uIf{$sit \geq w $}{
  	add $sit$ to $wt$ \;
  	$sit \leftarrow 0$ \;
  	increment $sic$ by 1 \;
  }
  \uIf{$sic \geq wi $}{
	break \;
  }
  }
 return $wt$ \; 
 
\end{algorithm}

\begin{algorithm}
\caption{Select performance measurements}
\footnotesize
\SetKwComment{Comment}{{$\triangleright$}\ }{}
\SetCommentSty{algcommentfont}
\label{alg:select_measurements}
 \KwData{overall warmup time $wt$, measurement iteration time $r$, no.~measurement iterations $i$,  fork measurements $M$
}
 \KwResult{Selected measurements according to configuration $M^{conf}$}
 $t\leftarrow 0$ \Comment*[r]{initialize simulated time}
 $sit\leftarrow 0$ \Comment*[r]{initialize simulated iteration time}
 $sic\leftarrow 0$ \Comment*[r]{initialize simulated iterations count}
 initialize empty list $M^{conf}$ \;

 \For {$e$ in $M$}{
  $ni\leftarrow \lceil e/100ms\rceil$ \Comment*[r]{ compute no. benchmark invocations}
  $it\leftarrow ni\cdot e$ \Comment*[r]{ estimate time spent in the iteration }
  
  \uIf{$t \geq wt $}{
  	add $it$ to $sit$ \;
  	append $e$ to $M^{conf}$ \;
    \uIf{$sit \geq r $}{
  		$sit \leftarrow 0$ \;
  		increment $sic$ by 1 \;
  		
  	}
  	\uIf{$sic \geq i $}{
		break \;
  	}
  }
  add $e$ to $t$ \;
  }
 return $M^{conf}$ \; 
 
\end{algorithm}

\subsubsection{Software developer configurations}

Software developers define JMH configurations in benchmark code through Java annotations.
When a benchmark is launched, JMH executes the benchmark according to developer configurations (\eg no. measurements and warmup iterations).
A trivial approach to obtain both \emph{wt} and the set of performance measurements $M^{conf}$ for each fork would be to simply run the benchmark, and extract this information from the execution logs and JSON result files produced by JMH.
Unfortunately, this approach would be extremely expensive in terms of time, and would not fit our own needs.
Instead, we use post-hoc analysis: We first obtain the JMH configurations as defined by developers, then we use this information to compute both \emph{wt} and $M^{conf}$ based on the performance measurements collected in our microbenchmarking setup (see \secref{sec:microbench}).

In order to obtain developer configurations, we leverage a JMH feature that allows to overwrite configurations on-the-fly via CLI arguments\footnote{ We use dynamic analysis (\ie running benchmarks by overwriting CLI arguments) instead of static analysis because this methodology ensures a better coverage and lower margin of errors. Indeed, developers may rely on other mechanisms than JMH annotations to configure benchmarks, (\eg see \texttt{OptionsBuilder} at \url{https://bit.ly/3OkxzJS}). Our approach  allows to safely retrieve configurations also in these cases, while this would have been impractical through static analysis.}.
We exploit this capability to reduce benchmark execution time and speed-up developer configurations retrieval.
We first execute each benchmark twice while reducing execution time through JMH CLI arguments.
Then, we retrieve developer configurations in the JSON result files of each individual execution.
Specifically, we first obtain the number of measurement and warmup iterations (\ie $wi$ and $i$), and the number of forks $f$ by executing each benchmark while setting the measurement and warmup time of each iteration to 1 nanoseconds (\verb|-w 1ns -r 1ns|).
Then, we retrieve the measurement and warmup time (\ie $w$ and $r$) of each iteration by running each benchmark while setting the number of forks, warmup and measurement iterations to 1 (\verb|-f 1 -wi 1 -i 1|).
At the end of this process, we obtain a tuple $(w, wi, r, i, f)$, where $w$ denotes the time of a warmup iteration, $wi$ denotes the number of warmup iterations, $r$ denotes the time of a measurement iteration, $i$ the number of measurement iterations and $f$ the number of forks.
We exploit $w$, $wi$, $r$ and $i$ along with the fork measurements $M$ (as collected in our microbenchmarking setup) to compute, for each fork, the \emph{estimated warmup time} (\emph{wt}) and the set of performance measurements $M^{conf}$.
Specifically, we estimate the time spent in each warmup/measurement iteration using the average execution time of each iteration as observed in our microbenchmarking setup; then, we derive $wt$ and $M^{conf}$ based on the JMH configuration.
We report the detailed process to obtain both \emph{wt} and $M^{conf}$ in \algoref{alg:estimate_wt} and \algoref{alg:select_measurements}, respectively.

Based on the above, we can exploit the number of configured forks $f$ defined by software developers to obtain the whole set of performance measurements gathered from the entire benchmark execution, namely $\mathcal{M}^{conf}$. In other words, we derive $\mathcal{M}^{conf}$ by joining all the sets of measurements $M^{conf}$ gathered from the first $f$ forks of the benchmark.

\subsubsection{Dynamic configurations}
In order to obtain $(w, wi, r, i, f)$ for each dynamic reconfiguration variant, we leverage the replication package provided by \cite{Laaber2020}.
Specifically, we use the scripts provided for post-hoc analysis, which take as input JMH result JSON files, and return, for each fork, the number of warmup iterations ($wi$) according to stability criteria.
The warmup iteration time $w$, the measurement time $r$, and the number of measurement iterations $i$ are fixed to $w=1s$, $r=1s$ and $i=10$, respectively, according to the experimental setup defined in \citep{Laaber2020}.

We then obtain $wt$ and $M^{conf}$ using the same approach adopted for developer configurations.

\section{Results} \label{sec:results}

This section presents the results of the experiments and provides answers to the RQs formulated in \secref{sec:rqs}.

\subsection{RQ$_1$ - Steady state assessment}\label{sec:results:rq1}

As described in Section~\ref{sec:exp_steadystate}, we classify the benchmarks in our study on the basis of their ability to eventually reach a steady state.
Such classification is first performed at fork level, and then at benchmark level by combining results from the steady state detection on forks.

\begin{figure}[h]
     \center
     \begin{subfigure}[b]{5.8cm}
		\center
		\includegraphics[width=\linewidth]{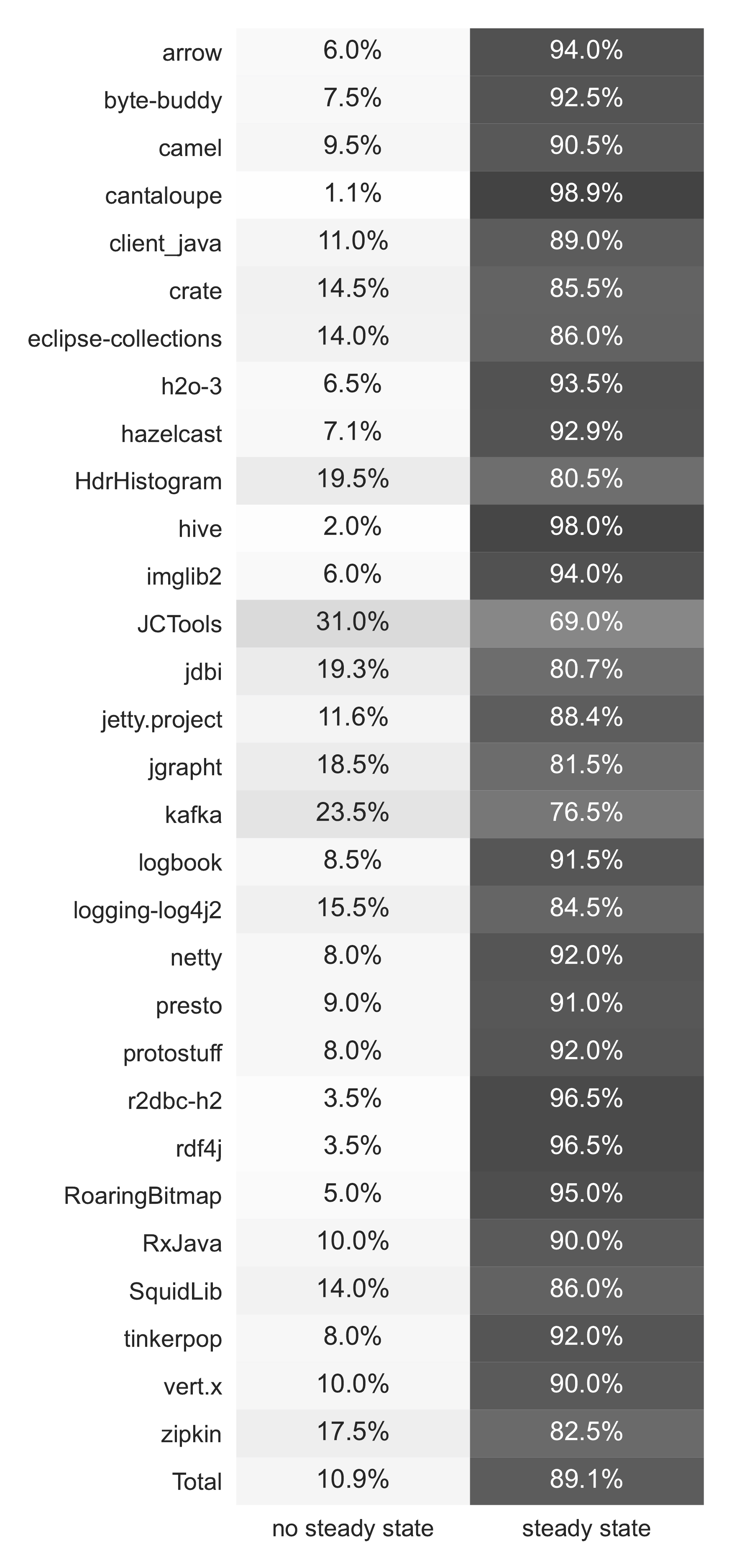}
  		\caption{Forks classification.}
  		\label{fig:rq1_fork_class}
     \end{subfigure}
     \hfill
     \begin{subfigure}[b]{5.8cm}
		\center
		\includegraphics[width=\linewidth]{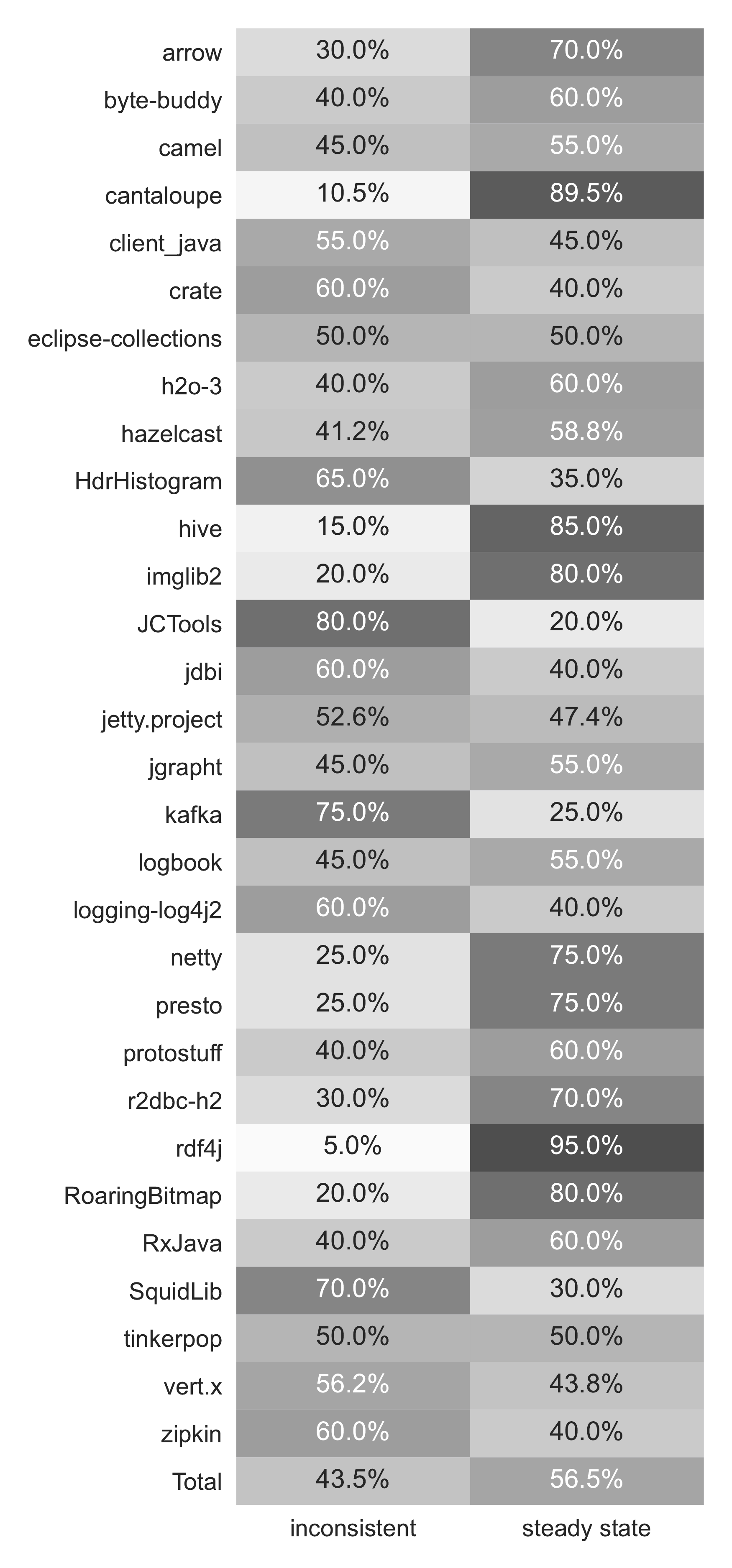}
  		\caption{Benchmarks classification.}
  		\label{fig:rq1_bench_perf}
     \end{subfigure}
     \caption{RQ$_1$. Steady state classification.}
\end{figure}

\begin{figure}
    \center
	\includegraphics[width=12cm]{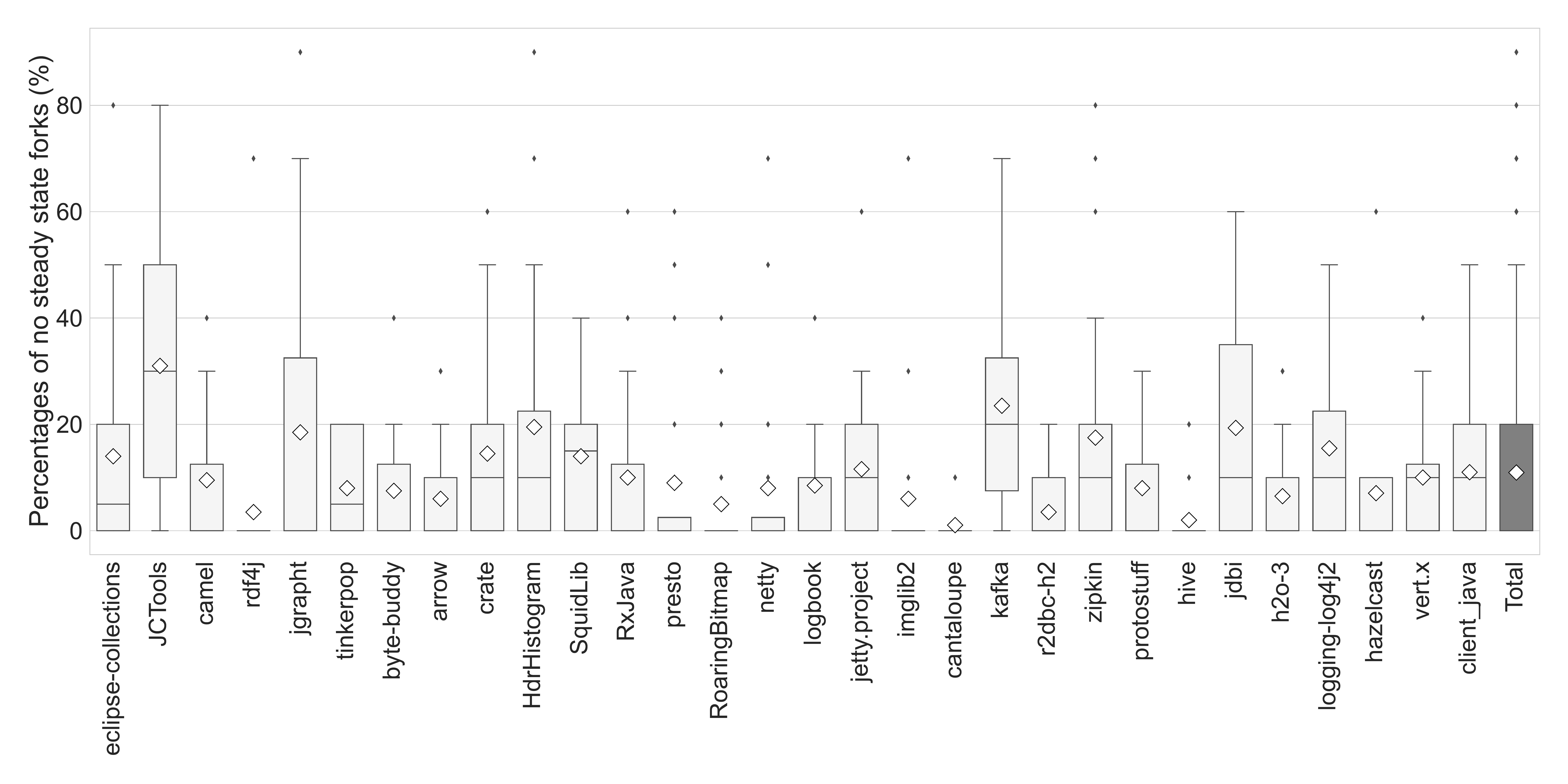}
  	\caption{RQ$_1$. Percentages of no steady state forks within each benchmark, grouped by subject system.}
  	\label{fig:perc_no_steady_state}
\end{figure}

In order to provide an overview of how many forks reached a steady state, Figure~\ref{fig:rq1_fork_class} reports the percentage of forks classified as \emph{steady state} or otherwise, as grouped by systems.
The percentage of forks that reached steady state varies between 69\% (\texttt{JCTools}) and 98.9\% (\texttt{cantaloupe}).
Even if there is some variability among systems, 28 of them, out of the 30 in our study, show a percentage of steady state forks above 80\%, with 18 of them above 90\%.
Globally, in most cases (89.1\% in the last row of Figure~\ref{fig:rq1_fork_class}), individual forks were able to reach a steady state according to our detection technique.

When we examine how the classified forks are distributed among the benchmarks, we get a less obvious outlook.
In Figure~\ref{fig:rq1_bench_perf}, we report the percentage of benchmarks classified as \emph{steady state}, or \emph{inconsistent} in the respective systems (note we do not report percentages for ``no steady state'' classification, since we didn't find any benchmark classified as such).
The first clear result is that there are no cases in which all the forks of a benchmark did not reach a steady state, since the totality of benchmarks is always distributed among the \emph{steady state} and \emph{inconsistent} columns.
On the one hand, this might encourage the assumption that, in the vast majority of cases, benchmarks reach and measure steady state performance.
On the other hand, we can assess that the percentage of benchmarks in which all the forks reached a steady state is subject to large variability depending on the specific system.
In fact, the percentage of \emph{steady state} benchmarks varies between 20\% (\texttt{JCTools}) and 95\%(\texttt{rdf4j}).
Only 5 systems overcome a 80\%  percentage and, as opposed to what one would expect, \emph{no steady state} forks are unevenly distributed among benchmarks, thus causing most systems to have a low percentage of \emph{steady state} benchmarks even though the percentage of \emph{steady state} forks was higher.
Since there are no benchmarks in which all the forks did not reach a steady state, all the remaining benchmarks are classified as \emph{inconsistent} (43.5\%), which means that their forks showed mixed behavior.

Another viewpoint on the classification of benchmarks is provided in Figure~\ref{fig:perc_no_steady_state}, where we report the distribution of the percentages of no steady state forks in benchmarks, as grouped by systems, which further clarifies what contributes to the percentages of inconsistent benchmarks.
We can notice that, in most cases (with very few exceptions like \texttt{JCTools}), the systems tend to exhibit inconsistent benchmarks with small percentages of no steady state forks.
It is worth recalling that a single fork (\ie 10\% in a benchmark) is enough to flip the classification from \emph{steady state} to \emph{inconsistent}.
This is, in fact, the most common case, as we can see in the distribution computed over all the systems (\ie \emph{Total} in Figure~\ref{fig:perc_no_steady_state}) that shows a mean around 10\%.

\begin{figure}[]
    \centering
    \includegraphics[width=.98\linewidth]{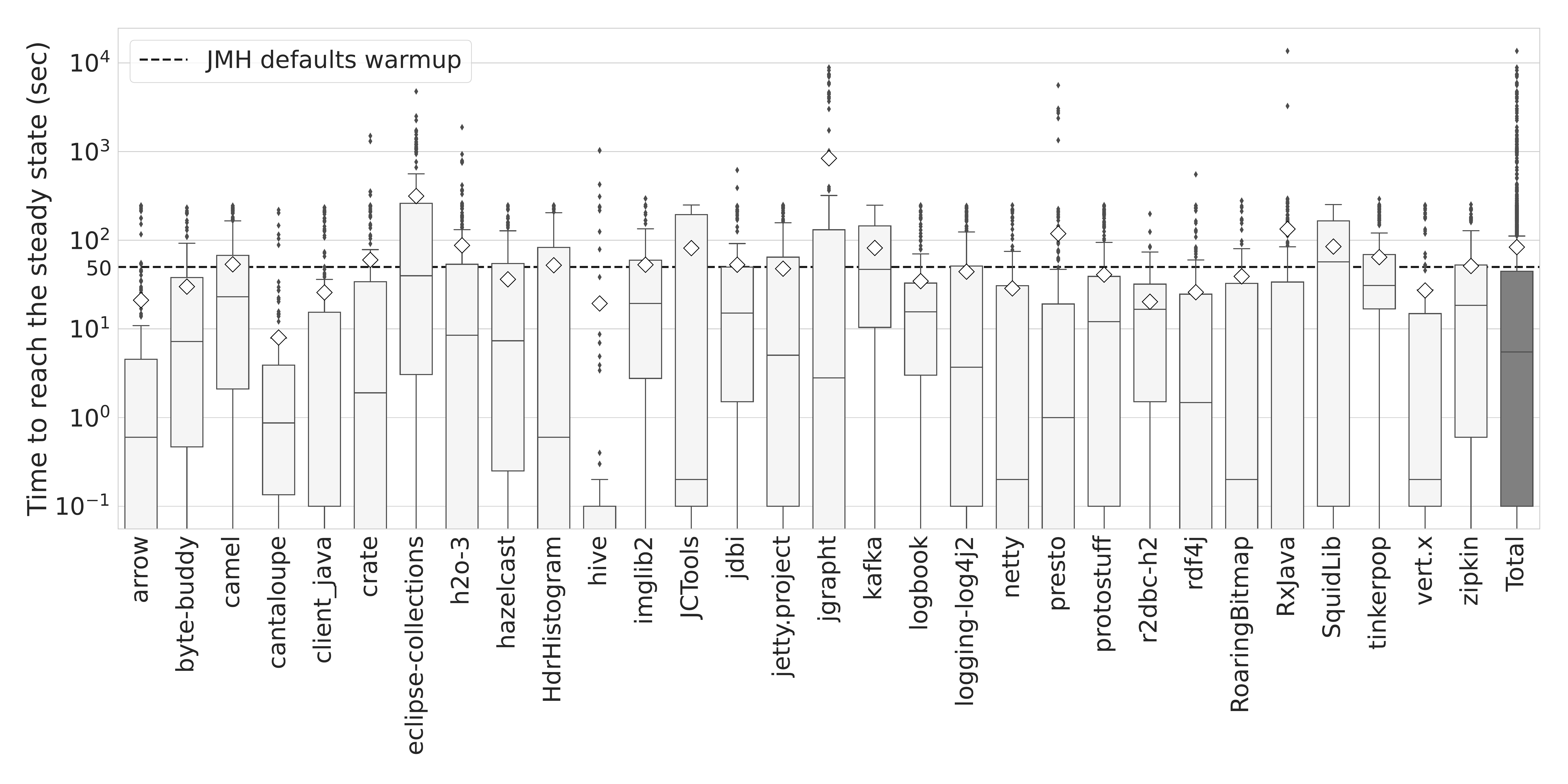}
    \caption{RQ$_1$. Time to reach the steady states, in seconds. The y-axis uses a logarithmic scale.}%
    \label{fig:steady_states_time_by_project_boxplot}
\end{figure}

From a practical perspective, it is also important to estimate how long it takes to reach a steady state, in the cases in which it is reached.
This provides a better view on how the time budget could be spent when executing the benchmarks.
\figref{fig:steady_states_time_by_project_boxplot} shows the distributions of the time to reach a steady state, as grouped by system and in total.
We can observe that the time spent considerably varies, even within a single system, therefore it can hardly be generalized.
This result is not surprising, because the attainment of a steady state inherently depends on the nature of the benchmark.
As most of the instability during the warmup is due to the JIT activity, we can imagine that, beside the size of the benchmark method itself, also the number of loaded classes plays a crucial role, since it will induce a different amount of compilation.

\figref{fig:steady_states_time_by_project_boxplot} also shows how the time spent to reach a steady state compares to the JMH default setting of 50 seconds for the warmup phase (dashed horizontal line in the figure).
We can observe that the difference between the detected end of the warmup phase and the JMH defaults largely differs from one system to the other, by extreme values for \emph{eclipse-collections} and \emph{jgrapht}.
The clear picture that emerges from this is that, in most cases, by using the JMH defaults we would overestimate the time needed to warm the benchmark up, therefore wasting a considerable amount of time dedicated to performance testing.
While the amount of time that would be wasted considerably varies from one system to another, the percentage of overestimated forks is quite consistent across all the systems.
This leads to the conclusion that, more often than not, the JMH defaults for the warmup time should not be used, rather one should rely on techniques to assess the actual amount of time a specific benchmark requires to reach a steady state.
\\

\begin{tcolorbox}[colback=black!3!white,colframe=black!33!white]
\textbf{RQ$_1$ summary} - When we only look at individual forks, measurements appear to reach a steady state in the majority of cases.
However, when combining forks at the benchmark level, we obtain mixed results.
These results provide evidence that benchmarks do not always reach a steady state of performance, thus showing, on a large corpus of JMH benchmarks, that the ``\emph{two-phase assumption}'' does not always hold.
Moreover, in most cases, the JMH defaults for the warmup time tend to overestimate the time needed to reach a steady state.
\end{tcolorbox}

\subsection{RQ$_2$ - Steady state impact}

In this subsection, we present results of our analysis on the impact of steady state on performance.

 \begin{figure}[h]
    \centering
    \includegraphics[width=4cm]{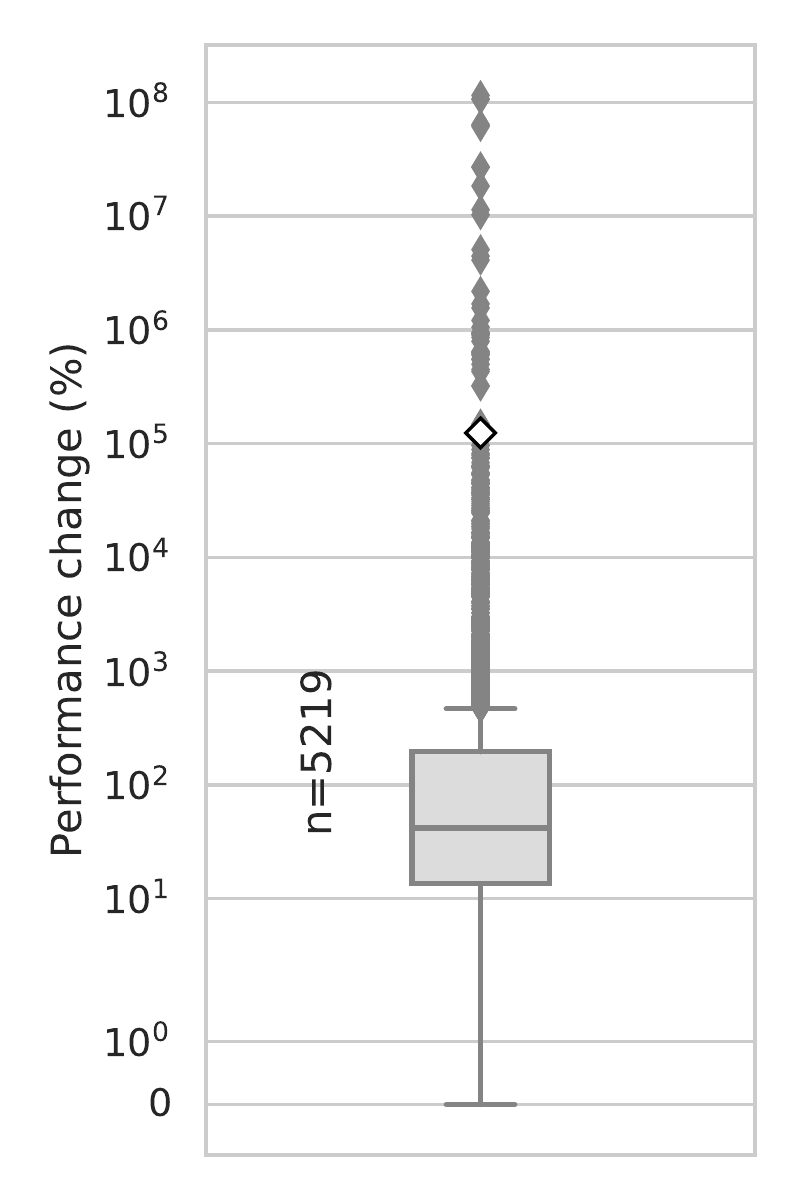}
    \caption{RQ$_2$. Steady state impact within forks. The box plot reports the distribution of RPD between steady and non-steady phases of the execution across all the steady forks. The y-axis is in logarithmic scale.}%
    \label{fig:impact-within-fork}
\end{figure}

\begin{table}[h]
\centering
\scriptsize

\begin{tabular}{llrrr}
\toprule
            project &    n &      mean &        std &  median \\
\midrule
              arrow &   188 &       599.6 &      1,814.9 &   72.1 \\
         byte-buddy &   185 &       276.1 &        469.1 &  113.4 \\
              camel &   181 &    10,277.3 &     51,821.1 &   49.7 \\
         cantaloupe &   188 &     2,237.9 &      3,534.1 &  506.4 \\
        client\_java &   178 &       270.8 &        323.2 &  274.2 \\
              crate &   171 &    20,919.6 &    125,119.1 &   26.5 \\
eclipse-collections &   172 &        70.4 &        185.0 &   17.4 \\
              h2o-3 &   187 &    42,129.6 &    342,798.8 &   18.9 \\
          hazelcast &   158 &       279.1 &        966.1 &   30.0 \\
       HdrHistogram &   161 &        74.3 &        143.7 &   25.9 \\
               hive &   196 &       386.4 &        325.0 &  344.1 \\
            imglib2 &   188 &       103.1 &        187.1 &   68.2 \\
            JCTools &   138 &        69.3 &         67.0 &   61.7 \\
               jdbi &   121 &     2,305.6 &     14,856.2 &   40.2 \\
      jetty.project &   168 &       387.0 &        649.1 &  106.8 \\
            jgrapht &   163 &        35.9 &         65.0 &   16.3 \\
              kafka &   153 &       196.9 &      2,063.9 &   16.0 \\
            logbook &   183 &       169.0 &        304.6 &   42.4 \\
     logging-log4j2 &   169 &       181.6 &        356.6 &   72.8 \\
              netty &   184 &     3,457.2 &     36,818.1 &  204.7 \\
             presto &   182 & 3,439,550.1 & 16,220,992.1 &   27.6 \\
         protostuff &   184 &       170.6 &        237.2 &   41.5 \\
           r2dbc-h2 &   193 &     3,460.6 &     10,268.1 &  303.7 \\
              rdf4j &   193 &        97.7 &        211.6 &   35.2 \\
      RoaringBitmap &   190 &        39.6 &         56.4 &   21.0 \\
             RxJava &   180 &       113.9 &        278.0 &   35.9 \\
           SquidLib &   172 &        57.1 &        118.8 &    8.2 \\
          tinkerpop &   184 &     1,330.7 &      7,177.5 &   24.9 \\
             vert.x &   144 &       654.1 &        951.8 &  207.0 \\
             zipkin &   165 &    28,604.1 &    152,763.8 &   21.6 \\
\midrule   
              Total & 5,219 &   123,972.4 &  3,087,067.7 &   41.8 \\
\bottomrule
\end{tabular}

\caption{RQ$_2$. Steady state impact within forks grouped by project. The first and second columns report, respectively, the name of the project and number of steady forks per project. The last three columns report RPD statistics (\ie mean, standard deviation and median). }
\label{tab:impact-within-fork}
\end{table}

\figref{fig:impact-within-fork} and \tabref{tab:impact-within-fork} report results of the analysis that investigates how performance changes (within each fork) when the steady state is reached. 
 In particular, \figref{fig:impact-within-fork} depicts the distribution of the \emph{relative performance deviation} (RPD) across all the steady forks of our study. The figure highlights strong performance deviations when the steady state is reached, with an average RPD of 123,937\% and a median of 41\% (IQR 14-195\%).
 
 By looking at the detailed results reported in \tabref{tab:impact-within-fork}, we can observe that the large mean is highly influenced by some specific projects (\eg \texttt{camel}, \texttt{crate}, \texttt{h2o-3} and \texttt{presto}), which report extremely high RPD (up to 3.5 billion \%). Nonetheless, even when considering projects with smaller RPD (\eg \texttt{jgrapht}, \texttt{RoaringBitmap} and \texttt{SquidLib}), we can observe considerable performance deviations between steady and non-steady measurements (respectively, 36\%, 40\% and 57\% on average).
  Besides the diversity across projects, \tabref{tab:impact-within-fork} also highlights a substantial diversity within each project. Indeed, performance deviations substantially differ across benchmarks of the same projects, as they report extremely high standard deviations (maximum standard deviation of $\sim$16 billion (\texttt{presto}), and minimum of 56 (\texttt{RoaringBitmap})).
  
  The above results suggest that performance substantially changes when forks reach a steady state of performance, and provide empirical evidence on the danger of using non-steady measurements during performance assessment.
  Indeed, given these large magnitudes, even a tiny portion of non-steady measurements can substantially distort performance indices, with significant implications on performance assessment.
   
\begin{figure}[h]
    \centering
    \includegraphics[width=4cm]{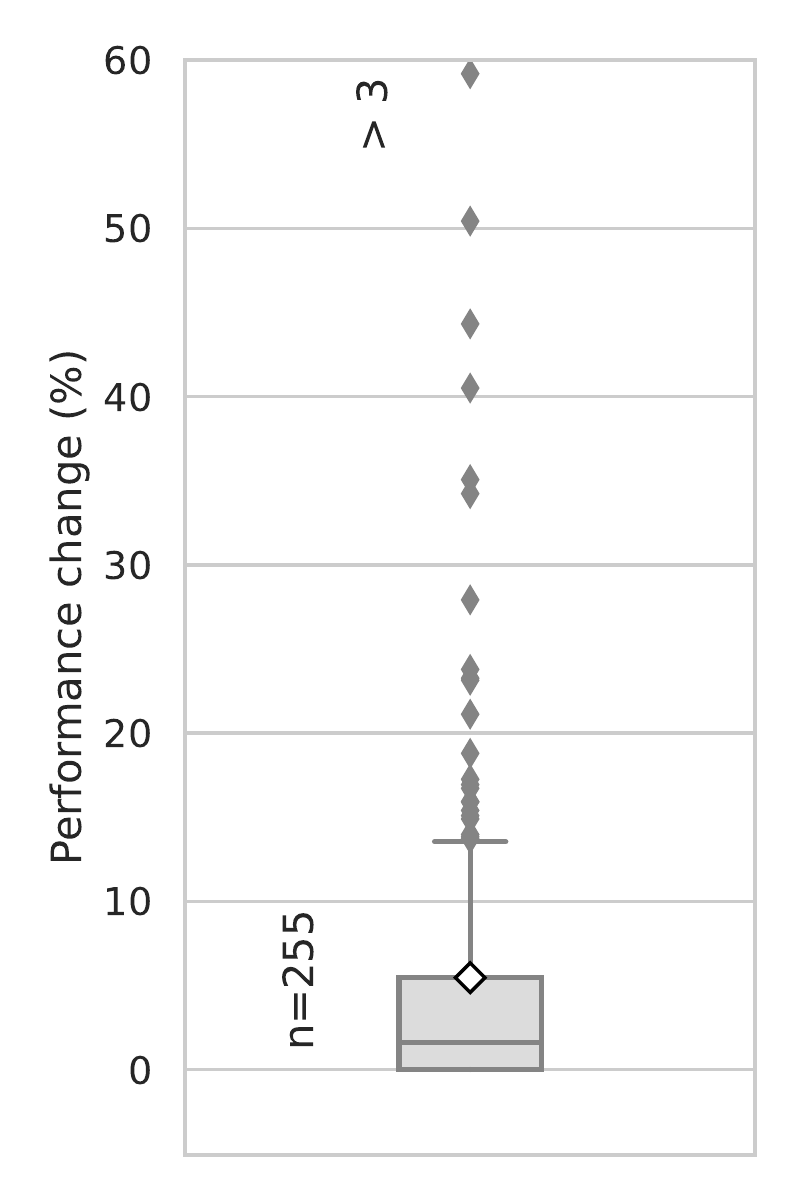}
    \caption{RQ$_2$. Steady state impact across forks. The box plot reports the distribution of RPDs between steady and non-steady forks across all the inconsistent benchmarks.}%
    \label{fig:impact-across-fork}
\end{figure}
\begin{table}[h]
\centering
\scriptsize
\begin{tabular}{llrrr}
\toprule
            project &   n &  mean &  std &  median \\
\midrule
              arrow &   6 &   8.2 & 10.2 &     3.5 \\
         byte-buddy &   8 &  14.5 & 27.0 &     0.0 \\
              camel &   9 &   2.7 &  3.9 &     1.2 \\
        client\_java &  11 &   3.4 &  4.0 &     2.8 \\
              crate &  12 &   3.3 &  3.9 &     2.4 \\
eclipse-collections &  10 &   2.5 &  4.3 &     0.2 \\
              h2o-3 &   8 &   3.7 &  4.9 &     1.7 \\
          hazelcast &   7 &   6.3 & 15.1 &     0.0 \\
       HdrHistogram &  13 &   7.3 & 19.8 &     0.0 \\
            JCTools &  16 &  19.2 & 17.7 &    13.3 \\
               jdbi &   9 &   7.4 &  7.1 &     6.0 \\
      jetty.project &  10 &   6.3 &  5.8 &     5.4 \\
            jgrapht &   9 &   6.6 &  5.7 &     4.2 \\
              kafka &  15 &   1.8 &  4.2 &     0.0 \\
            logbook &   9 &   3.2 &  4.8 &     0.7 \\
     logging-log4j2 &  12 &   4.4 &  3.8 &     3.9 \\
         protostuff &   8 &   0.6 &  1.2 &     0.0 \\
           r2dbc-h2 &   6 &   1.7 &  1.4 &     2.1 \\
             RxJava &   8 &  13.2 & 35.7 &     0.2 \\
           SquidLib &  14 &   2.1 &  2.7 &     1.4 \\
          tinkerpop &  10 &   1.6 &  1.4 &     1.6 \\
             vert.x &   9 &   1.8 &  2.9 &     0.0 \\
             zipkin &  12 &   5.5 &  5.5 &     3.7 \\
             others &  24 &   3.3 &  5.0 &     1.0 \\
\midrule
              Total & 255 &   5.5 & 11.6 &     1.6 \\
\bottomrule
\end{tabular}
\caption{RQ$_2$. Steady state impact across forks grouped by project. The first and second columns report, respectively, the name of the project and the number of inconsistent benchmarks per project. The last three columns report RPD statistics for each project (\ie mean, standard deviation and median).}
\label{tab:impact-across-fork}
\end{table}

\figref{fig:impact-across-fork} and \tabref{tab:impact-across-fork} report the results of our second analysis, which investigates performance deviations between steady and non-steady forks. By observing  \figref{fig:impact-across-fork}, we can notice that the reported performance deviations are significantly smaller than those within forks (see \figref{fig:impact-within-fork}).
 The average RPD between steady and non-steady forks is 5\%, while the median RPD is 2\% (IQR 0-5\%).
 Although these deviations may appear negligible at a first glance, they are still significant if placed in the context of Java microbenchmarking.
 Indeed, in these contexts, even relatively small performance regressions (\eg 5\%) may lead to rejections of code revisions~\footnote{As an example, see \texttt{netty} pull request \texttt{8614} at \url{https://bit.ly/33MqlMZ}.}, as they can have significant impact at system level.
 Moreover, if we look at some specific projects, such as \texttt{byte-buddy}, \texttt{JCTools} and \texttt{RxJava}, the reported deviations are even more conspicuous (average RPDs of respectively 14\%, 19\% and 13\%, see \tabref{tab:impact-across-fork}).
 Still, the deviations between steady and non-steady forks are substantially smaller than those within forks. 
  In that, it is worth to remark that we compare steady and non-steady forks using on purpose the same measurement window (\ie we discard all measurements collected before $st$, where $st$ denotes the steady starting time of the paired steady fork).
  Indeed, this methodology may potentially discard measurements that are related to the most unstable phases of the (non-steady) fork execution, and it can consequently smooth deviations from steady measurements.
Nonetheless, this fact may also suggest that performance deviations tend to improve over time during benchmark execution, even in forks that do not reach a steady state of performance.

\begin{figure}[htpb]
    \centering
    \includegraphics[width=12cm]{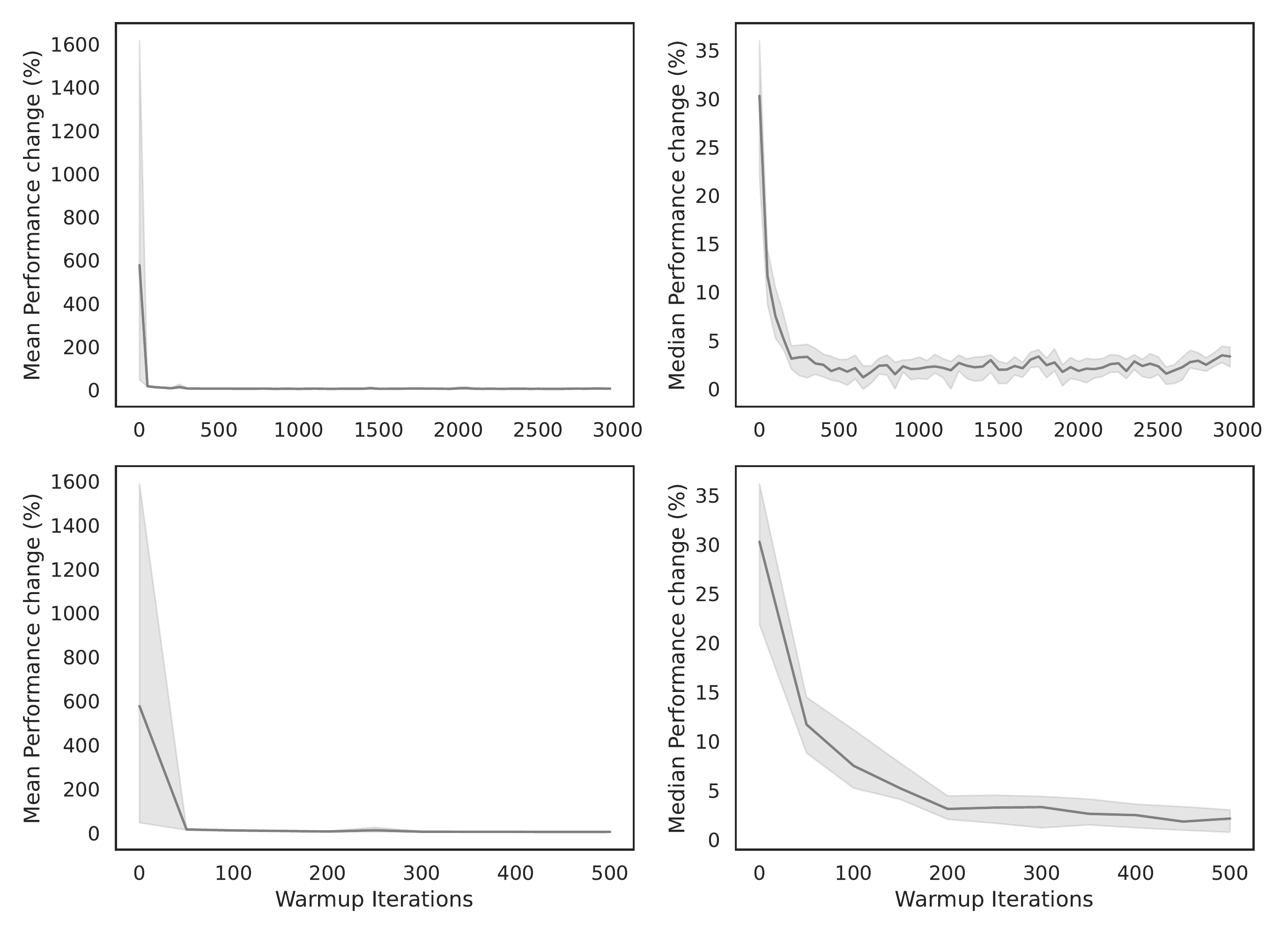}
    \caption{RQ$_2$. The impact of warmup iterations on RPD.  In each plot the x-axis denotes the number of warmup iterations, while the y-axis report RPD statistics. The dark shadow around the line represents the 95\% confidence interval. The plots in the first column report the mean RPD, while those in the second column report the median RPD.
    Plots in the first row present the overall results, while those in the second row zoom on the first 500 iterations.
     }%
    \label{fig:iteration-impact}
\end{figure}

\begin{figure}[htpb]
    \centering
    \includegraphics[width=12cm]{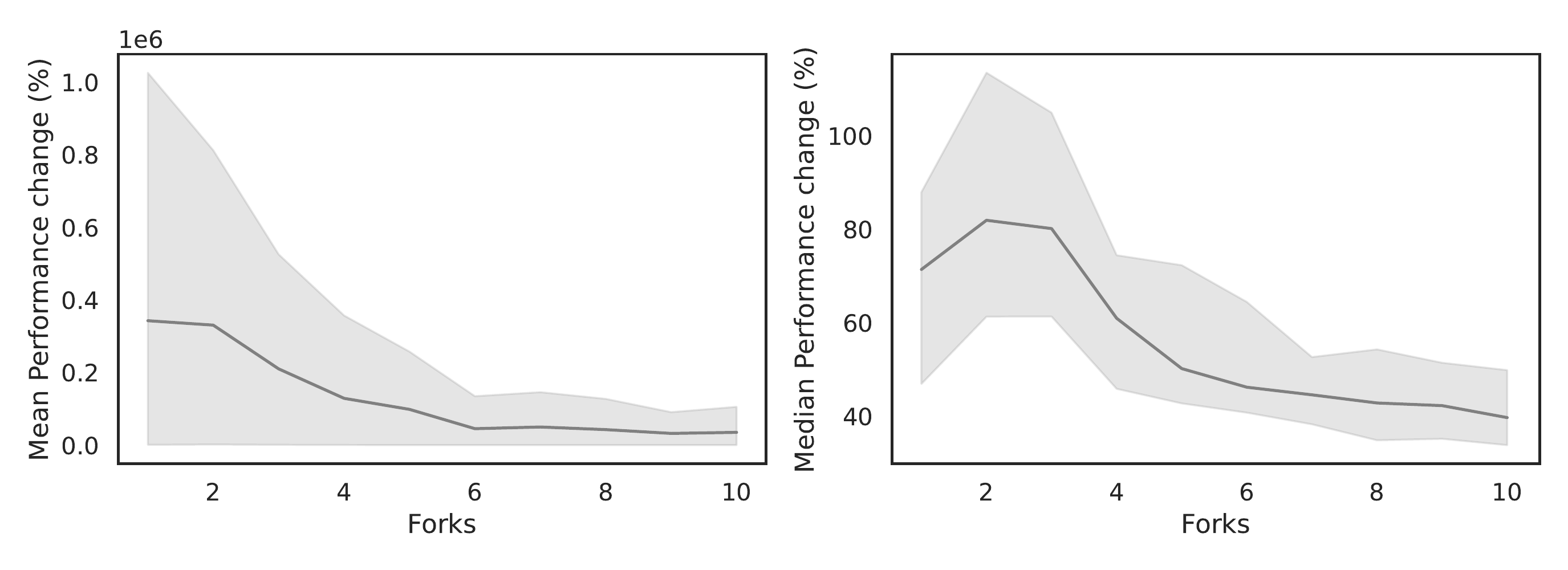}
    \caption{RQ$_2$. The impact of forks on RPD. In each plot the x-axis denotes the number of non-steady forks considered in the comparison, while the y-axis report RPD statistics. The dark shadow around the line represents the 95\% confidence interval. The left plot reports the mean RPD, while the right plot reports the median RPD.}%
    \label{fig:fork-impact}
\end{figure}

To further investigate this aspect, we assess how warmup iterations impact performance deviation. In particular, we investigate to what extent warmup iterations mitigate performance deviations (RPD) of non-steady forks. \figref{fig:iteration-impact} reports the results of our analysis. As it can be seen from the figure, RPDs seem to be considerably influenced by the number of warmup iterations, especially at the early stages of benchmark execution. Indeed, we can see a substantial drop in the first 50 iterations, where the mean RPD decreases from 578\% to 17\% and the median from 30\% to 12\%. Moreover, we found a clear trend toward RPD reduction in the first 300 iterations. After this point, the decreasing trend seems to disappear, and RPDs begin to show fluctuations with means ranging from a minimum of 6\% to a maximum of 9\%, and medians ranging from 1\% to 3\%.

The above results suggest that performance deviations of non-steady forks may be substantially mitigated through a reasonable amount of warmup iterations. In particular, a significant reduction of performance deviations can be obtained (\ie -561\% on average and -18\% on median) by using at least 50 warmup iterations, which correspond to 5 seconds of continuous benchmark execution and no less than 50 invocations. These deviations can be further reduced by using a higher number of warmup iterations up to 300 (\ie 30 seconds of continuous execution and at least 300 invocations). After this point, increasing the number of warmup iterations barely affect RPDs.

Besides investigating the impact of warmup iterations, we also analyze the impact of forks. Specifically, we study whether using a higher number of forks reduces performance deviations of non-steady measurements. As it can be observed in \figref{fig:fork-impact}, the analysis shows an overall trend toward RPD reduction when increasing the number of forks. The average RPD is reduced by each additional fork, while the median shows a swinging trend only in the first two forks, and then it constantly decreases. If we compare RPDs of individual forks to those of 5 forks, we observe a significant RPD reduction, \ie -244,237\% on the mean and -21\% on the median. This trend leads to an RPD reduction of -307,555\% (mean) and -31\% (median) when using 10 forks.
Still, the reported RPDs remain extremely high, \ie mean of 35,797\% and median of 40\%.
The latter results are not surprising if we consider that our analysis deliberately targets non-steady measurements, which are typically subject to severe performance deviations (as we have shown in the first analysis of this RQ, see \figref{fig:impact-within-fork}).
In fact, in this analysis we were not particularly interested on the absolute RPD, rather we wanted to assess how RPDs change with respect to the number of forks.
In this regard, our results suggest that increasing the number of forks can effectively mitigate the impact of non-steady measurements.
\\

\begin{tcolorbox}[colback=black!3!white,colframe=black!33!white]
\textbf{RQ$_2$ summary} - The attainment of steady state has relevant effects on software performance.
Performance substantially changes within forks when their execution reaches steady state.  
This difference in performance is less pronounced when comparing forks that never reach steady state against those that consistently reach it. Nonetheless, the reported performance deviations are still considerable and potentially harmful for performance assessment.
The use of an appropriate number of warmup iterations can significantly mitigate performance deviations induced by non-steady forks.
In addition, the use of an adequate number of forks can alleviate deviations that are induced by unstable measurements, which are collected before steady state execution occurs.
\end{tcolorbox}
\newpage
\subsection{RQ$_3$ - Developer configuration assessment}

In this subsection, we first present results of the assessment of developer static configurations, thereafter we provide answer to RQ$_3$.

\begin{figure}[hbtp]
  \centering
  \begin{subfigure}[b]{5.5cm}
                  \includegraphics[width=\linewidth]{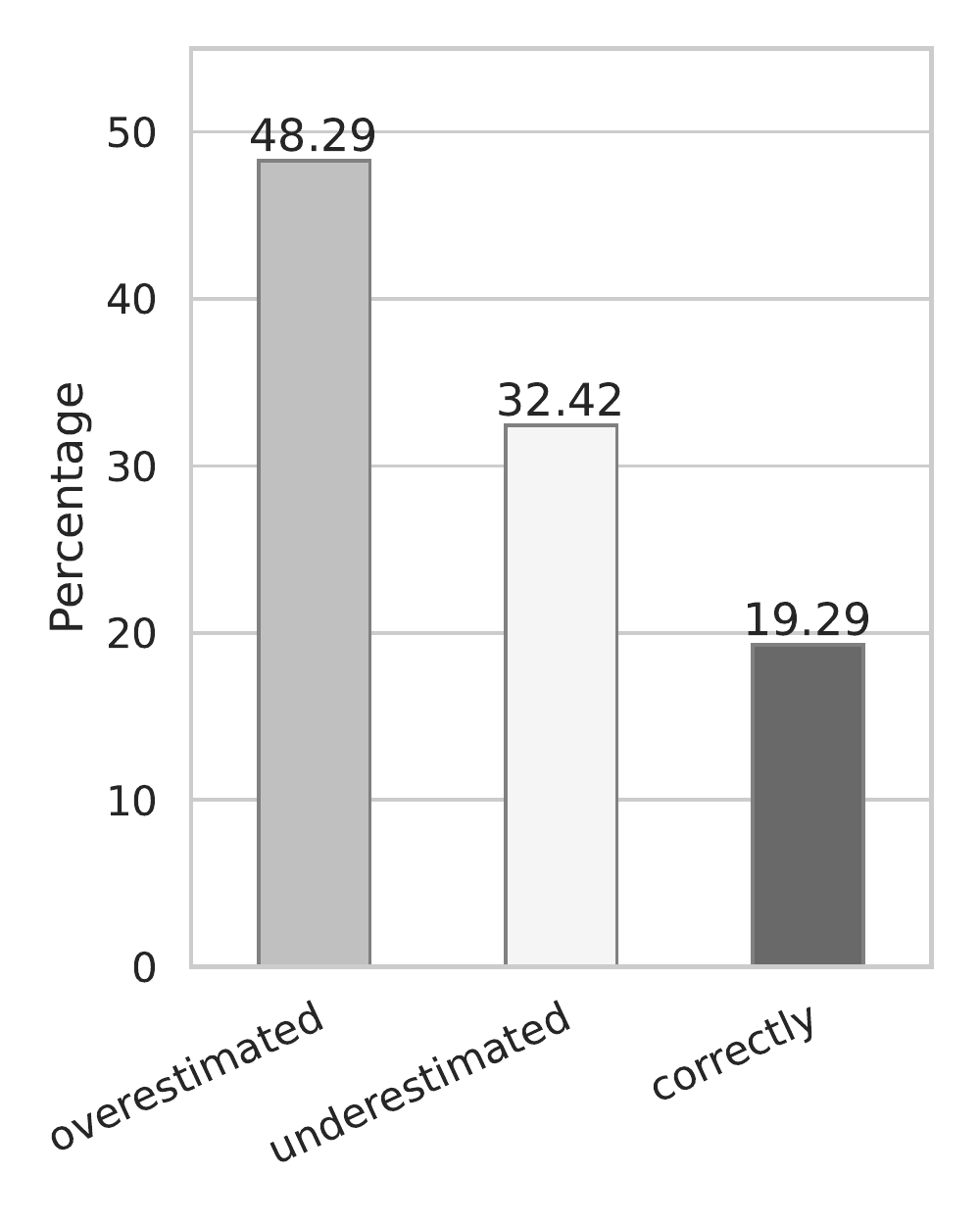}
          \caption{}\label{fig:developer_configurations_barplot}
  \end{subfigure}
  \hfill
  \begin{subfigure}[b]{5.5cm}
                  \includegraphics[width=\linewidth]{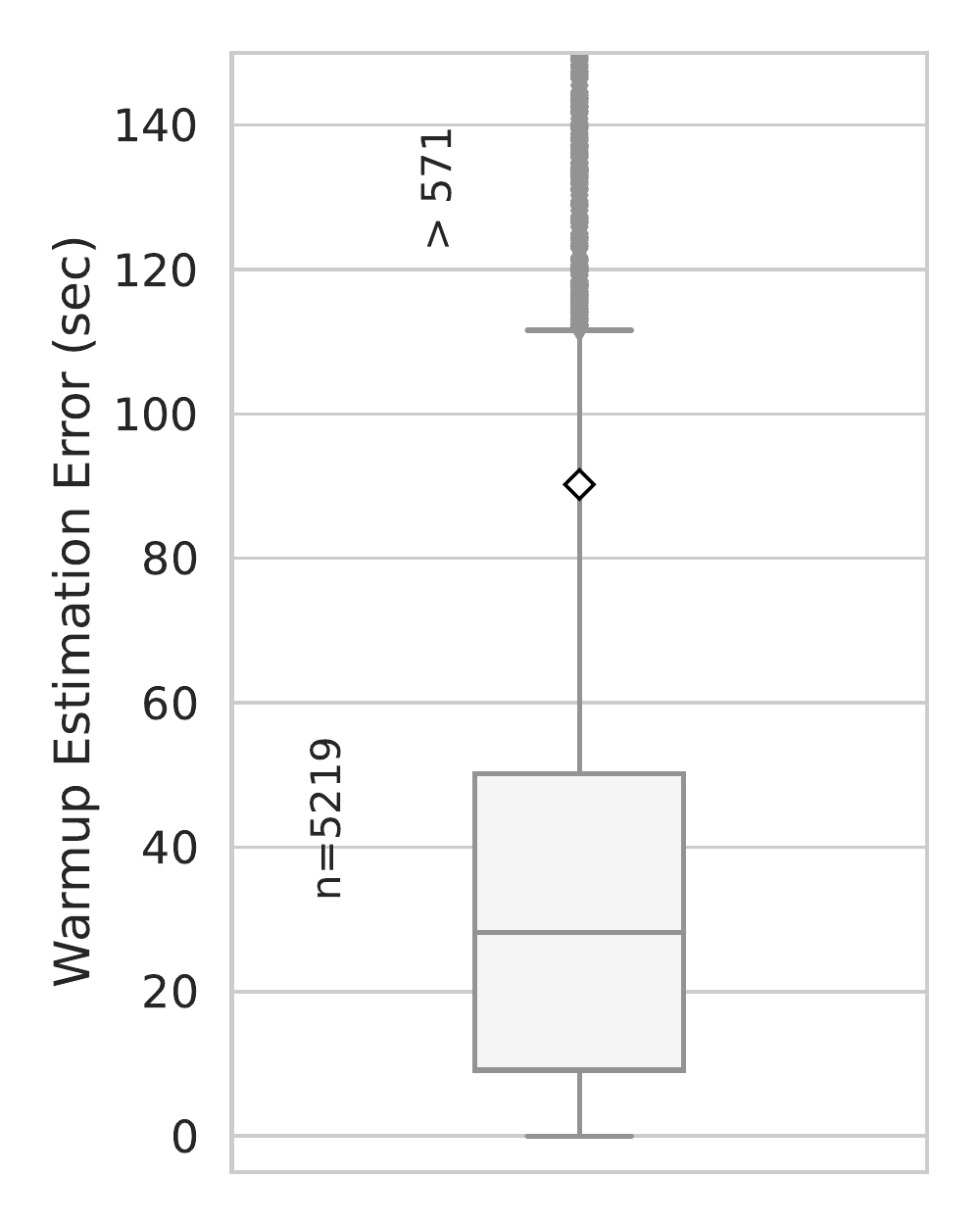}
          \caption{}\label{fig:developer_configurations_boxplot}
  \end{subfigure}
  \caption{RQ$_3$. Developer configurations - Warmup estimation accuracy.
          The left plot reports the percentages of overestimated, underestimated, and correctly estimated forks.
The right plot depicts the distribution of $WEE$ across all benchmark forks.
     $n$ is the total amount of data points, and the number on the top of the plot is the amount of outliers not drawn in the figure.}
  \label{fig:developer_configurations}
\end{figure}

\subsubsection{Warmup Estimation Accuracy}
In the first part of this subsection, we investigate to what extent developers accurately estimate \emph{steady state starting time} ($st$), \ie the end of the warmup phase. 

\figref{fig:developer_configurations_boxplot} depicts the distribution of \emph{warmup estimation error} ($WEE$) across all benchmark forks. 
Developers configurations lead to a $WEE$ that ranges between 9 and 50 seconds in half of the cases (\ie the interquartile range (IQR)), and they lead to a median and mean $WEE$ of 28 and 90 seconds, respectively.
Interestingly, we found that the estimation error is approximately as large as the steady state starting time (or more) in half of the forks, \ie the median of the ratios between $WEE$ and $st$ is 0.997 (IQR: 0.79-43.5).
 
It might be easier for some systems to assess the time required to reach a steady state than for other ones.
With respect to the steady state we detected, most systems (27 out of 30) show an estimation error of less than 100 seconds.
However, the developer configurations in \texttt{eclipse-collections}, \texttt{presto}, and \texttt{RxJava} seem to be less effective in estimating the warmup phase.
The most evident case is represented by \texttt{eclipse-collections} with an error, on average, of around 5 minutes.

Overall, these results suggest that, in most of the cases, \emph{software developers fail to accurately estimate the end of the warmup phase}, and often \emph{with a non-trivial estimation error}.

Besides investigating the estimation accuracy of developers, we also check whether they  provide more accurate estimates than JMH defaults\footnote{In our evaluation, we use the default configuration defined for JMH versions $\geq 1.21$, \ie 5 warmup and measurements iterations ($wi=5$ and $i=5$), and iteration time of 10 seconds for both measurement and warmup ($w=10s$ and $r=10s$).} (\ie the default configuration provided by JMH developers). To do so, we compare the $WEE$s provided by developer configurations against those provided by JMH defaults using the Wilcoxon Rank-Sum test~\citep{cohen2013}, and the Vargha Delaney's effect size measure~\citep{Vargha2000}. We found that developer configurations outperform JMH defaults with statistical significance ($p<0.001$) and small effect size (\vda=0.64). The median and the mean $WEE$ provided by JMH defaults are larger than those provided by developer configurations, \ie mean of 97 seconds and median of 50 seconds (IQR: 36-50 seconds). These results suggest that the estimates provided by developers are more accurate than those provided by JMH defaults.

\figref{fig:developer_configurations_barplot} reports the percentages of overestimated, underestimated, and correctly estimated forks  across all forks (\ie the estimated warmup time $wt$ is respectively smaller or larger than $st$ by at least 5s).
The bar chart shows that overestimation is more common than underestimation.
Developers overestimate the end of the warmup phase in 48\% of the forks (median WEE: 33 seconds, IQR: 19-50 seconds), whereas underestimation is reported in 32\% of the cases (median WEE: 150 seconds , IQR: 36-240 seconds). 
Developers accurately estimate it in only 19\% of the cases.

In the following subsections, we investigate side effects in both overestimated and underestimated forks.

\subsubsection{Overestimation side effects}

\figref{fig:overestimation_dev} shows the \emph{time waste} due to overestimation.
87\% of the overestimated forks waste more than 10 seconds (\ie 37\% of all the forks).
As can be observed by \figref{fig:overestimation_dev_barplot}, overestimation leads to a time waste between 10 and 25 seconds in 30\% of the cases, between 25 and 50 seconds in 38\% of the cases, and it leads to a time waste higher than 50 seconds in 19\% of the cases.
\figref{fig:overestimation_dev_boxplot} depicts the distribution of \emph{time waste} across overestimated forks. The box plot shows that the average \emph{time waste} is 37 seconds, and the median is 33 seconds (IQR: 19-49 seconds).

\begin{figure}[hbtp]
 \centering
 \begin{subfigure}[b]{5.5cm}
   \includegraphics[width=\textwidth]{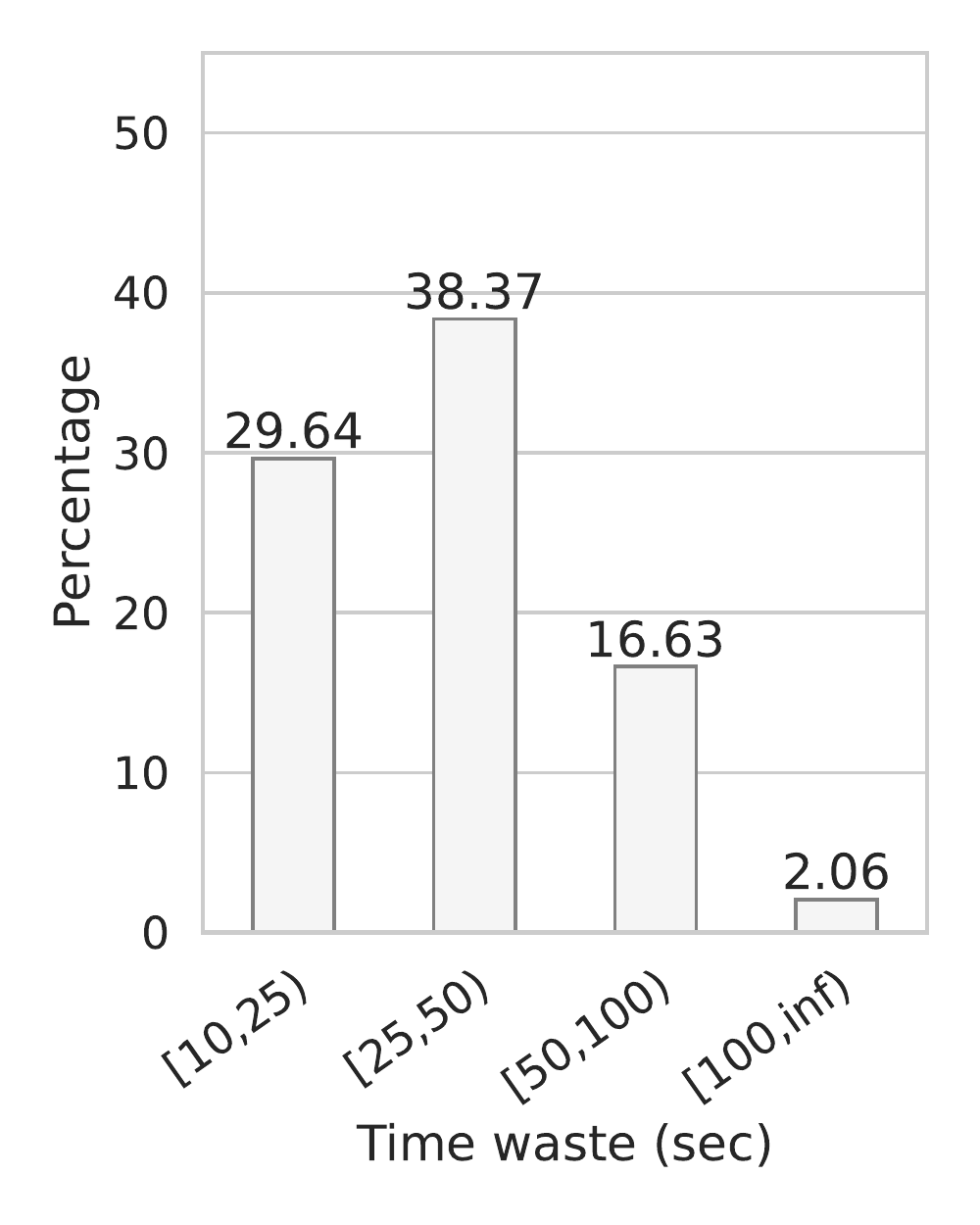}
  \caption{}\label{fig:overestimation_dev_barplot}
 \end{subfigure}
 \hfill
 \begin{subfigure}[b]{5.5cm}
                 \includegraphics[width=\linewidth]{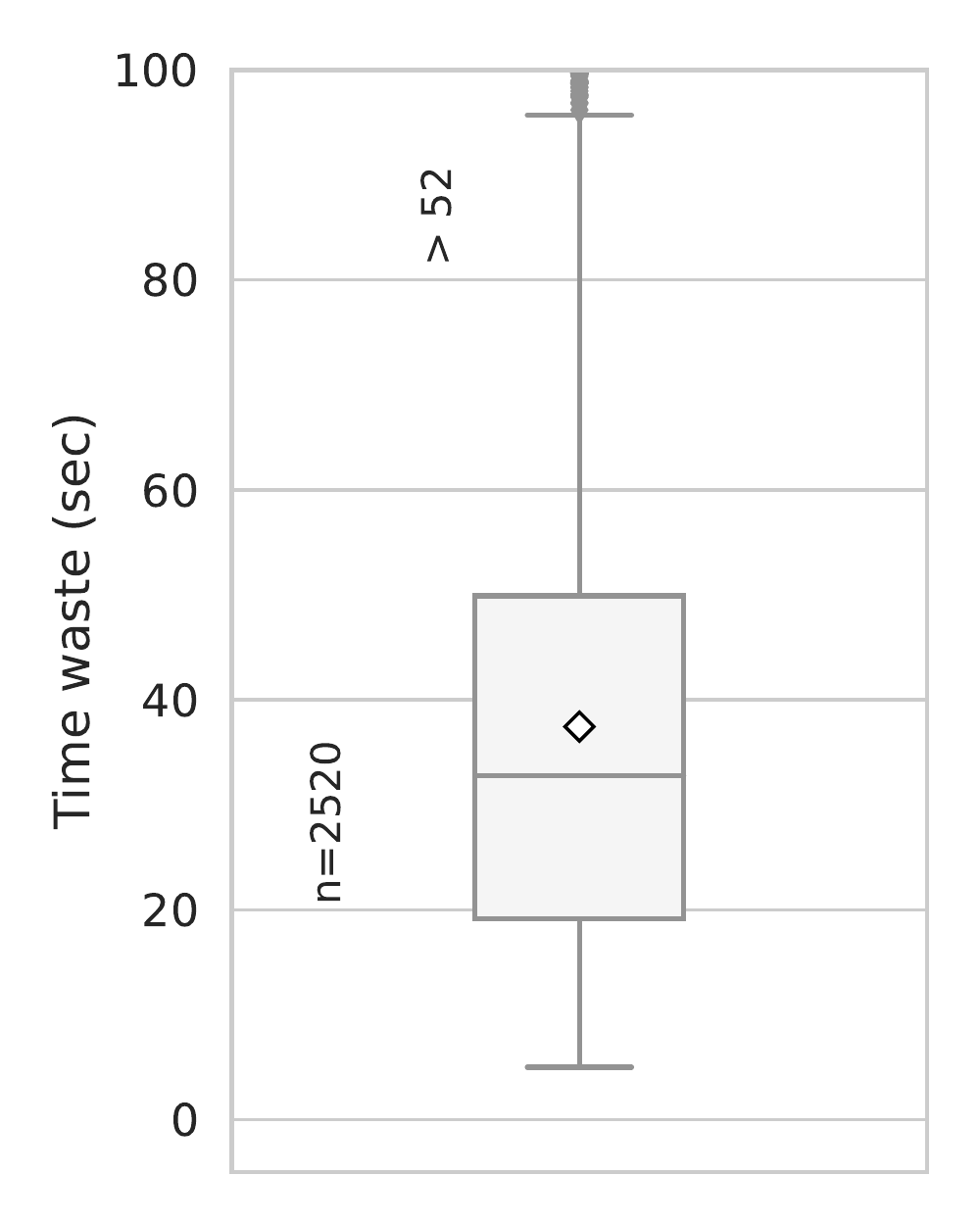}
         \caption{}\label{fig:overestimation_dev_boxplot}
 \end{subfigure}
 \caption{RQ$_3$. Developer configurations - Overestimation side effects (\emph{Time waste}).
  The left plot reports the percentages of overestimated forks where $\emph{Time waste}\, (sec) \in \{[10,25), [25,50), [50,100), [100, inf)\}$.
     The right plot depicts the distribution of \emph{Time waste} across all overestimated forks.
     $n$ is the total amount of data points, and the number on the top of the plot is the amount of outliers not drawn in the figure.}
   \label{fig:overestimation_dev}
\end{figure}

The amount of time wasted on warmup, after reaching a steady state, considerably varies from one system to another.
In most systems, by using the developers configurations, each fork wastes, on average, more than 20 seconds.
More extremes behaviors can be found, for instance, in \texttt{jdbi}, \texttt{netty}, and \texttt{rdf4j}, with an average wasted time of more than 50 seconds per fork.
On the contrary, \texttt{cantaloupe}, \texttt{JCTools}, \texttt{r2dbc-h2}, and \texttt{zipkin} might waste just a few seconds, and probably their configurations do not need any adjustment.

Such absolute \emph{time wastes}, when contextualized within concrete performance assurance processes (that involve multiple benchmarks), can have substantial effects on the overall execution time and, as consequence, can hamper microbenchmarks adoption for continuous performance assessment. 
For example, a time waste of 33 seconds in a relatively small performance testing suite (\eg \texttt{r2dbc-h2}), which involves a typical number of 5 forks\footnote{The default number of forks in JMH is 5 (see \url{https://bit.ly/3mOBvHy})}, could lead to an overall time waste of approximately one hour\footnote{\label{fn:timewaste}To compute the overall time waste, we multiplied the time waste (33 seconds) by the number of forks (5) and the number of benchmarks in the testing suite (20 for \texttt{r2dbc-h2} and 1,302 for \texttt{RxJava}).}.
In larger testing suites, such as \texttt{RxJava}, the same \emph{time waste} could lead to an overall waste of about 2 days and a half.
Besides this, $wt$ is mostly composed by time waste in a large number of cases. In fact, we have measured that the median of the ratios between the time waste and the estimated warmup time $wt$ is approximately 0.97 (IQR: 0.77-0.99), \ie in half of overestimated forks, at least 97\% of the estimated warmup time $wt$ consists of time waste.

The reported results highlight a substantial portion of time wasted during microbenchmarking, and stress the need for better microbenchmark configuration approaches that reduce their execution time. These findings further motivate prior efforts in reducing execution time through dynamic reconfiguration \citep{Laaber2020}, and highlight huge opportunities for execution time reduction in microbenchmarks.

\subsubsection{Underestimation side effects}
Although less frequent than overestimation, underestimation can have relevant side effects on microbenchmarking.
Indeed, it can lead to consider performance measurements that significantly differ from steady state performance, as in practice they fall within the warmup phase.

\figref{fig:underestimation_dev} reports the distribution of the relative performance deviation ($RPD$) across underestimated forks.
Underestimation leads to an $RPD$ of at least 5\% in 57\% of cases (\ie 22\% of all the forks).
\figref{fig:underestimation_dev_barplot} depicts that underestimation induces an $RPD$ between 5\% and 10\% in 15\% of the cases, between 10\% and 25\% in 20\% of the cases, and it induces an $RPD$ greater than 50\% in 9\% of the cases.
The box plot (\figref{fig:underestimation_dev_boxplot}) shows a mean $RPD$ of 17\%, and a median of 7\%~(IQR: 1-21\%).

\begin{figure}[hbtp]
  \centering
 \begin{subfigure}[b]{5.5cm}
   \includegraphics[width=\textwidth]{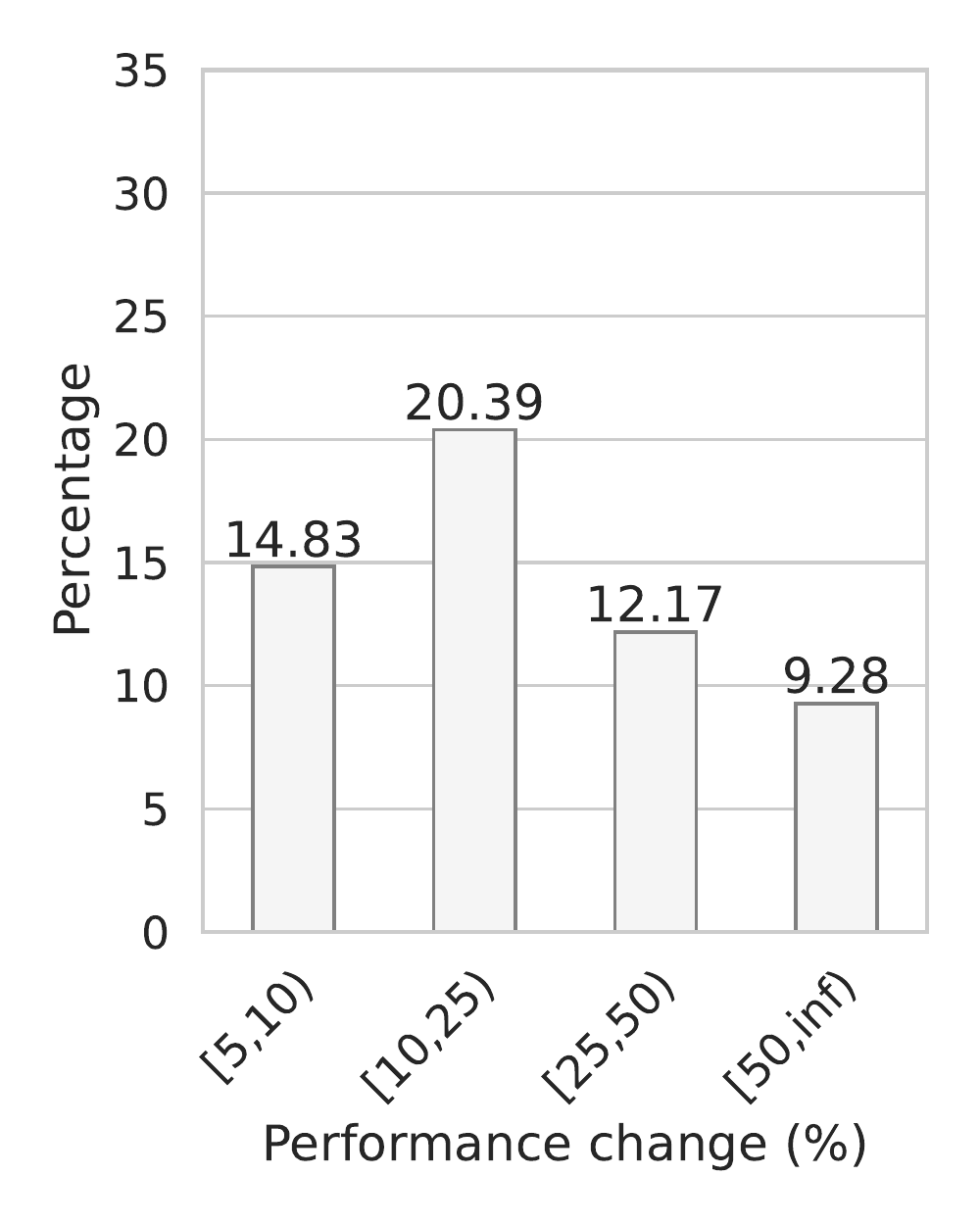}
  \caption{}\label{fig:underestimation_dev_barplot}
 \end{subfigure}
 \hfill
 \begin{subfigure}[b]{5.5cm}
                 \includegraphics[width=\linewidth]{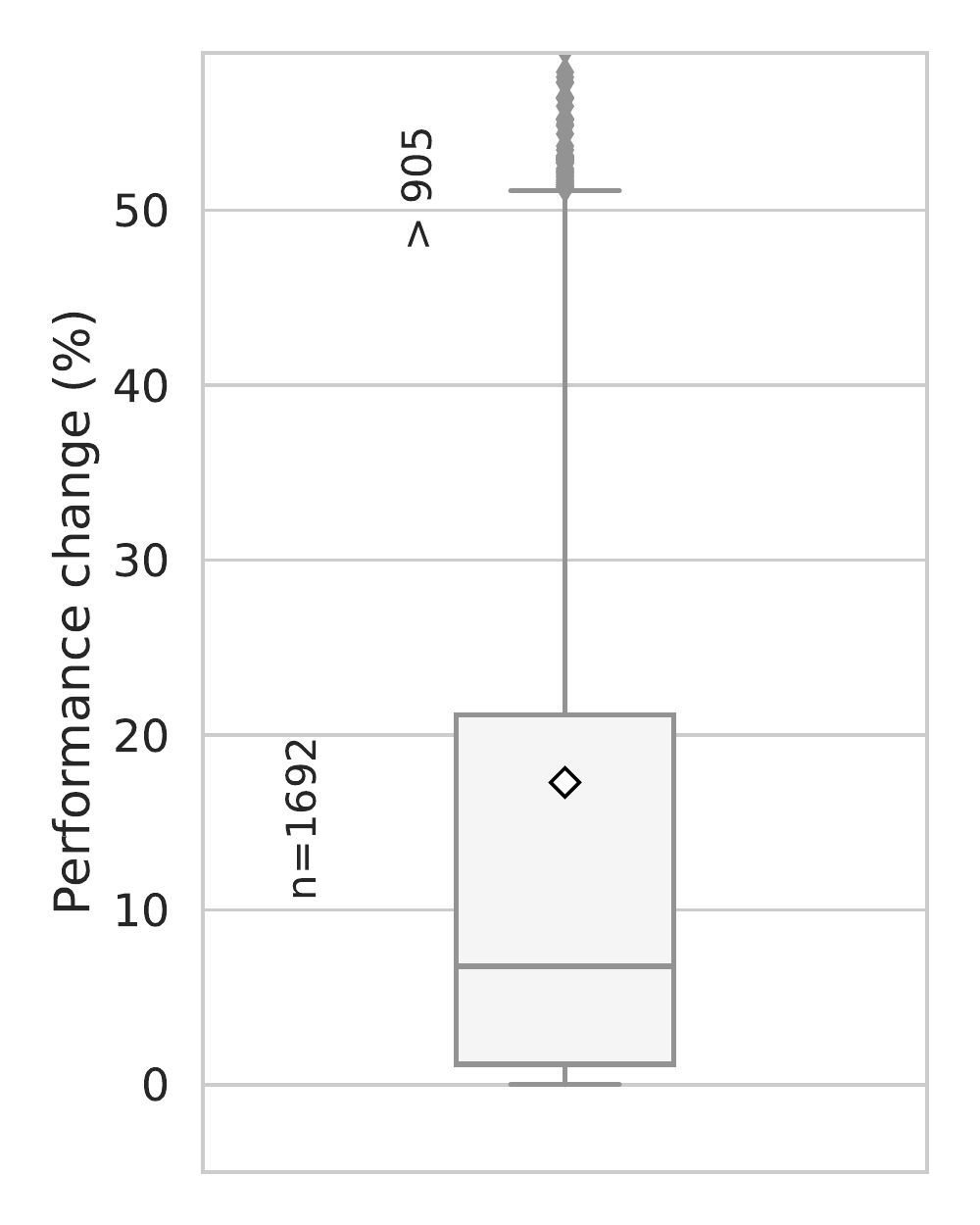}
	 \caption{}\label{fig:underestimation_dev_boxplot}
 \end{subfigure}
 \caption{RQ$_3$. Developer configurations - Underestimation side effects (\emph{performance change}).
  The left plot reports the percentages of underestimated forks where $\emph{performance change}\, (\%) \in \{[5,10), [10,25), [25,50),[50,inf) \}$.
  The right plot depicts the distribution of \emph{performance change} across all underestimated forks.
  $n$ is the total amount of data points, and the number on the top of the plot is the amount of outliers not drawn in the figure.}
   \label{fig:underestimation_dev}
\end{figure}

Some systems show a large performance deviation when underestimating the warmup time.
In nine systems, developers will observe measurements that deviate at least by 20\% from the performance reached in the steady state.
Such large deviations prevent the benchmarks to spot smaller performance changes, in fact defeating their purpose in practical performance testing scenarios.

These results highlight significant performance deviations due to underestimation.
In Java systems, even relatively small performance regressions (\eg 5\%) may lead to rejections of code revisions. 
In fact, a microbenchmark regression, for example due to software refactoring \citep{traini}, can have huge impact at system level, as microbenchmarks measure performance at fine-grained level \citep{Laaber2018, Leitner2017}.
For this reason, the reported $RPD$s can have severe consequences on steady state performance assessment, as they can easily lead to faulty judgments of code revisions.\\
According to our results, developers may rely on measurements that significantly differ from those collected during steady state execution.
This finding sheds a light on the perils of underestimation, and on how such inaccuracy can disrupt performance assessment.
Indeed, the reported results show that underestimation is not rare in the current developer practice~(32\% of forks), and it often leads to potentially misleading results~(57\% of underestimated forks lead to a performance deviation $\geq$5\%). 

\subsubsection{Benchmark level assessment}

\begin{figure}[htpb]
    \centering
    \includegraphics[width=9cm]{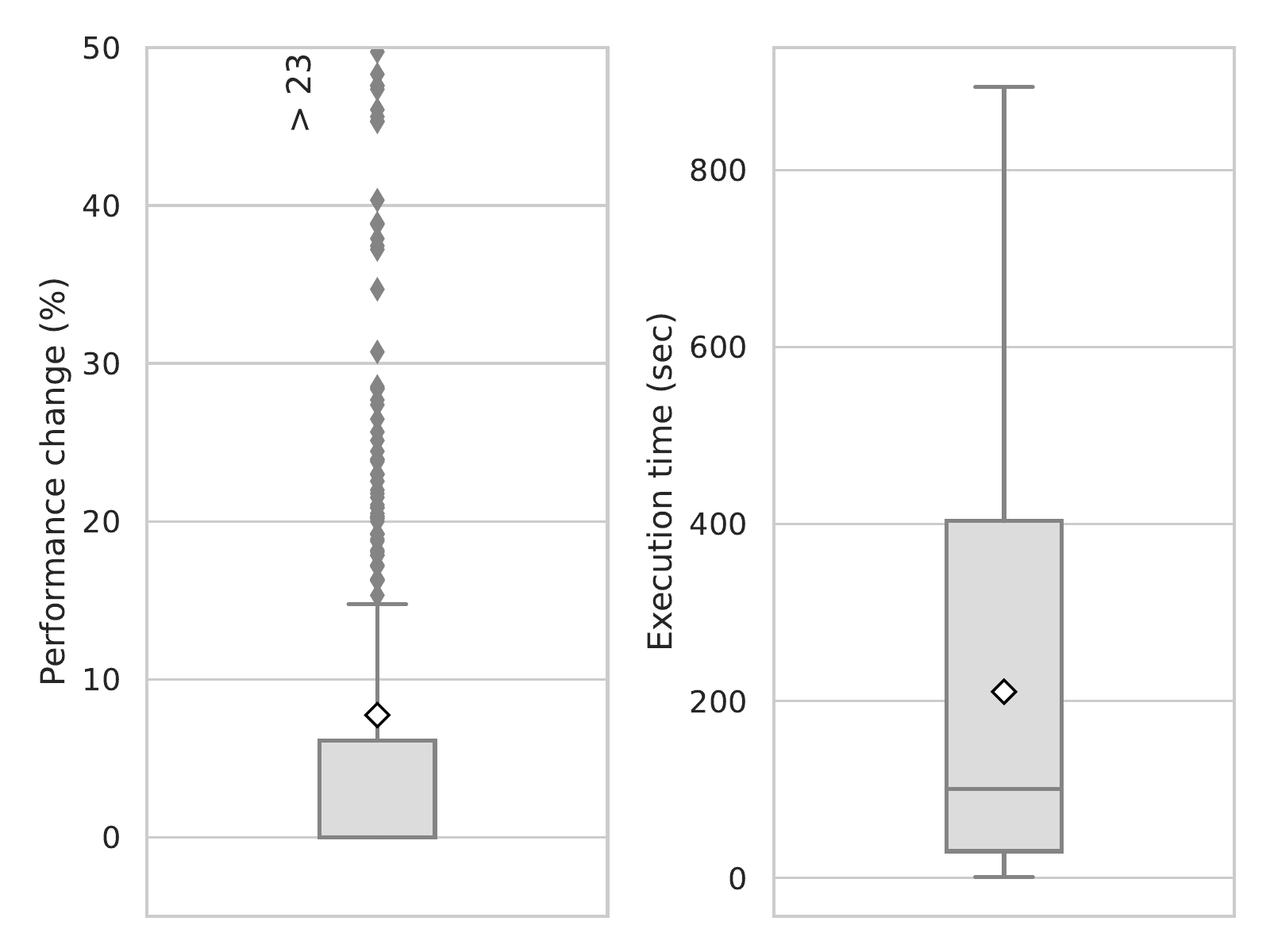}
    \caption{RQ$_3$. Developer configurations: Benchmark level assessment. The left plot depicts the RPD distribution across all benchmarks. The right plot depicts the execution time distribution. }
    \label{fig:dev_perfdev}
\end{figure}

Here, we discuss the results of the analysis of developer configurations at benchmark level. The left plot of \figref{fig:dev_perfdev} reports the distribution of performance deviations of  measurements gathered by software developers when compared to steady state measurements. The first aspect that we can notice by observing the plot is that the impact of underestimation seems to be considerably mitigated when considering aggregated measurements. This finding is in line with our previous analysis on forks (see RQ$_2$), in which we have shown that the number of forks can substantially mitigate deviations of non-steady measurements. Performance deviations are not statistically significant in about half of the benchmarks (\ie 53\% of the cases), the median RPD is 0\% and the IQR is 0-6\%. Nonetheless, 47\% of benchmarks report statistically significant performance deviations (\ie the confidence interval does not contain zero), the mean RPD is 8\%, and a quarter of benchmarks report a deviation higher 6\%.  As we have already discussed in our prior analysis on underestimation, these magnitudes of deviation can be harmful in Java microbenchmarking, as they can mislead performance assessment and lead to wrong judgements of software revisions.

The right plot of \figref{fig:dev_perfdev} reports the distribution of benchmark execution times based on developer configurations.
Developer configurations lead to extremely different execution times, with durations ranging from a minimum of 1 second to a maximum of 893 seconds.
The average execution time is 210 seconds, and the median is 100 seconds (IQR 30-403 seconds).
Our previous analysis on overestimation has already highlighted large opportunities for execution time reduction.
In the light of the above results, these opportunities appear even more significant.
For example, our prior analysis has reported a median time waste of 33 seconds in overestimated forks.
If we compare this result with the median execution time of a benchmark, \ie 100 seconds, the time waste appears extremely relevant, \ie approximately one third of the entire benchmark execution.
In that, it is worth to remark that time wastes are measured on individual forks, while the execution times (reported in \figref{fig:dev_perfdev}) measure the entire duration of a benchmark, which typically involves multiple forks (3 on average in the case of developer configurations).
This finding further remarks on the need for better techniques to reduce benchmark execution time without affecting result quality.
\\

\begin{tcolorbox}[colback=black!3!white,colframe=black!33!white]
  \textbf{RQ$_3$ summary} -
 Developer configurations have limited effectiveness for steady state performance assessment. Developers fail to accurately estimate the end of the warmup phase, often with a non-trivial estimation error. In a large number of cases, this error leads to a substantial increase in the execution time (\ie overestimation).
Nevertheless, underestimation is not rare in the current developer practice, and when this happens it significantly distorts performance assessment.
\end{tcolorbox}

\subsection{RQ$_4$  - Dynamic reconfiguration assessment}\label{sec:results:rq3}
In this subsection, we first present results of the assessment of dynamic reconfiguration techniques, then we answer to RQ$_4$ .

\subsubsection{Warmup Estimation Accuracy}


\begin{figure}
\begin{subfigure}[b]{5.7cm}
        \centering
                \includegraphics[width=\linewidth]{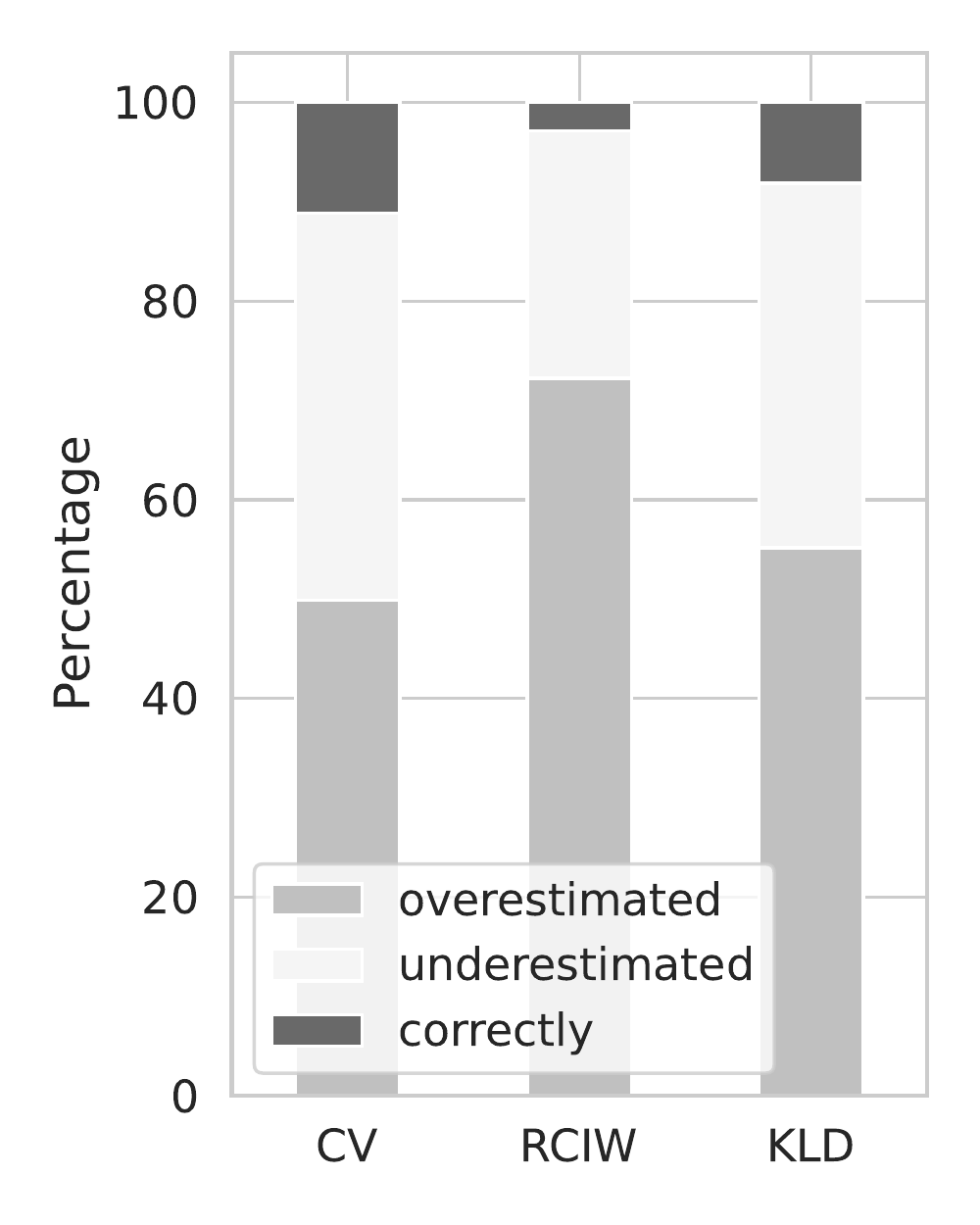}
        \caption{}
        \label{fig:warmup_estimation_DyCo_barplot}
\end{subfigure}
\hfill
\begin{subfigure}[b]{5.7cm}
        \centering
                \includegraphics[width=\linewidth]{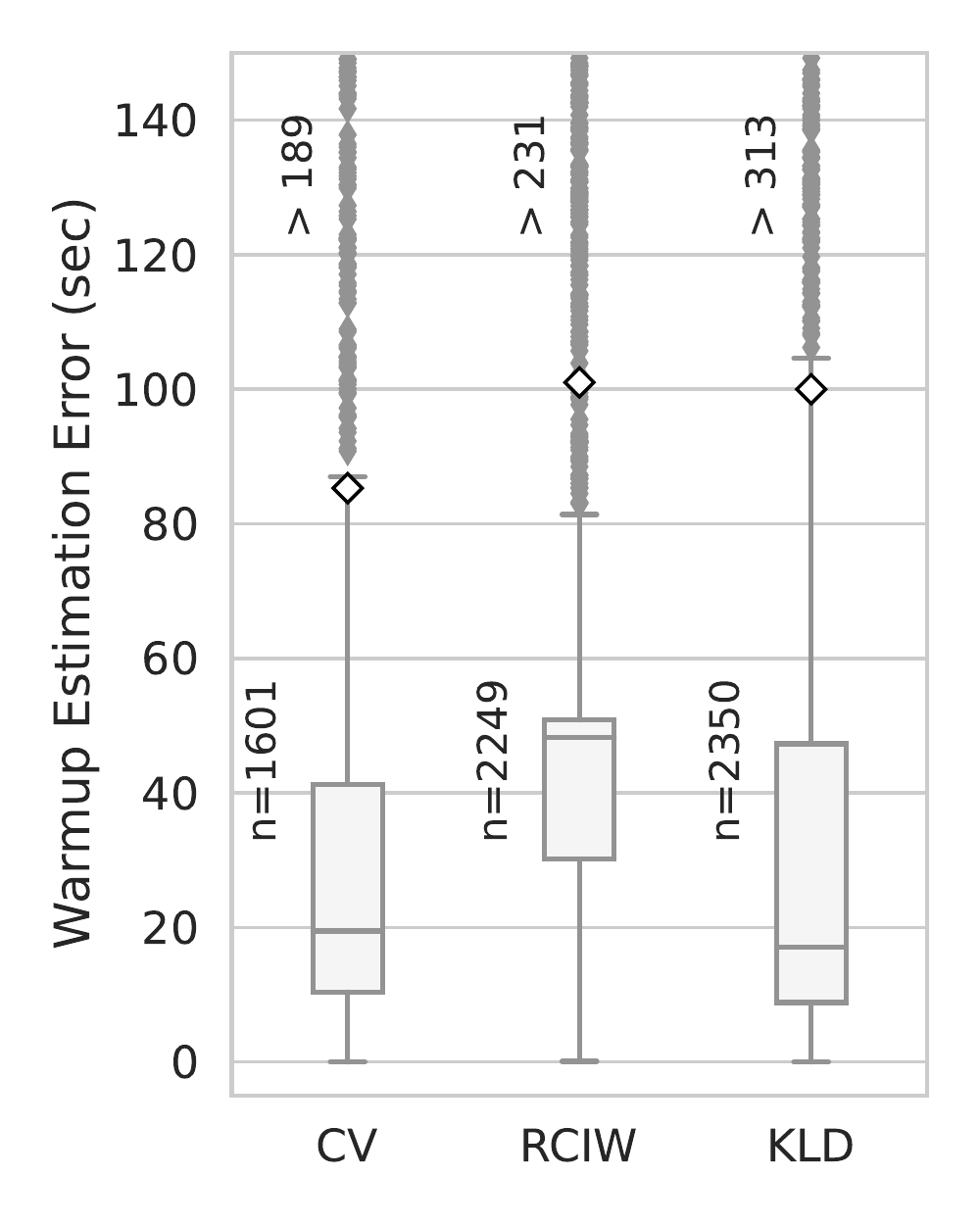}
        \caption{}
        \label{fig:warmup_estimation_DyCo_boxplot}
\end{subfigure}
\caption{RQ$_4$. Dynamic reconfiguration - Warmup estimation accuracy.
     The left plot reports the percentages of overestimated, underestimated, and correctly estimated forks per dynamic reconfiguration technique.
The right plot depicts the distribution of $WEE$ across all benchmark forks per dynamic reconfiguration technique.
     $n$ is the total amount of data points, and the number on the top of the plot is the amount of outliers not drawn in the figure.}
\label{fig:warmup_estimation_DyCo}
\end{figure}

\figref{fig:warmup_estimation_DyCo_boxplot} depicts the distributions of warmup estimation errors ($WEE$) per dynamic reconfiguration technique across all benchmark forks.
As it can be seen from the figure, \emph{RCIW} leads to higher $WEE$ (median: 48 seconds, IQR: 30-50 seconds) when compared to other techniques, whereas $CV$ and $KLD$ report similar distributions.
\emph{CV} leads to a median $WEE$ of 19 seconds (IQR: 10–41 seconds), while \emph{KLD} reports a median of 17 seconds (IQR: 9–47 seconds).
Also, we have measured that, in half of the forks, all dynamic techniques report a $WEE$ approximately as large as the steady state starting time $st$ (or more), \ie the medians of the ratios between $WEE$ and $st$ for \emph{CV}, \emph{RCIW} and \emph{KLD} are respectively 0.97 (IQR: 0.79-45.7), 2.04 (IQR: 0.78-75) and 0.97 (IQR:  0.79-36.1).
These results indicate that all dynamic reconfiguration techniques lead to a substantial error in the estimate of steady state starting time.
Nevertheless, \emph{CV} and \emph{KLD} clearly provide more accurate estimates than \emph{RCIW}.

The bar chart in \figref{fig:warmup_estimation_DyCo_barplot} confirms the specificity of \emph{RCIW}.
As it can be observed from the bar chart, \emph{RCIW} reports a strong tendency toward overestimation, while \emph{CV} and \emph{KLD} show similar frequencies both in terms of underestimation and overestimation. \emph{RCIW} overestimates  72\% of forks (median WEE: 46 seconds, IQR: 31-50 seconds), and it underestimates only 25\% forks (median WEE: 181 seconds, IQR: 86-236 seconds). 
On the other hand, \emph{CV} reports 50\% of overestimations (median WEE: 17 seconds, IQR: 12-24 seconds) and 39\% of underestimations (median WEE: 131 seconds, IQR: 29-249 seconds). Similarly, \emph{KLD} reports 55\% overestimated forks (median WEE: 13 seconds, IQR: 9-21 seconds) and 37\% underestimated forks (median WEE: 145 seconds, IQR: 29-255 seconds).

\subsubsection{Overestimation side effects}


\begin{figure}
     \centering
     \begin{subfigure}[b]{5.7cm}
         \centering
                 \includegraphics[width=\linewidth]{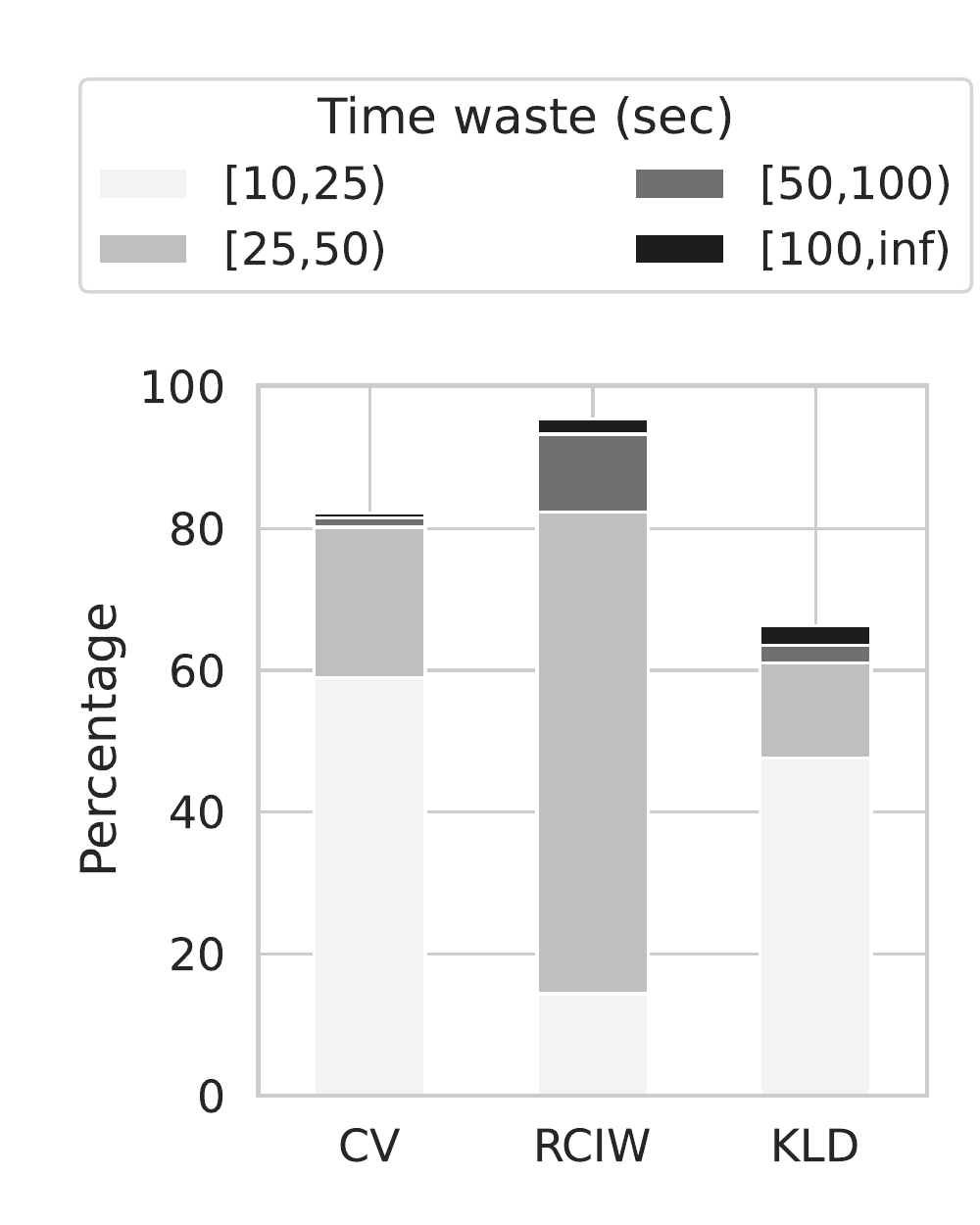}
         \caption{}
         \label{fig:overestimation_stacked}
     \end{subfigure}
     \hfill
     \begin{subfigure}[b]{5.7cm}
         \centering
                \includegraphics[width=\linewidth]{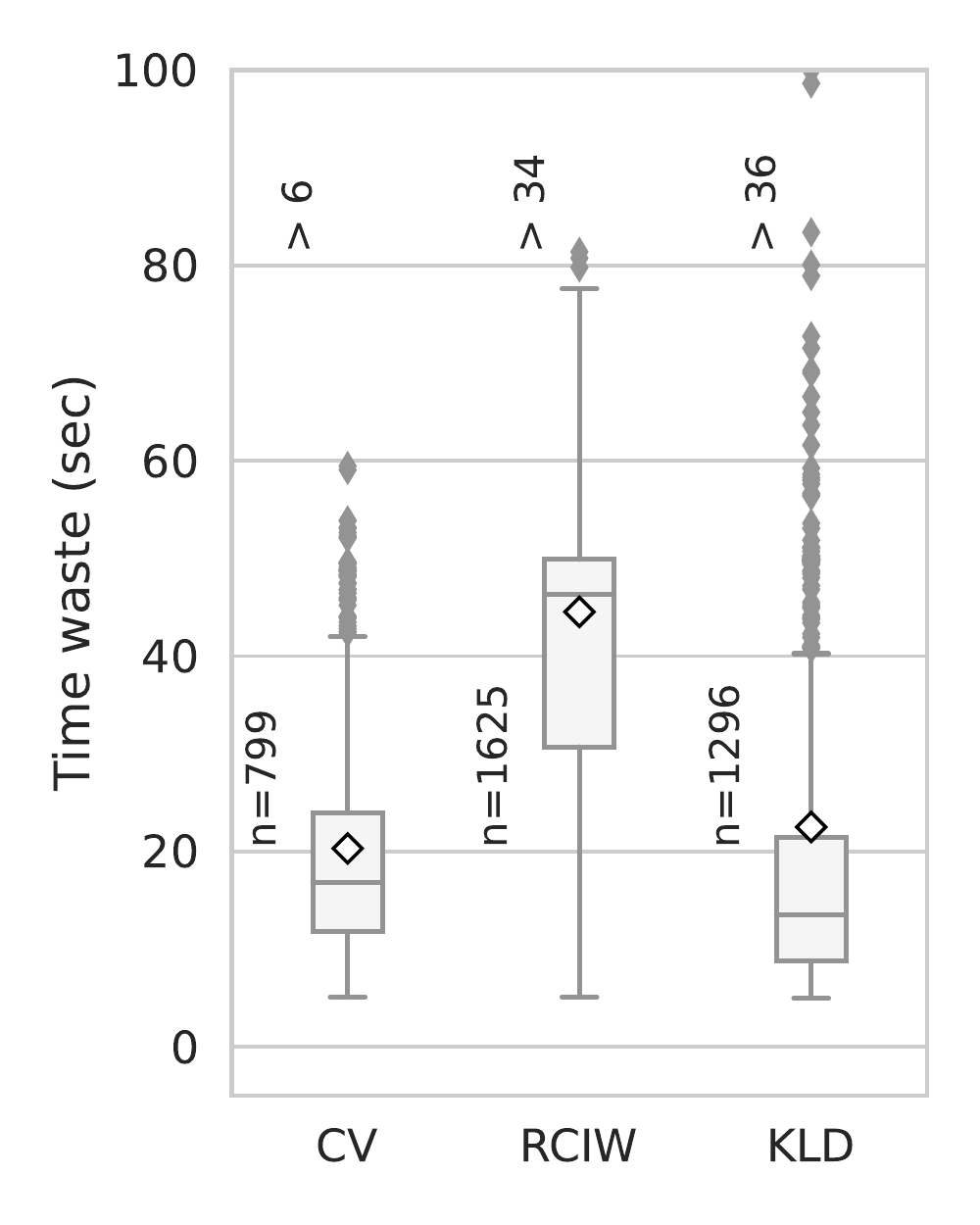}
         \caption{}
         \label{fig:overestimation_boxplot}
     \end{subfigure}
     \caption{RQ$_4$. Dynamic reconfiguration - Overestimation side effects (\emph{time waste}).
             The left plot reports the percentages of overestimated forks where $\emph{time waste}\, (sec) \in \{[10,25), [25,50), [50,100), [100, inf)\}$.
     The right plot depicts the distribution of \emph{time waste} across all overestimated forks per dynamic reconfiguration technique.
     $n$ is the total amount of data points, and the number on the top of the plot is the amount of outliers not drawn in the figure).}
     \label{fig:overestimation_DyCo}
\end{figure}

Besides the large number of overestimated forks, \emph{RCIW} also causes higher \emph{time wastes} when compared to other techniques.
As it can be observed in \figref{fig:overestimation_DyCo}, \emph{RCIW} reports a mean \emph{time waste} of 45 seconds and a median of 46 seconds (IQR: 31-55 seconds). 
95\% of overestimated forks lead to a \emph{time waste} of at least 10 seconds, 81\% lead to a time waste of at least 25 seconds, and 13\% to a time waste of at least 50 seconds. 
On the other hand, \emph{CV} and \emph{KLD} report a median time waste of 17 seconds (IQR: 12-24 seconds) and 14 seconds (IQR: 9-21 seconds), respectively (see \figref{fig:overestimation_boxplot}). 
Nonetheless, we have measured that in half of the overestimated forks, all techniques lead to an estimated warmup time that mostly consists of time waste, \ie the medians of the ratios between the \emph{time waste} and \emph{wt} for \emph{CV}, \emph{RCIW}, and \emph{KLD} are respectively 0.98 (IQR: 0.88-0.99), 0.96 (IQR: 0.7-0.99), and 0.97 (IQR: 0.88-0.99).
Overall, these results indicate that different techniques lead to diverse outcomes in terms of overestimation.
For example, RCIW induces more frequent overestimations and higher time wastes when compared to other techniques (see \figref{fig:overestimation_stacked}).
Despite this diversity, overestimation is frequent across all dynamic reconfiguration techniques, and it often leads to a non-trivial time waste, which can hamper continuous performance assessment.

\subsubsection{Underestimation side effects}


\begin{figure}
     \centering
     \begin{subfigure}[b]{5.7cm}
         \centering
                 \includegraphics[width=\linewidth]{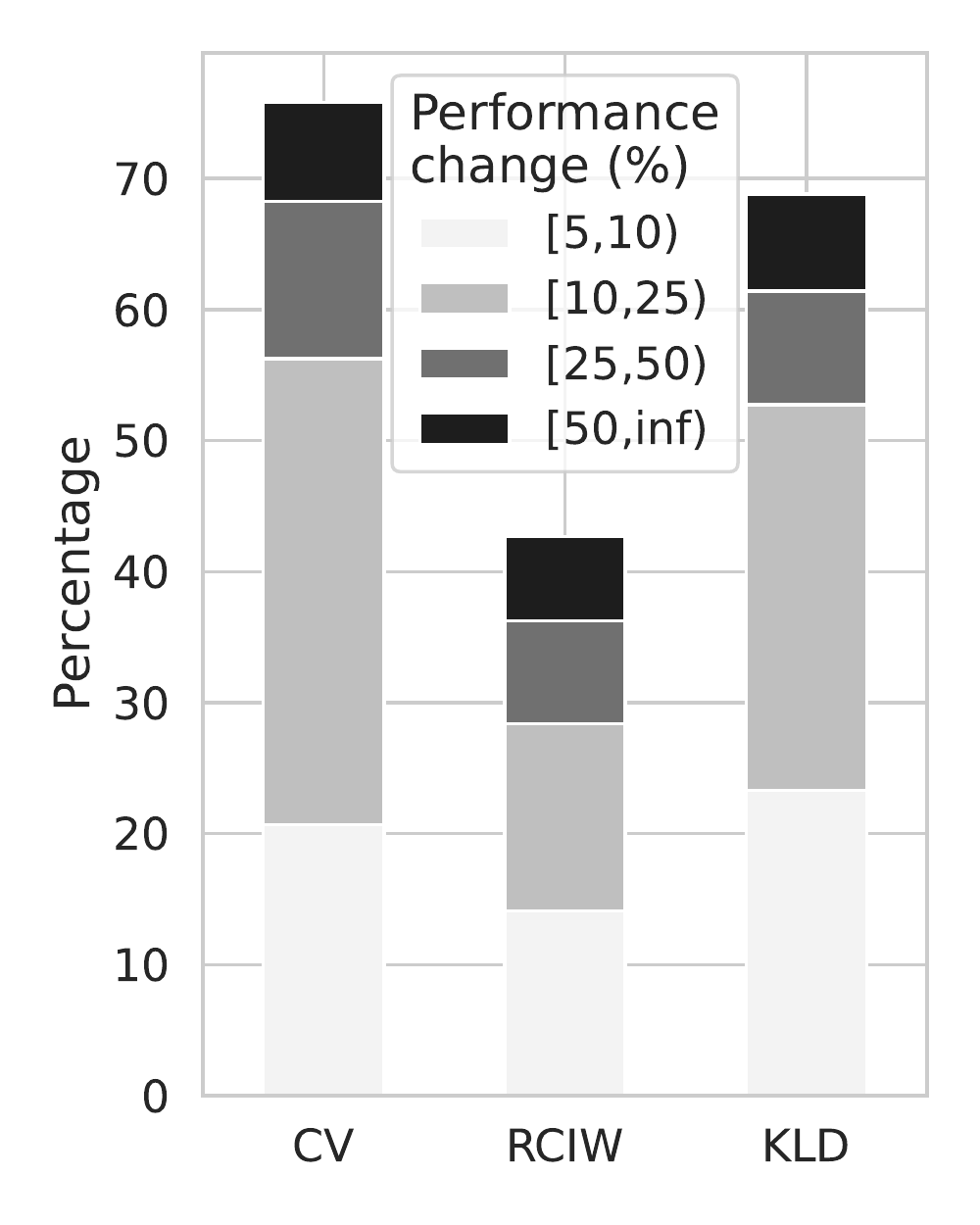}
         \caption{}
         \label{fig:underestimation_stacked}
     \end{subfigure}
     \hfill
     \begin{subfigure}[b]{5.7cm}
         \centering
                \includegraphics[width=\linewidth]{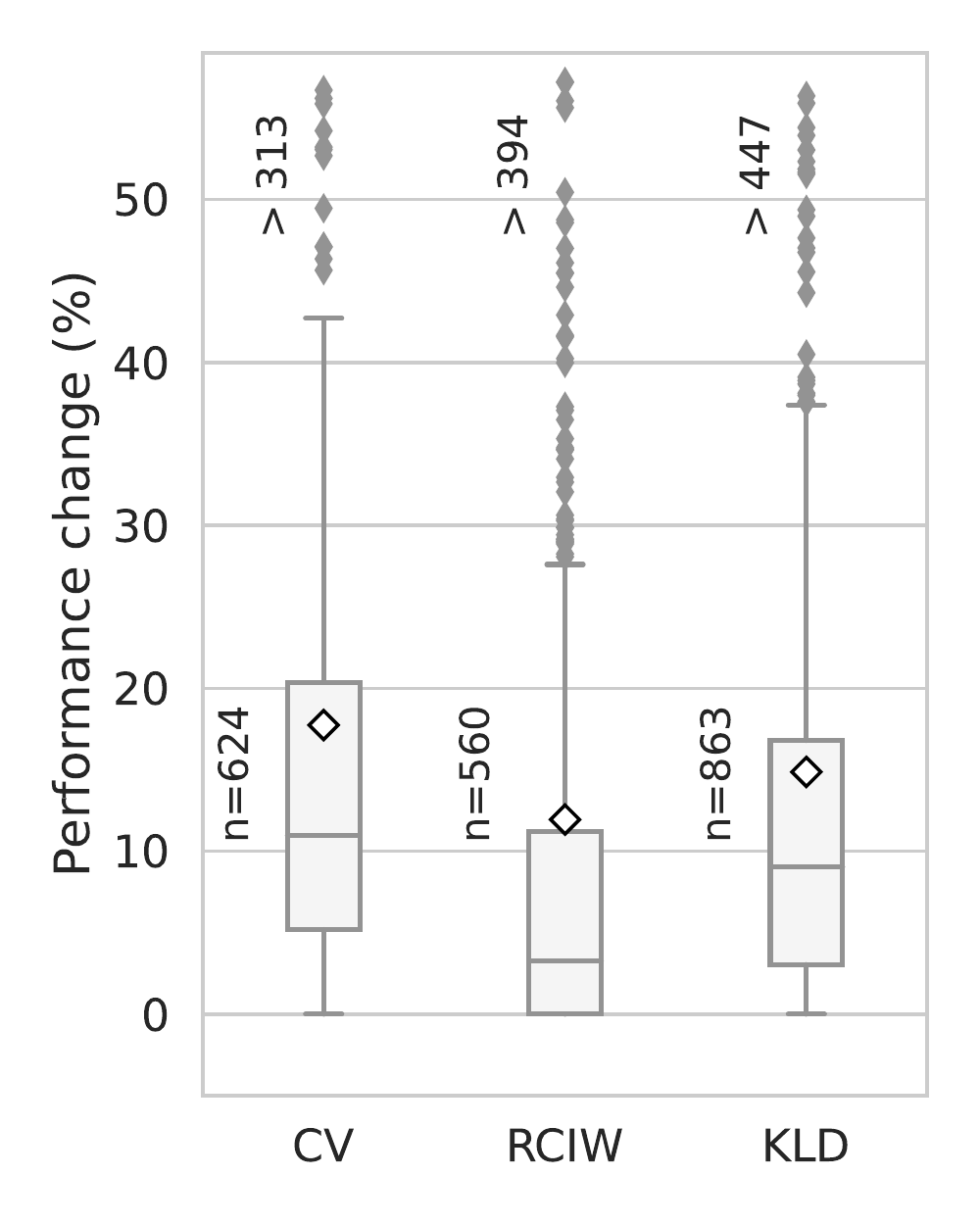}
         \caption{}
         \label{fig:underestimation_boxplot}
     \end{subfigure}
     \caption{RQ$_4$. Dynamic reconfiguration - Underestimation side effects (\emph{performance change}).
             The left plot reports the percentages of underestimated forks where $\emph{performance change}\, (\%) \in \{[5,10), [10,25), [25,50),[50,inf) \}$.
     The right plot depicts the distribution of \emph{performance change} across all underestimated forks per dynamic reconfiguration technique.
     $n$ is the total amount of data points, and the number on the top of the plot is the amount of outliers not drawn in the figure).}
     \label{fig:underestimation_DyCo}
\end{figure}

 
\emph{RCIW} is less prone to underestimation than other techniques, and it also has less marked side effects due to underestimation.
As it can be observed in \figref{fig:underestimation_DyCo}, \emph{CV} and \emph{KLD} lead to higher performance changes when compared to \emph{RCIW}.
\emph{RCIW} reports a median $RPD$ of 3\% (IQR: 0-11\%), whereas \emph{CV} and \emph{KLD} report medians of 10\% (IQR: 5-20\%) and 9\% (IQR: 3\%-17\%), respectively (see \figref{fig:underestimation_boxplot}).
Additionally, \emph{CV} and \emph{KLD} cause performance changes of at least 5\%, respectively, in 76\% and 69\% of the cases, while \emph{RCIW} achieves a similar deviation in only 43\% of forks (see \figref{fig:underestimation_stacked}).
These results suggest that underestimation side effect varies depending on the dynamic reconfiguration technique.
Some dynamic reconfiguration techniques (\ie CV and KLD) are more prone to induce underestimation, and they often lead to a non-trivial performance deviation which can potentially mislead steady state performance assessment.
Other techniques, such as \emph{RCIW}, are instead ``safer'' in terms of underestimation, and ensure higher results quality in terms of performance assessment.

 \subsubsection{Benchmark level assessment}
\begin{figure}[htpb]
    \centering
    \includegraphics[width=12cm]{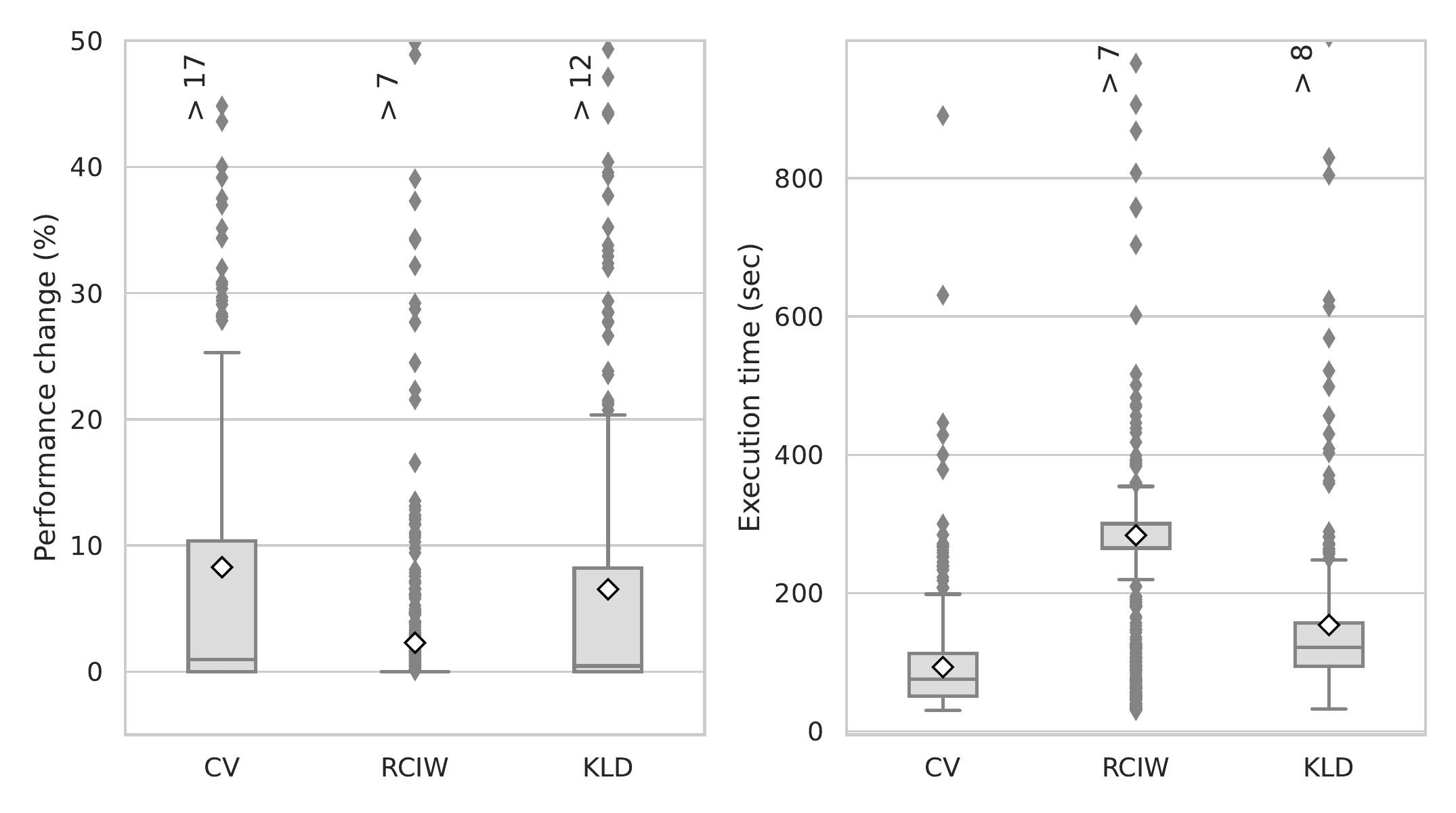}
    \caption{RQ$_4$. Dynamic reconfiguration: Benchmark level assessment. The left plot depicts the RPD distribution across all benchmarks. The right plot depicts the execution time distribution. }
    \label{fig:dyn_perfdev}
\end{figure}

The analysis at benchmark level confirms the trend observed in our prior analysis.
\figref{fig:dyn_perfdev} reports the deviations ($RPD$) of measurements gathered through dynamic reconfiguration techniques when compared to steady measurements (left plot), along with the distributions of benchmark execution times (right plot).
 As it can be observed, \emph{RCIW} is by far the most reliable technique in terms of performance deviation. It reports an average $RPD$ of 2\%, a median of 0\% and an IQR of 0-0\%. About 91\% of the benchmarks report an $RPD$ smaller than 5\%.
 On the other hand, \emph{RCIW} provides also the most time consuming process with a mean execution time of 283 seconds and a median of 300 seconds (IQR 265-300 seconds).
 This is not surprising, as our prior analysis has highlighted a large overestimation rate (72\%) and relevant time wastes (45 seconds on average) in \emph{RCIW}.

Contrariwise, \emph{CV} and \emph{KLD} are less reliable in terms of results quality, by reporting a mean $RPD$ of respectively 8\% (median: 1\%, IQR: 0-10\%) and 7\% (median: 0.5\%, IQR 0-8\%), but they are also less demanding in terms of execution time. \emph{CV} reports a mean execution time of 93 seconds (median: 75 seconds, IQR: 51-113 seconds), while \emph{KLD} reports a mean of 154 seconds (median: 121 seconds, IQR: 94-156 seconds).
 \emph{CV} and \emph{KLD} show similar behaviors in terms of execution times and performance deviations, however, as it emerges by the box plots in \figref{fig:dyn_perfdev}, they show two opposite tendencies. In particular, \emph{CV} shows a slight tendency towards faster execution times and less reliable results, while \emph{KLD} shows the opposite behavior, \ie better results quality and more time-consuming executions.
\\

\begin{tcolorbox}[colback=black!3!white,colframe=black!33!white]
\textbf{RQ$_4$ summary} -
Dynamic reconfiguration techniques provide far from optimal estimates of the warmup phase, often with a non-trivial error.
The side effects vary depending on the technique.
\emph{RCIW} is more prone to overestimation than other techniques, and it induces more time-consuming benchmark executions (\ie higher time waste).
On the other hand, \emph{CV} and \emph{KLD} often lead to performance measurements that differ from those collected during the steady state, while \emph{RCIW} provides a ``safer'' assessment of steady state performance.
\end{tcolorbox}

\subsection{RQ$_5$ - Dynamic reconfiguration \emph{vs} Developer configuration }
In this subsection, we first present results of the comparison between dynamic reconfiguration techniques and  developer static configurations for each considered metric: (i) \emph{warmup estimation error} ($WEE$), (ii) \emph{estimated warmup time} ($wt$), and (iii) \emph{relative performance deviation} ($RPD$). Then, we provide answer to RQ$_5$.
\subsubsection{Warmup Estimation Error}\label{sec:dyn_vs_dev_wee}
\begin{table}[p]
\centering
\scriptsize
\begin{tabular}{lrrr}
\toprule
 System &   CV\tiny{$^{^{\textcolor{white}{***}}}$} &  RCIW\tiny{$^{^{\textcolor{white}{***}}}$} & KLD\tiny{$^{^{\textcolor{white}{***}}}$} \\
\midrule
arrow               &  \textbf{$<$0.001(0.95)}\tiny{$^{^{***}}$} &  \textbf{0.003(0.86)}\tiny{$^{^{***}}$} &  \textbf{$<$0.001(0.92)}\tiny{$^{^{***}}$} \\
byte-buddy          &  0.321(-)\tiny{$^{^{\textcolor{white}{***}}}$} &  \textbf{$<$0.001(0.64)}\tiny{$^{^{*\textcolor{white}{**}}}$} &  \textbf{0.01(0.8)}\tiny{$^{^{***}}$} \\
camel               &  0.25(-)\tiny{$^{^{\textcolor{white}{***}}}$} &  0.264(-)\tiny{$^{^{\textcolor{white}{***}}}$} &  \textbf{$<$0.001(0.86)}\tiny{$^{^{***}}$} \\
cantaloupe          &  $<$0.001(0.46)\tiny{$^{^{\textcolor{white}{***}}}$} &  $<$0.001(0.36)\tiny{$^{^{*\textcolor{white}{**}}}$} &  $<$0.001(0.36)\tiny{$^{^{*\textcolor{white}{**}}}$} \\
client\_java         &  \textbf{0.019(0.88)}\tiny{$^{^{***}}$} &  \textbf{0.01(0.66)}\tiny{$^{^{**\textcolor{white}{*}}}$} &  0.154(-)\tiny{$^{^{\textcolor{white}{***}}}$} \\
crate               &  0.405(-)\tiny{$^{^{\textcolor{white}{***}}}$} &  0.097(-)\tiny{$^{^{\textcolor{white}{***}}}$} &  0.063(-)\tiny{$^{^{\textcolor{white}{***}}}$} \\
eclipse-collections &  \textbf{0.007(0.85)}\tiny{$^{^{***}}$} &  0.163(-)\tiny{$^{^{\textcolor{white}{***}}}$} &  0.43(-)\tiny{$^{^{\textcolor{white}{***}}}$} \\
h2o-3               &  \textbf{0.025(0.86)}\tiny{$^{^{***}}$} &  0.321(-)\tiny{$^{^{\textcolor{white}{***}}}$} &  \textbf{0.003(0.79)}\tiny{$^{^{***}}$} \\
hazelcast           &  \textbf{0.014(0.73)}\tiny{$^{^{***}}$} &  0.147(-)\tiny{$^{^{\textcolor{white}{***}}}$} &  0.527(-)\tiny{$^{^{\textcolor{white}{***}}}$} \\
HdrHistogram        &  \textbf{0.023(0.65)}\tiny{$^{^{**\textcolor{white}{*}}}$} &  0.282(-)\tiny{$^{^{\textcolor{white}{***}}}$} &  0.346(-)\tiny{$^{^{\textcolor{white}{***}}}$} \\
hive                &  $<$0.001(0.38)\tiny{$^{^{*\textcolor{white}{**}}}$} &  $<$0.001(0.36)\tiny{$^{^{*\textcolor{white}{**}}}$} &  $<$0.001(0.33)\tiny{$^{^{**\textcolor{white}{*}}}$} \\
imglib2             &  0.094(-)\tiny{$^{^{\textcolor{white}{***}}}$} &  0.279(-)\tiny{$^{^{\textcolor{white}{***}}}$} &  0.225(-)\tiny{$^{^{\textcolor{white}{***}}}$} \\
JCTools             &  \textbf{$<$0.001(0.7)}\tiny{$^{^{**\textcolor{white}{*}}}$} &  0.719(-)\tiny{$^{^{\textcolor{white}{***}}}$} &  \textbf{$<$0.001(0.65)}\tiny{$^{^{**\textcolor{white}{*}}}$} \\
jdbi                &  0.091(-)\tiny{$^{^{\textcolor{white}{***}}}$} &  0.136(-)\tiny{$^{^{\textcolor{white}{***}}}$} &  0.052(-)\tiny{$^{^{\textcolor{white}{***}}}$} \\
jetty.project       &  0.867(-)\tiny{$^{^{\textcolor{white}{***}}}$} &  0.265(-)\tiny{$^{^{\textcolor{white}{***}}}$} &  0.357(-)\tiny{$^{^{\textcolor{white}{***}}}$} \\
jgrapht             &  0.261(-)\tiny{$^{^{\textcolor{white}{***}}}$} &  0.492(-)\tiny{$^{^{\textcolor{white}{***}}}$} &  \textbf{$<$0.001(0.77)}\tiny{$^{^{***}}$} \\
kafka               &  \textbf{0.019(0.82)}\tiny{$^{^{***}}$} &  0.655(-)\tiny{$^{^{\textcolor{white}{***}}}$} &  0.225(-)\tiny{$^{^{\textcolor{white}{***}}}$} \\
logbook             &  \textbf{0.002(0.89)}\tiny{$^{^{***}}$} &  \textbf{$<$0.001(0.67)}\tiny{$^{^{**\textcolor{white}{*}}}$} &  \textbf{0.001(0.84)}\tiny{$^{^{***}}$} \\
logging-log4j2      &  \textbf{$<$0.001(0.89)}\tiny{$^{^{***}}$} &  0.304(-)\tiny{$^{^{\textcolor{white}{***}}}$} &  \textbf{$<$0.001(0.86)}\tiny{$^{^{***}}$} \\
netty               &  \textbf{$<$0.001(0.86)}\tiny{$^{^{***}}$} &  \textbf{$<$0.001(0.77)}\tiny{$^{^{***}}$} &  \textbf{$<$0.001(0.84)}\tiny{$^{^{***}}$} \\
presto              &  0.181(-)\tiny{$^{^{\textcolor{white}{***}}}$} &  \textbf{$<$0.001(0.59)}\tiny{$^{^{*\textcolor{white}{**}}}$} &  \textbf{$<$0.001(0.56)}\tiny{$^{^{\textcolor{white}{***}}}$} \\
protostuff          &  0.14(-)\tiny{$^{^{\textcolor{white}{***}}}$} &  0.645(-)\tiny{$^{^{\textcolor{white}{***}}}$} &  0.294(-)\tiny{$^{^{\textcolor{white}{***}}}$} \\
r2dbc-h2            &  \textbf{$<$0.001(0.72)}\tiny{$^{^{***}}$} &  \textbf{$<$0.001(0.63)}\tiny{$^{^{*\textcolor{white}{**}}}$} &  0.988(-)\tiny{$^{^{\textcolor{white}{***}}}$} \\
rdf4j               &  0.361(-)\tiny{$^{^{\textcolor{white}{***}}}$} &  0.076(-)\tiny{$^{^{\textcolor{white}{***}}}$} &  0.534(-)\tiny{$^{^{\textcolor{white}{***}}}$} \\
RoaringBitmap       &  \textbf{$<$0.001(0.93)}\tiny{$^{^{***}}$} &  0.204(-)\tiny{$^{^{\textcolor{white}{***}}}$} &  \textbf{$<$0.001(0.88)}\tiny{$^{^{***}}$} \\
RxJava              &  \textbf{0.034(0.91)}\tiny{$^{^{***}}$} &  \textbf{0.002(0.87)}\tiny{$^{^{***}}$} &  \textbf{$<$0.001(0.86)}\tiny{$^{^{***}}$} \\
SquidLib            &  \textbf{0.001(0.79)}\tiny{$^{^{***}}$} &  \textbf{0.009(0.74)}\tiny{$^{^{***}}$} &  \textbf{0.01(0.77)}\tiny{$^{^{***}}$} \\
tinkerpop           &  0.215(-)\tiny{$^{^{\textcolor{white}{***}}}$} &  0.131(-)\tiny{$^{^{\textcolor{white}{***}}}$} &  0.803(-)\tiny{$^{^{\textcolor{white}{***}}}$} \\
vert.x              &  0.7(-)\tiny{$^{^{\textcolor{white}{***}}}$} &  \textbf{$<$0.001(0.57)}\tiny{$^{^{*\textcolor{white}{**}}}$} &  0.355(-)\tiny{$^{^{\textcolor{white}{***}}}$} \\
zipkin              &  0.53(-)\tiny{$^{^{\textcolor{white}{***}}}$} &  0.064(-)\tiny{$^{^{\textcolor{white}{***}}}$} &  0.372(-)\tiny{$^{^{\textcolor{white}{***}}}$} \\
\midrule
Total               &  \textbf{0.047(0.79)}\tiny{$^{^{***}}$} &  \textbf{$<$0.001(0.68)}\tiny{$^{^{**\textcolor{white}{*}}}$} &  \textbf{$<$0.001(0.74)}\tiny{$^{^{***}}$} \\
\bottomrule
\end{tabular}

\caption{RQ$_5$. $WEE$ comparison. Results of the Wilcoxon test (with \vda effect sizes in brackets) that compare $WEE$ of dynamic reconfiguration techniques to the ones obtained using developer configurations. Systems where dynamic configurations perform better than developer configurations ($p\leq0.05$ and \vda $> 0.5$) are highlighted in bold. Asterisks denote interpretation of the \vda effect size: small ($^{*}$), medium ($^{**}$), large ($^{***}$).}
\label{tab:rq4_time}
\end{table}

\begin{figure}[p]
    \centering
	\includegraphics[width=7.5cm]{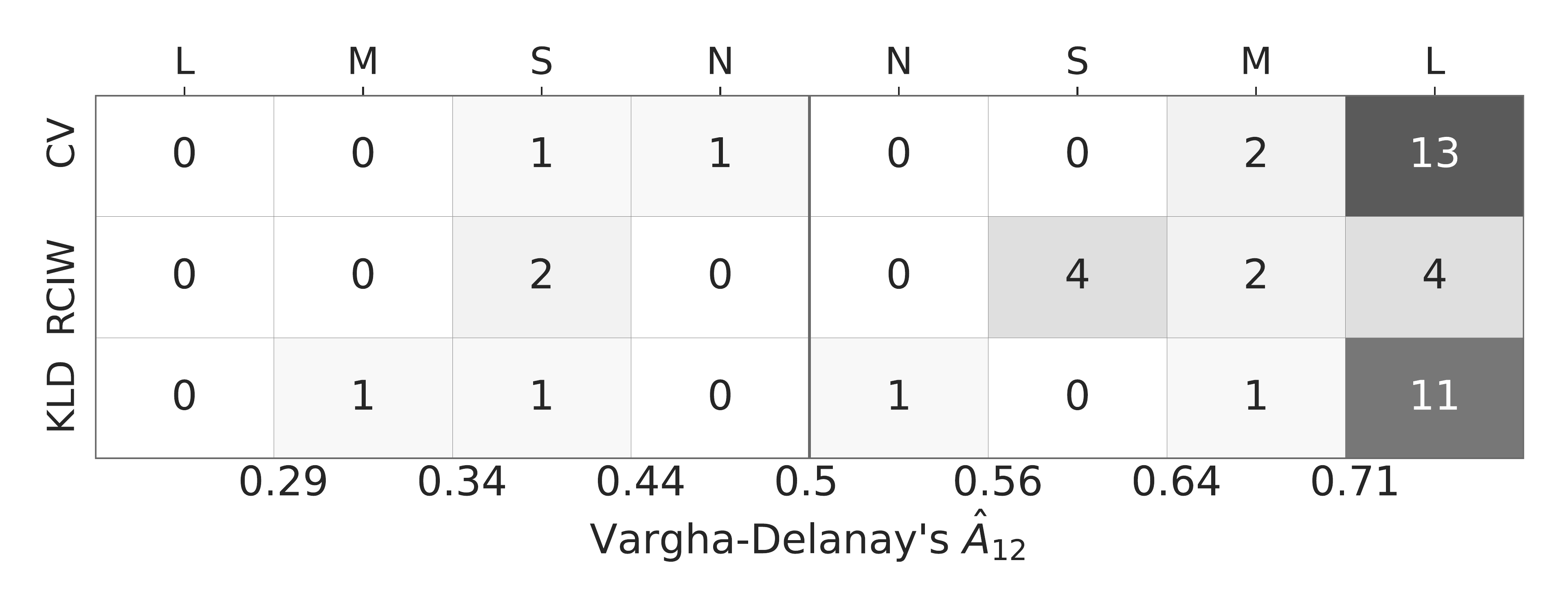}
  	\caption{RQ$_5$. $WEE$ comparison summary. Each cell reports the number of systems whose comparison leads to a statistically significant change ($p\leq~0.05$) within a specific \vda effect size range: negligible (N), small (S), medium (M) and large (L). \vda~$>0.5$ indicates that dynamic configurations perform better than developer configurations.}
  	\label{fig:rq4_timescore}
\end{figure}
 We performed the Wilcoxon test to check the significance of the difference between $WEE$ of dynamic reconfiguration techniques and developer static configurations. The detailed results of the comparison within and across systems are reported in \tabref{tab:rq4_time}. As it can be observed by the last row of the table, the differences are statistically significant for all dynamic reconfiguration techniques (p$\leq$0.05), with two techniques reporting extremely small p-values (p$<$0.001). The comparison leads to a medium effect size in $RCIW$ (\vda $\geq$ 0.64), and to a large effect size in $CV$ and $KLD$ (\vda $\geq$ 0.71). These results indicate that, when compared to developer static configurations, dynamic reconfigurations techniques overall provide more accurate estimates of the end of the warmup phase.

Nevertheless, \tabref{tab:rq4_time} also shows that the difference between $WEE$ of dynamic reconfiguration techniques and developer configurations varies across systems.
\figref{fig:rq4_timescore} shows a summary of these results, where each cell reports the number of projects whose comparison leads to a statically significant difference (p$\leq$0.05) within a specific \vda effect size range.

As it can be observed by the figure, $CV$ provides better warmup estimates than developers in 15 of the \systems systems (13 with large effect sizes, and 2 with medium effect sizes). Conversely, developer configurations provide lowers estimation errors than $CV$ in only 2 systems (respectively, negligible and small effect size).

$KLD$ shows a similar trend to that observed for $CV$. $KLD$ provides better warmup estimates than developers in 13 out of the \systems systems (11 with large effect sizes, 1 with medium effect size, and 1 with negligible effect size), whereas developers outperform $KLD$ in only 2 systems (respectively, small and medium effect size).

$RCIW$ also shows improvement over static configurations in a considerable number of systems (10 out of \systems), though with lower effect sizes (4 large, 2 medium, and 4 small). Again, developer configurations provide better estimations in only 2 systems \footnote{Interestingly, \texttt{cantaloupe} and \texttt{hive} are the same two systems where developers provide better warmup estimations than whatever dynamic reconfiguration technique.}.

Overall, we can observe that dynamic reconfiguration techniques provide more accurate warmup estimates than software developer ones.
In particular, $CV$ and $KLD$ outperform developer configurations in terms of $WEE$ on a considerable number of projects with high effect sizes.

\subsubsection{Estimated Warmup Time}\label{sec:dyn_vs_dev_wt}
\begin{table}[p]
\centering
\scriptsize
\begin{tabular}{lrrr}
\toprule
 System &   CV\tiny{$^{^{\textcolor{white}{***}}}$} &  RCIW\tiny{$^{^{\textcolor{white}{***}}}$} & KLD\tiny{$^{^{\textcolor{white}{***}}}$} \\
\midrule
arrow               &  \textbf{$<$0.001(0.87)}\tiny{$^{^{***}}$} &  \textbf{$<$0.001(0.63)}\tiny{$^{^{*\textcolor{white}{**}}}$} &  \textbf{$<$0.001(0.87)}\tiny{$^{^{***}}$} \\
byte-buddy          &  \textbf{0.002(0.56)}\tiny{$^{^{*\textcolor{white}{**}}}$} &  $<$0.001(0.21)\tiny{$^{^{***}}$} &  \textbf{$<$0.001(0.59)}\tiny{$^{^{*\textcolor{white}{**}}}$} \\
camel               &  \textbf{$<$0.001(0.75)}\tiny{$^{^{***}}$} &  0.002(0.49)\tiny{$^{^{\textcolor{white}{***}}}$} &  \textbf{$<$0.001(0.77)}\tiny{$^{^{***}}$} \\
cantaloupe          &  $<$0.001(0.22)\tiny{$^{^{***}}$} &  $<$0.001(0.05)\tiny{$^{^{***}}$} &  $<$0.001(0.06)\tiny{$^{^{***}}$} \\
client\_java         &  \textbf{$<$0.001(0.66)}\tiny{$^{^{**\textcolor{white}{*}}}$} &  $<$0.001(0.21)\tiny{$^{^{***}}$} &  \textbf{$<$0.001(0.66)}\tiny{$^{^{**\textcolor{white}{*}}}$} \\
crate               &  \textbf{$<$0.001(0.62)}\tiny{$^{^{*\textcolor{white}{**}}}$} &  $<$0.001(0.41)\tiny{$^{^{*\textcolor{white}{**}}}$} &  \textbf{$<$0.001(0.65)}\tiny{$^{^{**\textcolor{white}{*}}}$} \\
eclipse-collections &  \textbf{$<$0.001(0.59)}\tiny{$^{^{*\textcolor{white}{**}}}$} &  $<$0.001(0.42)\tiny{$^{^{*\textcolor{white}{**}}}$} &  \textbf{$<$0.001(0.58)}\tiny{$^{^{*\textcolor{white}{**}}}$} \\
h2o-3               &  \textbf{$<$0.001(0.62)}\tiny{$^{^{*\textcolor{white}{**}}}$} &  $<$0.001(0.44)\tiny{$^{^{*\textcolor{white}{**}}}$} &  \textbf{0.005(0.57)}\tiny{$^{^{*\textcolor{white}{**}}}$} \\
hazelcast           &  $<$0.001(0.39)\tiny{$^{^{*\textcolor{white}{**}}}$} &  $<$0.001(0.32)\tiny{$^{^{**\textcolor{white}{*}}}$} &  0.078(-)\tiny{$^{^{\textcolor{white}{***}}}$} \\
HdrHistogram        &  $<$0.001(0.34)\tiny{$^{^{*\textcolor{white}{**}}}$} &  $<$0.001(0.25)\tiny{$^{^{***}}$} &  $<$0.001(0.33)\tiny{$^{^{**\textcolor{white}{*}}}$} \\
hive                &  $<$0.001(0.06)\tiny{$^{^{***}}$} &  $<$0.001(0.06)\tiny{$^{^{***}}$} &  $<$0.001(0.06)\tiny{$^{^{***}}$} \\
imglib2             &  $<$0.001(0.4)\tiny{$^{^{*\textcolor{white}{**}}}$} &  $<$0.001(0.34)\tiny{$^{^{**\textcolor{white}{*}}}$} &  $<$0.001(0.39)\tiny{$^{^{*\textcolor{white}{**}}}$} \\
JCTools             &  $<$0.001(0.36)\tiny{$^{^{*\textcolor{white}{**}}}$} &  $<$0.001(0.27)\tiny{$^{^{***}}$} &  0.002(0.42)\tiny{$^{^{*\textcolor{white}{**}}}$} \\
jdbi                &  \textbf{$<$0.001(0.65)}\tiny{$^{^{**\textcolor{white}{*}}}$} &  0.004(0.47)\tiny{$^{^{\textcolor{white}{***}}}$} &  \textbf{0.002(0.63)}\tiny{$^{^{*\textcolor{white}{**}}}$} \\
jetty.project       &  $<$0.001(0.37)\tiny{$^{^{*\textcolor{white}{**}}}$} &  $<$0.001(0.28)\tiny{$^{^{***}}$} &  $<$0.001(0.42)\tiny{$^{^{*\textcolor{white}{**}}}$} \\
jgrapht             &  \textbf{$<$0.001(0.67)}\tiny{$^{^{**\textcolor{white}{*}}}$} &  $<$0.001(0.37)\tiny{$^{^{*\textcolor{white}{**}}}$} &  0.363(-)\tiny{$^{^{\textcolor{white}{***}}}$} \\
kafka               &  \textbf{$<$0.001(0.65)}\tiny{$^{^{**\textcolor{white}{*}}}$} &  0.18(-)\tiny{$^{^{\textcolor{white}{***}}}$} &  \textbf{$<$0.001(0.64)}\tiny{$^{^{*\textcolor{white}{**}}}$} \\
logbook             &  \textbf{$<$0.001(0.76)}\tiny{$^{^{***}}$} &  $<$0.001(0.33)\tiny{$^{^{**\textcolor{white}{*}}}$} &  \textbf{$<$0.001(0.75)}\tiny{$^{^{***}}$} \\
logging-log4j2      &  \textbf{$<$0.001(0.75)}\tiny{$^{^{***}}$} &  0.525(-)\tiny{$^{^{\textcolor{white}{***}}}$} &  \textbf{$<$0.001(0.75)}\tiny{$^{^{***}}$} \\
netty               &  \textbf{$<$0.001(0.69)}\tiny{$^{^{**\textcolor{white}{*}}}$} &  \textbf{$<$0.001(0.58)}\tiny{$^{^{*\textcolor{white}{**}}}$} &  \textbf{$<$0.001(0.74)}\tiny{$^{^{***}}$} \\
presto              &  0.075(-)\tiny{$^{^{\textcolor{white}{***}}}$} &  $<$0.001(0.15)\tiny{$^{^{***}}$} &  $<$0.001(0.3)\tiny{$^{^{**\textcolor{white}{*}}}$} \\
protostuff          &  $<$0.001(0.34)\tiny{$^{^{**\textcolor{white}{*}}}$} &  $<$0.001(0.23)\tiny{$^{^{***}}$} &  $<$0.001(0.35)\tiny{$^{^{*\textcolor{white}{**}}}$} \\
r2dbc-h2            &  $<$0.001(0.2)\tiny{$^{^{***}}$} &  $<$0.001(0.05)\tiny{$^{^{***}}$} &  $<$0.001(0.35)\tiny{$^{^{*\textcolor{white}{**}}}$} \\
rdf4j               &  \textbf{$<$0.001(0.7)}\tiny{$^{^{**\textcolor{white}{*}}}$} &  0.383(-)\tiny{$^{^{\textcolor{white}{***}}}$} &  0.25(-)\tiny{$^{^{\textcolor{white}{***}}}$} \\
RoaringBitmap       &  \textbf{$<$0.001(0.83)}\tiny{$^{^{***}}$} &  \textbf{$<$0.001(0.71)}\tiny{$^{^{**\textcolor{white}{*}}}$} &  \textbf{$<$0.001(0.8)}\tiny{$^{^{***}}$} \\
RxJava              &  \textbf{$<$0.001(0.72)}\tiny{$^{^{***}}$} &  \textbf{0.021(0.6)}\tiny{$^{^{*\textcolor{white}{**}}}$} &  \textbf{$<$0.001(0.72)}\tiny{$^{^{***}}$} \\
SquidLib            &  \textbf{$<$0.001(0.67)}\tiny{$^{^{**\textcolor{white}{*}}}$} &  \textbf{$<$0.001(0.53)}\tiny{$^{^{\textcolor{white}{***}}}$} &  \textbf{$<$0.001(0.64)}\tiny{$^{^{**\textcolor{white}{*}}}$} \\
tinkerpop           &  0.33(-)\tiny{$^{^{\textcolor{white}{***}}}$} &  $<$0.001(0.32)\tiny{$^{^{**\textcolor{white}{*}}}$} &  \textbf{0.008(0.52)}\tiny{$^{^{\textcolor{white}{***}}}$} \\
vert.x              &  0.958(-)\tiny{$^{^{\textcolor{white}{***}}}$} &  $<$0.001(0.13)\tiny{$^{^{***}}$} &  \textbf{0.03(0.62)}\tiny{$^{^{*\textcolor{white}{**}}}$} \\
zipkin              &  $<$0.001(0.31)\tiny{$^{^{**\textcolor{white}{*}}}$} &  $<$0.001(0.2)\tiny{$^{^{***}}$} &  $<$0.001(0.41)\tiny{$^{^{*\textcolor{white}{**}}}$} \\\midrule
Total               &  \textbf{$<$0.001(0.55)}\tiny{$^{^{\textcolor{white}{***}}}$} &  $<$0.001(0.36)\tiny{$^{^{*\textcolor{white}{**}}}$} &  \textbf{$<$0.001(0.55)}\tiny{$^{^{\textcolor{white}{***}}}$} \\
\bottomrule
\end{tabular}

\caption{RQ$_5$. Estimated warmup time ($wt$) comparison. Results of the Wilcoxon test (with \vda effect sizes in brackets) that compare the $wt$ of dynamic reconfiguration techniques to the ones obtained using developers configurations. Systems where dynamic configurations report shorter $wt$ than developer configurations ($p\leq0.05$ and \vda $> 0.5$) are highlighted in bold. Asterisks denote interpretation of the \vda effect size: small ($^{*}$), medium ($^{**}$), large ($^{***}$).}
\label{tab:rq4_wt}
\end{table}

\begin{figure}[p]
    \centering
	\includegraphics[width=7.5cm]{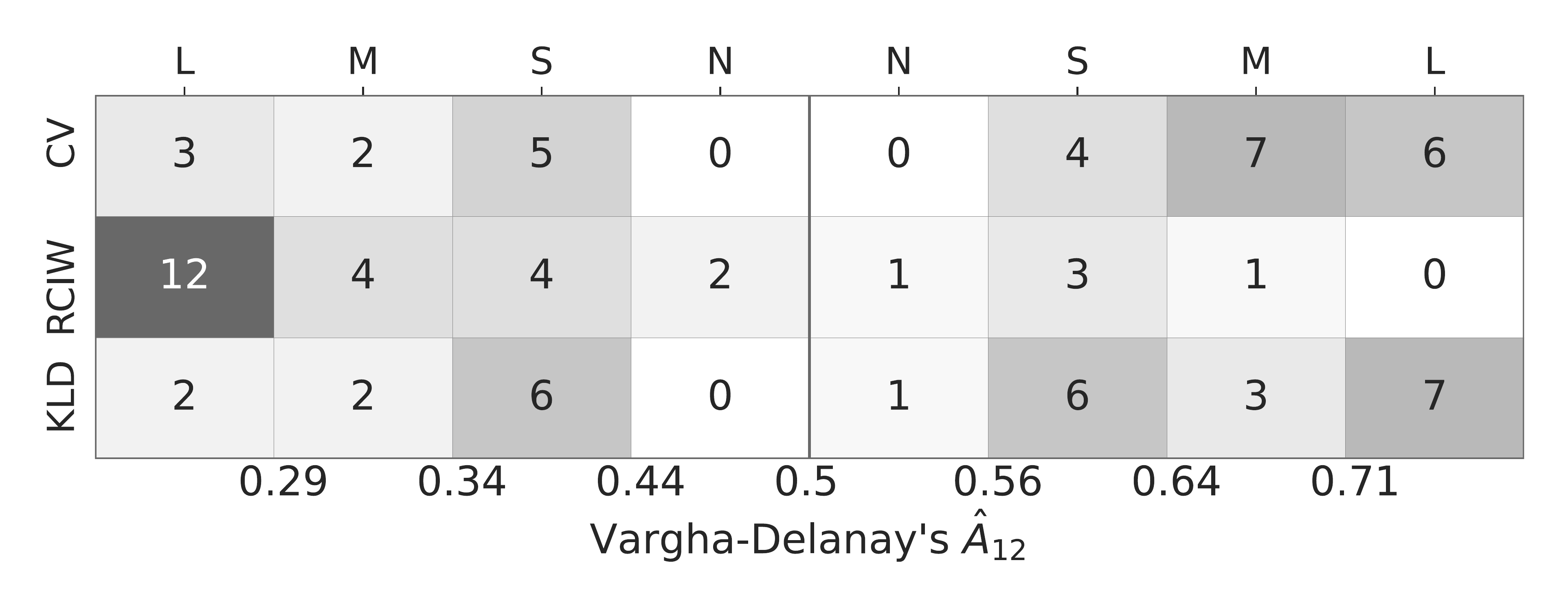}
  	\caption{RQ$_5$. Summary $wt$ comparison. Each cell reports the number of systems whose comparison leads to a statistically significant change ($p\leq~0.05$) within a specific \vda effect size range: negligible (N), small (S), medium (M) and large (L). \vda~$>0.5$ indicates that dynamic configurations lead to shorter $wt$ than developer configurations.}
  	\label{fig:rq4_wt}
\end{figure}


In this subsection, we investigate the difference between the estimated warmup time ($wt$) provided by dynamic reconfiguration techniques and developer static configurations.
\tabref{tab:rq4_wt} reports results of Wilcoxon tests for each system, and across all systems.

As it can be observed by the last row of the table, $CV$ and $KLD$ report a statistically significant difference (p$<$0.001) with tendency toward improvement (\ie smaller $wt$ values) but with negligible effect size. 
$RCIW$ also reports statistically significant difference (p$<$0.001), but with the opposite tendency, \ie larger $wt$ (\vda $<$ 0.5), and a small effect size.

If we look at project-level results, we can observe a remarkable diversity among projects for $CV$ and $KLD$.
As it can be seen in \figref{fig:rq4_wt}, $CV$ leads to higher $wt$ than those defined by developers (\vda $<$ 0.5) in 10 of the \systems systems, and it reports lower $wt$ (\vda $>$ 0.5) in 17 systems, thus not showing a clear trend.
A similar behavior can be observed for $KLD$: 10 of the 28 systems that report a statically significant difference (p $\leq$ 0.05) have \vda$<$0.5, while 17 of them have \vda$>$0.5.

On the other hand, $RCIW$ performs worse than developer configurations in most of the systems. $RCIW$ leads to higher $wt$ than developers in 22 of the \systems systems: 12 with large effect sizes, 4 with medium effect sizes, and 4 with small effect sizes, 2 with negligible effect sizes. It leads to shorter $wt$ in only 5 projects (1 negligible, 3 small, and 1 medium effect sizes).

Furthermore, if we compare overestimation frequency/side effect of $RCIW$ (see Figs.~\ref{fig:warmup_estimation_DyCo_barplot} and \ref{fig:overestimation_DyCo}) and developer configurations (see Figs. \ref{fig:developer_configurations} and \ref{fig:overestimation_dev}), we can observe that $RCIW$ reports more frequent overestimations, namely 72\% \emph{vs} 48\%, and higher \emph{time wastes}, namely median of 46 seconds (IQR: 31-55 seconds) \emph{vs} median of 33 seconds (IQR: 19-49 seconds).

Overall, our results indicate that $CV$ and $KLD$ can lead to different behaviors based on the context, that is they can provide either higher or lower $wt$ than developers depending on the system.
On the other hand, $RCIW$ induces higher estimates of the warmup time $wt$ when compared to developer configurations, thus increasing microbenchmark execution time.

\subsubsection{Relative Performance Deviation}
\begin{table}[p]
\centering
\scriptsize
\begin{tabular}{lrrr}
\toprule
 System &   CV\tiny{$^{^{\textcolor{white}{***}}}$} &  RCIW\tiny{$^{^{\textcolor{white}{***}}}$} & KLD\tiny{$^{^{\textcolor{white}{***}}}$} \\
 \midrule
arrow               &  $<$0.001(0.48)\tiny{$^{^{\textcolor{white}{***}}}$} &  0.145(-)\tiny{$^{^{\textcolor{white}{***}}}$} &  $<$0.001(0.46)\tiny{$^{^{\textcolor{white}{***}}}$} \\
byte-buddy          &  \textbf{$<$0.001(0.53)}\tiny{$^{^{\textcolor{white}{***}}}$} &  \textbf{0.005(0.6)}\tiny{$^{^{*\textcolor{white}{**}}}$} &  \textbf{$<$0.001(0.53)}\tiny{$^{^{\textcolor{white}{***}}}$} \\
camel               &  $<$0.001(0.44)\tiny{$^{^{\textcolor{white}{***}}}$} &  $<$0.001(0.5)\tiny{$^{^{\textcolor{white}{***}}}$} &  $<$0.001(0.4)\tiny{$^{^{*\textcolor{white}{**}}}$} \\
cantaloupe          &  \textbf{$<$0.001(0.66)}\tiny{$^{^{**\textcolor{white}{*}}}$} &  \textbf{$<$0.001(0.67)}\tiny{$^{^{**\textcolor{white}{*}}}$} &  \textbf{$<$0.001(0.67)}\tiny{$^{^{**\textcolor{white}{*}}}$} \\
client\_java         &  \textbf{0.001(0.54)}\tiny{$^{^{\textcolor{white}{***}}}$} &  \textbf{0.048(0.55)}\tiny{$^{^{\textcolor{white}{***}}}$} &  \textbf{$<$0.001(0.53)}\tiny{$^{^{\textcolor{white}{***}}}$} \\
crate               &  $<$0.001(0.48)\tiny{$^{^{\textcolor{white}{***}}}$} &  0.346(-)\tiny{$^{^{\textcolor{white}{***}}}$} &  $<$0.001(0.46)\tiny{$^{^{\textcolor{white}{***}}}$} \\
eclipse-collections &  \textbf{$<$0.001(0.53)}\tiny{$^{^{\textcolor{white}{***}}}$} &  0.295(-)\tiny{$^{^{\textcolor{white}{***}}}$} &  \textbf{$<$0.001(0.5)}\tiny{$^{^{\textcolor{white}{***}}}$} \\
h2o-3               &  $<$0.001(0.5)\tiny{$^{^{\textcolor{white}{***}}}$} &  \textbf{0.019(0.56)}\tiny{$^{^{\textcolor{white}{***}}}$} &  $<$0.001(0.5)\tiny{$^{^{\textcolor{white}{***}}}$} \\
hazelcast           &  \textbf{$<$0.001(0.76)}\tiny{$^{^{***}}$} &  \textbf{$<$0.001(0.77)}\tiny{$^{^{***}}$} &  \textbf{0.008(0.76)}\tiny{$^{^{***}}$} \\
HdrHistogram        &  0.296(-)\tiny{$^{^{\textcolor{white}{***}}}$} &  \textbf{0.007(0.61)}\tiny{$^{^{*\textcolor{white}{**}}}$} &  \textbf{0.012(0.6)}\tiny{$^{^{*\textcolor{white}{**}}}$} \\
hive                &  \textbf{$<$0.001(0.52)}\tiny{$^{^{\textcolor{white}{***}}}$} &  \textbf{$<$0.001(0.52)}\tiny{$^{^{\textcolor{white}{***}}}$} &  \textbf{0.002(0.52)}\tiny{$^{^{\textcolor{white}{***}}}$} \\
imglib2             &  \textbf{$<$0.001(0.8)}\tiny{$^{^{***}}$} &  \textbf{$<$0.001(0.78)}\tiny{$^{^{***}}$} &  \textbf{$<$0.001(0.76)}\tiny{$^{^{***}}$} \\
JCTools             &  0.739(-)\tiny{$^{^{\textcolor{white}{***}}}$} &  \textbf{0.031(0.59)}\tiny{$^{^{*\textcolor{white}{**}}}$} &  \textbf{0.042(0.57)}\tiny{$^{^{*\textcolor{white}{**}}}$} \\
jdbi                &  \textbf{$<$0.001(0.54)}\tiny{$^{^{\textcolor{white}{***}}}$} &  0.282(-)\tiny{$^{^{\textcolor{white}{***}}}$} &  \textbf{$<$0.001(0.54)}\tiny{$^{^{\textcolor{white}{***}}}$} \\
jetty.project       &  \textbf{0.006(0.69)}\tiny{$^{^{**\textcolor{white}{*}}}$} &  \textbf{$<$0.001(0.69)}\tiny{$^{^{**\textcolor{white}{*}}}$} &  \textbf{0.019(0.66)}\tiny{$^{^{**\textcolor{white}{*}}}$} \\
jgrapht             &  \textbf{0.002(0.52)}\tiny{$^{^{\textcolor{white}{***}}}$} &  0.114(-)\tiny{$^{^{\textcolor{white}{***}}}$} &  0.179(-)\tiny{$^{^{\textcolor{white}{***}}}$} \\
kafka               &  $<$0.001(0.44)\tiny{$^{^{\textcolor{white}{***}}}$} &  0.076(-)\tiny{$^{^{\textcolor{white}{***}}}$} &  $<$0.001(0.41)\tiny{$^{^{*\textcolor{white}{**}}}$} \\
logbook             &  $<$0.001(0.44)\tiny{$^{^{\textcolor{white}{***}}}$} &  \textbf{0.004(0.51)}\tiny{$^{^{\textcolor{white}{***}}}$} &  $<$0.001(0.39)\tiny{$^{^{*\textcolor{white}{**}}}$} \\
logging-log4j2      &  0.079(-)\tiny{$^{^{\textcolor{white}{***}}}$} &  0.334(-)\tiny{$^{^{\textcolor{white}{***}}}$} &  \textbf{0.007(0.51)}\tiny{$^{^{\textcolor{white}{***}}}$} \\
netty               &  $<$0.001(0.48)\tiny{$^{^{\textcolor{white}{***}}}$} &  0.059(-)\tiny{$^{^{\textcolor{white}{***}}}$} &  $<$0.001(0.44)\tiny{$^{^{*\textcolor{white}{**}}}$} \\
presto              &  0.753(-)\tiny{$^{^{\textcolor{white}{***}}}$} &  \textbf{$<$0.001(0.66)}\tiny{$^{^{**\textcolor{white}{*}}}$} &  \textbf{0.021(0.63)}\tiny{$^{^{*\textcolor{white}{**}}}$} \\
protostuff          &  \textbf{$<$0.001(0.79)}\tiny{$^{^{***}}$} &  \textbf{$<$0.001(0.82)}\tiny{$^{^{***}}$} &  \textbf{$<$0.001(0.78)}\tiny{$^{^{***}}$} \\
r2dbc-h2            &  \textbf{$<$0.001(0.84)}\tiny{$^{^{***}}$} &  \textbf{$<$0.001(0.87)}\tiny{$^{^{***}}$} &  \textbf{$<$0.001(0.82)}\tiny{$^{^{***}}$} \\
rdf4j               &  $<$0.001(0.45)\tiny{$^{^{\textcolor{white}{***}}}$} &  0.827(-)\tiny{$^{^{\textcolor{white}{***}}}$} &  0.007(0.47)\tiny{$^{^{\textcolor{white}{***}}}$} \\
RoaringBitmap       &  \textbf{$<$0.001(0.51)}\tiny{$^{^{\textcolor{white}{***}}}$} &  \textbf{$<$0.001(0.53)}\tiny{$^{^{\textcolor{white}{***}}}$} &  $<$0.001(0.48)\tiny{$^{^{\textcolor{white}{***}}}$} \\
RxJava              &  0.003(0.49)\tiny{$^{^{\textcolor{white}{***}}}$} &  0.35(-)\tiny{$^{^{\textcolor{white}{***}}}$} &  0.003(0.5)\tiny{$^{^{\textcolor{white}{***}}}$} \\
SquidLib            &  $<$0.001(0.41)\tiny{$^{^{*\textcolor{white}{**}}}$} &  0.003(0.45)\tiny{$^{^{\textcolor{white}{***}}}$} &  $<$0.001(0.42)\tiny{$^{^{*\textcolor{white}{**}}}$} \\
tinkerpop           &  \textbf{$<$0.001(0.51)}\tiny{$^{^{\textcolor{white}{***}}}$} &  \textbf{0.044(0.6)}\tiny{$^{^{*\textcolor{white}{**}}}$} &  $<$0.001(0.47)\tiny{$^{^{\textcolor{white}{***}}}$} \\
vert.x              &  \textbf{0.039(0.54)}\tiny{$^{^{\textcolor{white}{***}}}$} &  \textbf{0.001(0.57)}\tiny{$^{^{*\textcolor{white}{**}}}$} &  \textbf{$<$0.001(0.5)}\tiny{$^{^{\textcolor{white}{***}}}$} \\
zipkin              &  \textbf{$<$0.001(0.78)}\tiny{$^{^{***}}$} &  \textbf{$<$0.001(0.8)}\tiny{$^{^{***}}$} &  \textbf{$<$0.001(0.75)}\tiny{$^{^{***}}$} \\
\midrule
Total               &  \textbf{$<$0.001(0.57)}\tiny{$^{^{*\textcolor{white}{**}}}$} &  \textbf{$<$0.001(0.6)}\tiny{$^{^{*\textcolor{white}{**}}}$} &  \textbf{$<$0.001(0.55)}\tiny{$^{^{\textcolor{white}{***}}}$} \\
\bottomrule
\end{tabular}

\caption{RQ$_5$. $RPD$ fork level comparison. Results of the Wilcoxon test (with \vda effect sizes in brackets) that compare the $RPD$ of dynamic reconfiguration techniques to the ones obtained using developer configurations. Projects where dynamic configurations perform better than developer configurations ($p\leq0.05$ and \vda $> 0.5$) are highlighted in bold. Asterisks denote interpretation of the \vda effect size: small ($^{*}$), medium ($^{**}$), large ($^{***}$).}
\label{tab:rq4_perf}
\end{table}

\begin{figure}[p]
	\center
	\includegraphics[width=7.5cm]{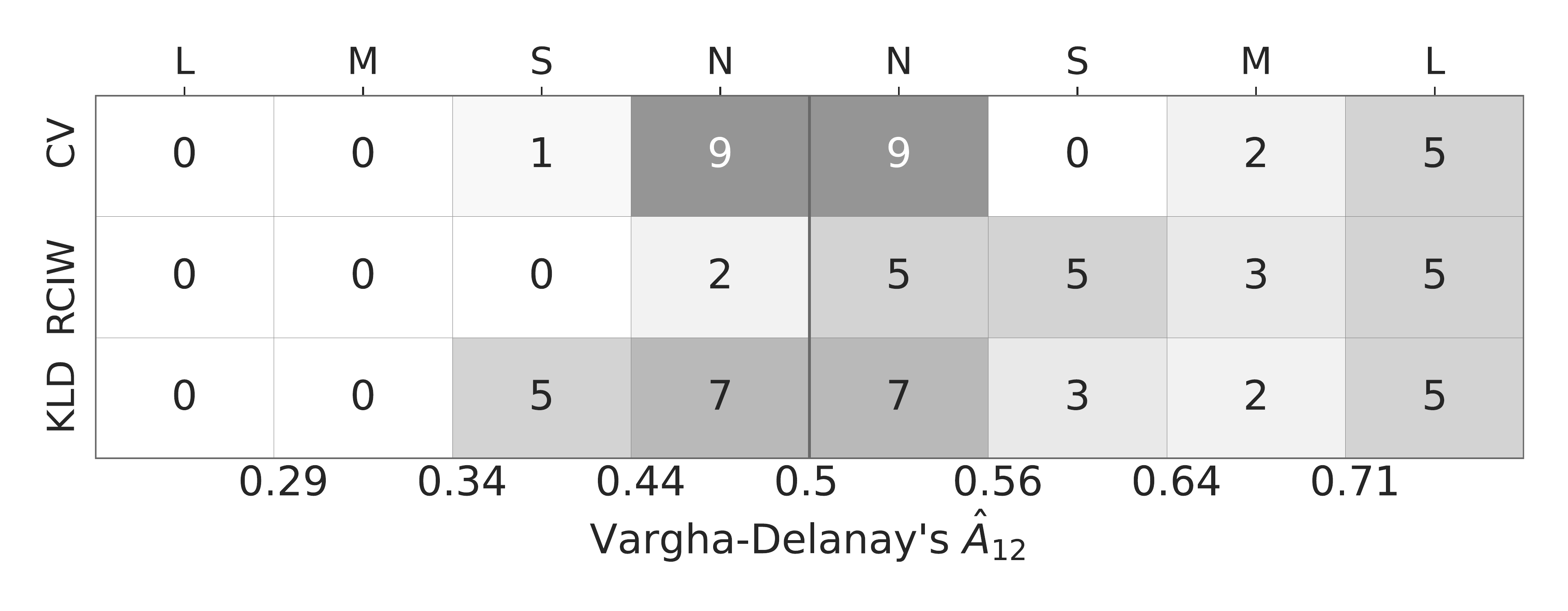}
  	\caption{RQ$_5$. Summary $RPD$ fork level comparison. Each cell reports the number of systems whose comparison leads to a statistically significant change ($p\leq~0.05$) within a specific \vda effect size range: negligible (N), small (S), medium (M) and large (L). \vda~$>0.5$ indicates that dynamic configurations perform better than developer configurations.}
  	\label{fig:rq4_perfscore}
\end{figure}

In this subsection, we assess the difference between $RPD$ of dynamic reconfiguration techniques and developer static configurations.
Overall, we can observe that dynamic reconfiguration techniques slightly improve developer configurations, as shown in the last row of \tabref{tab:rq4_perf}. The comparisons report statistically significant differences for all techniques (p$<$0.001), respectively with small ($CV$ and $RCIW$) and negligible ($KLD$) effect sizes.

If we look at project-level results (see \tabref{tab:rq4_perf} and \figref{fig:rq4_perfscore}), we can observe that $RCIW$ leads to statistically significant improvements over developer configurations in a large number of projects. In particular, $RCIW$ improves developer configurations in 18 of the \systems systems (5 with large, 3 with medium, 5 with small, and 5 with negligible effect sizes), while it degrades $RPD$ in only 2 systems (both with negligible effect sizes).
By comparing the $RPD$ distributions of $RCIW$ (\figref{fig:underestimation_DyCo}) and developer configurations (\figref{fig:underestimation_dev}), we can observe that the former produces lower deviations with respect to steady state measurements. For example, using developer configurations, about 57\% of the forks lead to an $RPD$ of at least 5\%, while the same performance deviation is achieved in 43\% of forks when using $RCIW$.
Even more, $RCIW$ provides a median $RPD$ of 3\% (IQR: 0-11\%), whereas developer configurations lead to a deviation of 7\% (IQR: 1-21\%).

These results demonstrate that performance measurements gathered through $RCIW$ deviate less from steady state measurements than those collected through developer static configurations, thereby ensuring better results quality.

$CV$ and $KLD$ also report statistically significant improvements over static configurations in 16 and 17 systems, respectively. Nonetheless, developer configurations perform better than $CV$ and $KLD$ in, respectively, 10 and 12 systems.
Despite this, by looking at \figref{fig:rq4_perfscore}, we can observe that systems where developer configurations perform better than $CV$ and $KLD$, tend to have lower effect sizes than those where they provide improvement.
For example, if we exclude negligible effect sizes, we can observe that $CV$ still leads to improvement (\vda $\geq$ 0.56) in 7 systems (5 large and 2 medium), while it performs worse than developer configurations (\vda $\leq$ 0.44) in only one case with small effect size. 
Similarly, $KLD$ leads to an effect size \vda $\geq$ 0.56 in 10 systems (5 large, 2 medium, 3 small), while it reports a small effect size \vda $\leq$ 0.44 in 5 systems.

These results suggest that the comparison between $RPD$ of $CV$/$KLD$ and developer configurations lead to different outcomes depending on the system, though with a slight overall tendency towards improvement.
The results across benchmarks of all systems confirm this tendency. As shown in the last row of \tabref{tab:rq4_perf}, $CV$ and $KLD$ report statistically significant improvements  (p$<$0.001), respectively with small and negligible effect sizes.
Given these results, we can safely state that $CV$ and $KLD$ only provide marginal improvements over developer static configurations in terms of performance deviation.


\subsubsection{Benchmark level assessment}
\begin{table}[p]
\centering
\scriptsize
\begin{tabular}{lrrr}
\toprule
 System &   CV\tiny{$^{^{\textcolor{white}{***}}}$} &  RCIW\tiny{$^{^{\textcolor{white}{***}}}$} & KLD\tiny{$^{^{\textcolor{white}{***}}}$} \\
\midrule
arrow               &  0.015(0.29)\tiny{$^{^{**\textcolor{white}{*}}}$} &  0.285(-)\tiny{$^{^{\textcolor{white}{***}}}$} &  0.046(0.37)\tiny{$^{^{*\textcolor{white}{**}}}$} \\
byte-buddy          &  0.084(-)\tiny{$^{^{\textcolor{white}{***}}}$} &  \textbf{0.015(0.73)}\tiny{$^{^{***}}$} &  0.196(-)\tiny{$^{^{\textcolor{white}{***}}}$} \\
camel               &  0.001(0.19)\tiny{$^{^{***}}$} &  0.18(-)\tiny{$^{^{\textcolor{white}{***}}}$} &  $<$0.001(0.11)\tiny{$^{^{***}}$} \\
cantaloupe          &  \textbf{0.001(0.79)}\tiny{$^{^{***}}$} &  \textbf{0.001(0.83)}\tiny{$^{^{***}}$} &  \textbf{0.002(0.81)}\tiny{$^{^{***}}$} \\
client\_java         &  0.008(0.4)\tiny{$^{^{*\textcolor{white}{**}}}$} &  \textbf{0.043(0.57)}\tiny{$^{^{*\textcolor{white}{**}}}$} &  $<$0.001(0.29)\tiny{$^{^{**\textcolor{white}{*}}}$} \\
crate               &  0.131(-)\tiny{$^{^{\textcolor{white}{***}}}$} &  0.735(-)\tiny{$^{^{\textcolor{white}{***}}}$} &  0.114(-)\tiny{$^{^{\textcolor{white}{***}}}$} \\
eclipse-collections &  0.002(0.2)\tiny{$^{^{***}}$} &  0.347(-)\tiny{$^{^{\textcolor{white}{***}}}$} &  0.005(0.29)\tiny{$^{^{***}}$} \\
h2o-3               &  0.002(0.25)\tiny{$^{^{***}}$} &  0.069(-)\tiny{$^{^{\textcolor{white}{***}}}$} &  0.05(0.39)\tiny{$^{^{*\textcolor{white}{**}}}$} \\
hazelcast           &  \textbf{0.004(0.76)}\tiny{$^{^{***}}$} &  \textbf{0.004(0.8)}\tiny{$^{^{***}}$} &  \textbf{0.009(0.71)}\tiny{$^{^{**\textcolor{white}{*}}}$} \\
HdrHistogram        &  0.776(-)\tiny{$^{^{\textcolor{white}{***}}}$} &  0.074(-)\tiny{$^{^{\textcolor{white}{***}}}$} &  \textbf{0.028(0.61)}\tiny{$^{^{*\textcolor{white}{**}}}$} \\
hive                &  \textbf{0.03(0.72)}\tiny{$^{^{***}}$} &  \textbf{0.031(0.73)}\tiny{$^{^{***}}$} &  0.056(-)\tiny{$^{^{\textcolor{white}{***}}}$} \\
imglib2             &  \textbf{0.035(0.59)}\tiny{$^{^{*\textcolor{white}{**}}}$} &  0.068(-)\tiny{$^{^{\textcolor{white}{***}}}$} &  \textbf{0.015(0.59)}\tiny{$^{^{*\textcolor{white}{**}}}$} \\
JCTools             &  0.144(-)\tiny{$^{^{\textcolor{white}{***}}}$} &  0.655(-)\tiny{$^{^{\textcolor{white}{***}}}$} &  0.317(-)\tiny{$^{^{\textcolor{white}{***}}}$} \\
jdbi                &  0.055(-)\tiny{$^{^{\textcolor{white}{***}}}$} &  \textbf{0.025(0.71)}\tiny{$^{^{***}}$} &  0.087(-)\tiny{$^{^{\textcolor{white}{***}}}$} \\
jetty.project       &  0.064(-)\tiny{$^{^{\textcolor{white}{***}}}$} &  \textbf{0.003(0.76)}\tiny{$^{^{***}}$} &  0.099(-)\tiny{$^{^{\textcolor{white}{***}}}$} \\
jgrapht             &  0.756(-)\tiny{$^{^{\textcolor{white}{***}}}$} &  0.311(-)\tiny{$^{^{\textcolor{white}{***}}}$} &  0.51(-)\tiny{$^{^{\textcolor{white}{***}}}$} \\
kafka               &  $<$0.001(0.15)\tiny{$^{^{***}}$} &  \textbf{0.018(0.68)}\tiny{$^{^{**\textcolor{white}{*}}}$} &  0.002(0.24)\tiny{$^{^{***}}$} \\
logbook             &  0.046(0.36)\tiny{$^{^{*\textcolor{white}{**}}}$} &  \textbf{0.037(0.66)}\tiny{$^{^{**\textcolor{white}{*}}}$} &  0.019(0.26)\tiny{$^{^{***}}$} \\
logging-log4j2      &  0.463(-)\tiny{$^{^{\textcolor{white}{***}}}$} &  0.18(-)\tiny{$^{^{\textcolor{white}{***}}}$} &  0.31(-)\tiny{$^{^{\textcolor{white}{***}}}$} \\
netty               &  0.091(-)\tiny{$^{^{\textcolor{white}{***}}}$} &  \textbf{0.028(0.65)}\tiny{$^{^{**\textcolor{white}{*}}}$} &  0.055(-)\tiny{$^{^{\textcolor{white}{***}}}$} \\
presto              &  0.861(-)\tiny{$^{^{\textcolor{white}{***}}}$} &  \textbf{0.026(0.67)}\tiny{$^{^{**\textcolor{white}{*}}}$} &  \textbf{0.041(0.61)}\tiny{$^{^{*\textcolor{white}{**}}}$} \\
protostuff          &  \textbf{$<$0.001(0.73)}\tiny{$^{^{***}}$} &  \textbf{$<$0.001(0.88)}\tiny{$^{^{***}}$} &  \textbf{0.001(0.74)}\tiny{$^{^{***}}$} \\
r2dbc-h2            &  \textbf{0.006(0.76)}\tiny{$^{^{***}}$} &  \textbf{$<$0.001(0.88)}\tiny{$^{^{***}}$} &  \textbf{0.013(0.68)}\tiny{$^{^{**\textcolor{white}{*}}}$} \\
rdf4j               &  0.015(0.38)\tiny{$^{^{*\textcolor{white}{**}}}$} &  0.48(-)\tiny{$^{^{\textcolor{white}{***}}}$} &  0.551(-)\tiny{$^{^{\textcolor{white}{***}}}$} \\
RoaringBitmap       &  0.001(0.24)\tiny{$^{^{***}}$} &  0.009(0.4)\tiny{$^{^{*\textcolor{white}{**}}}$} &  0.028(0.34)\tiny{$^{^{**\textcolor{white}{*}}}$} \\
RxJava              &  0.463(-)\tiny{$^{^{\textcolor{white}{***}}}$} &  \textbf{0.005(0.71)}\tiny{$^{^{**\textcolor{white}{*}}}$} &  0.959(-)\tiny{$^{^{\textcolor{white}{***}}}$} \\
SquidLib            &  0.011(0.31)\tiny{$^{^{**\textcolor{white}{*}}}$} &  0.861(-)\tiny{$^{^{\textcolor{white}{***}}}$} &  0.079(-)\tiny{$^{^{\textcolor{white}{***}}}$} \\
tinkerpop           &  $<$0.001(0.12)\tiny{$^{^{***}}$} &  0.155(-)\tiny{$^{^{\textcolor{white}{***}}}$} &  $<$0.001(0.14)\tiny{$^{^{***}}$} \\
vert.x              &  0.114(-)\tiny{$^{^{\textcolor{white}{***}}}$} &  \textbf{0.005(0.79)}\tiny{$^{^{***}}$} &  0.286(-)\tiny{$^{^{\textcolor{white}{***}}}$} \\
zipkin              &  0.148(-)\tiny{$^{^{\textcolor{white}{***}}}$} &  \textbf{0.005(0.73)}\tiny{$^{^{***}}$} &  0.14(-)\tiny{$^{^{\textcolor{white}{***}}}$} \\
\midrule
Total               &  $<$0.001(0.45)\tiny{$^{^{\textcolor{white}{***}}}$} &  \textbf{$<$0.001(0.64)}\tiny{$^{^{*\textcolor{white}{**}}}$} &  0.186(-)\tiny{$^{^{\textcolor{white}{***}}}$} \\
\bottomrule
\end{tabular}
\caption{RQ$_5$. $RPD$ benchmark level comparison. Results of the Wilcoxon test (with \vda effect sizes in brackets) that compare $RPD$ of dynamic reconfiguration techniques to the ones obtained using developer configurations. Systems where dynamic configurations perform better than developer configurations ($p\leq0.05$ and \vda $> 0.5$) are highlighted in bold. Asterisks denote interpretation of the \vda effect size: small ($^{*}$), medium ($^{**}$), large ($^{***}$).}
\label{tab:rq4_benchlevel}
\end{table}

\begin{figure}[p]
    \centering
	\includegraphics[width=7.5cm]{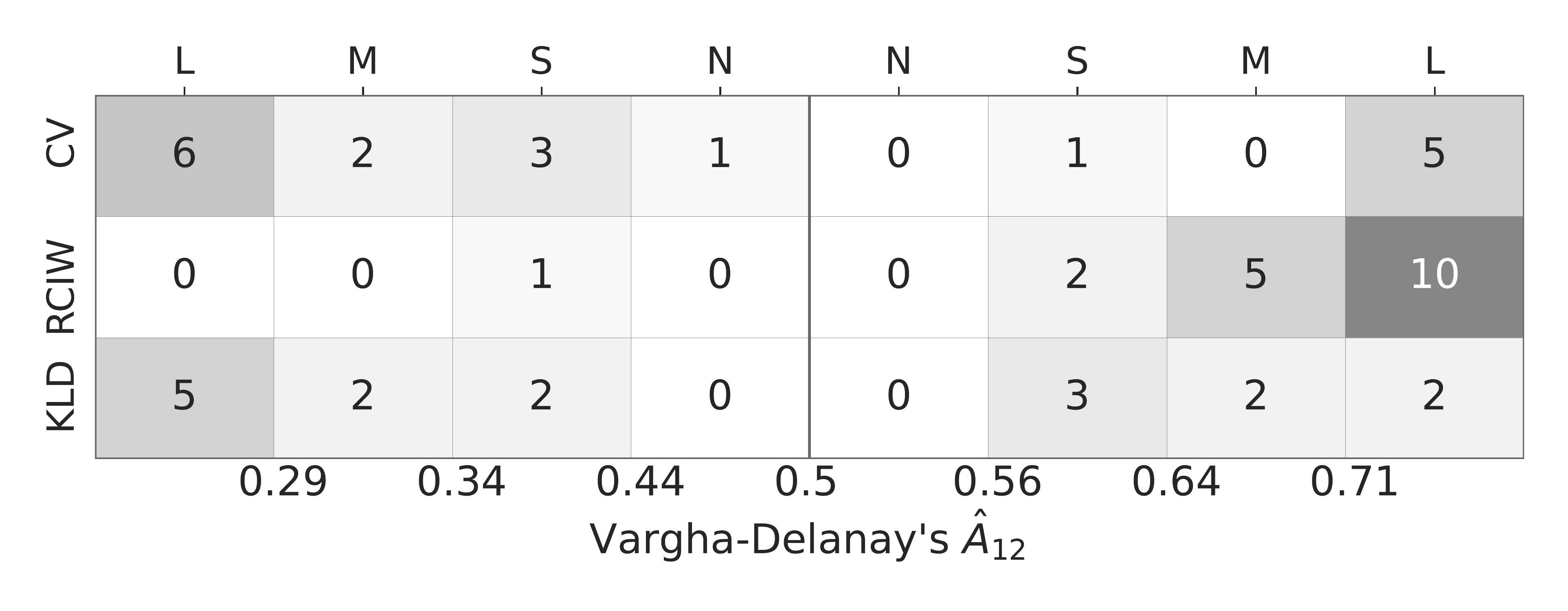}
  	\caption{RQ$_5$. Summary $RPD$ benchmark level comparison. Each cell reports the number of systems whose comparison leads to a statistically significant change ($p\leq~0.05$) within a specific \vda effect size range: negligible (N), small (S), medium (M) and large (L). \vda~$>0.5$ indicates that dynamic configurations perform better than developer configurations.}
  	\label{fig:rq4_benchlevel}
\end{figure}

Interestingly, when we look at benchmark level results, we can observe significant differences
 (see \figref{fig:rq4_benchlevel} and \tabref{tab:rq4_benchlevel}). \emph{CV} notably shifts from a tendency towards improvement to a tendency towards regression ($p$$<$$0.001$ and \vda=0.45).
\emph{KLD}, which reported statistically significant differences and tendency towards improvement at fork level, reports neither improvement nor regression ($p>0.05$) at benchmark level. By analyzing project level results, we can further appreciate this shift.
At fork level, \emph{CV} reports worse RPDs than developers in only one project (with non-negligible effect size). At benchmark level instead, if we exclude negligible effect sizes, it reports worse performance deviations than developer configurations in 11 projects (3 with small effect sizes, 2 with medium, and 6 with large).
Likewise, \emph{KLD} reports worse RPDs in 9 projects at benchmark level (2 small, 2 medium and 5 large effect sizes), while, at fork level, it reports worse RPDs in only 5 project with small effect sizes.
The only technique that seems to provide improvement over developer configurations both at benchmark and fork level is \emph{RCIW}, which reports statistically significant improvement ($p<0.001$) and small effect size (\vda=0.64). 
\emph{RCIW} outperforms developer configurations in 17 projects (10 with large effect sizes, 5 with medium and 2 with small), and the effect sizes provided at benchmark level are even better than those provided at fork level (\ie higher \vda) in 15 out of the 30 projects.
These results may suggest that the capability of \emph{RCIW} to dynamically stop forks at run-time may further improve performance deviations when compared to those of developers.
To further investigate this aspect, we analyzed the results of these 15 projects, and we found that, in 80\% of the benchmarks, \emph{RCIW} involves 5 fork executions, \ie the maximum number of forks for dynamic reconfiguration techniques (based on the original parameterization provided in \citep{Laaber2020}). That is, forks are not halted by stability criteria, rather they are stopped because the technique has reached the maximum number of allowed forks. This is somehow equivalent to statically fix the number of forks to 5.

Our analysis at fork level has shown that dynamic reconfiguration outperforms developer configurations in terms of performance deviation due to its capability to dynamically stop warmup iterations.
The same cannot be said when stability criteria are applied to halt forks.
Indeed, the analysis at benchmark level has shown that this capability of dynamic reconfiguration has only neutral or negative effects on performance deviations, and it is hard to perceive any improvement brought by this specific feature. 
\\

\begin{tcolorbox}[colback=black!3!white,colframe=black!33!white]
\textbf{RQ$_5$ summary} - When compared to developer configurations, dynamic reconfiguration techniques provide more accurate estimates of the warmup time and better results quality (in most of the cases). $CV$ and $KLD$ show the largest improvement in terms of estimation accuracy, but only provide slight improvement in terms of results quality. 
On the other hand, $RCIW$ outperforms developer configurations both in terms of estimation accuracy and results quality, but this improvement often comes at the expense of an increased warmup time. When time doesn't not represent a key concern, $RCIW$ should be the primary choice.
\end{tcolorbox}

\section{Discussion} \label{sec:discussion}
Overall, we can observe that Java microbenchmarking is still subject to some flaws.
The results for RQ$_1$ provide evidence that benchmarks do not always reach a steady state of performance.
About 11\% of benchmark forks never reach a steady state of performance, and 43\% of benchmark executions involve at least one fork that doesn't hit the steady state.
These results are consistent with the seminal study on VM microbenchmarking of \cite{Barrett2017}, thus showing, on a larger corpus of benchmarks and in the more defined scope of ``testing-oriented'' Java microbenchmarks, that the ``\emph{two-phase assumption}'' does not always hold.
With this finding, we aim to raise awareness among developers (and researchers) that deal with Java microbenchmarking. An important lesson here is that some benchmark forks (mean $\sim$10\%) may not be representative of ``actual'' steady state performance, since their performance may continuously fluctuate over time, with a non-negligible deviation from steady state performance. Unfortunately, the only way to avoid this issue is to execute each fork for a large number of iterations, and then run the \cite{Barrett2017} technique to determine if the steady state of performance is reached or not. While this methodology may be appropriate in a research context (like ours), it may be impractical in real-word performance assurance processes, where benchmarks are repeatedly executed against software evolution, and time/resources are subject to constraints~\citep{traini2021}. Nonetheless, there are certain measures that can be put in place to (partially) mitigate this problem. For example, the analyses performed for RQ$_2$ showed that performance deviations of non-steady forks can be significantly reduced by using a minimum of 50 warmup iterations that, based on our microbenchmarking setup, correspond to 5 seconds of continuous benchmark execution and no less than 50 invocations. The performance deviations can be further mitigated by increasing the number of warmup iterations up to 300 (\ie 30 seconds of continuous benchmark execution and no less than 300 invocations). It is worth to notice that these values are considerably different than those provided by JMH defaults, which define 50 seconds of continuous benchmark execution and no less than 5 invocations for warmup. On the basis of our results, our practical suggestion is to never execute a benchmark for less than 5 seconds (and less than 50 invocations) before starting to collect measurements. When time does not represent a major concern, warmup should last for at least 30 seconds of continuous benchmark execution, and no less than 300 invocations.
 

Microbenchmarking is far from trivial even when benchmarks consistently reach a steady state of performance. The results of RQ$_2$ sheds a light on the potential pitfalls of using non-steady measurements. Performance measurements gathered in non-steady phases of benchmark execution substantially deviate from those collected during steady phases ($\sim$124k\% on average). Hence, relying on them can significantly mislead performance assessment.
 To deal with this problem, the current practice mostly rely on developers' \emph{guesses} to estimate the end of the warmup phase and discard measurements subject to performance fluctuations.
Based on the results for RQ$_3$, this approach seems to drastically mitigate these large deviations, \ie developer configurations lead to an average deviation from steady measurements of 8\%. 
Nonetheless, warmup estimation remains challenging and subject to (large) errors.
The results for RQ$_3$ show that developer static configurations fail to accurately estimate the end of the warmup phase, often with a non-trivial estimation error (median: 28 seconds). Developers tend to overestimate warmup time more frequently than underestimating it (48\% \emph{vs} 32\%). Nonetheless, both of these kinds of estimation errors produce relevant (though diverse) side effects.
For example, we showed that overestimation produces severe time wastes (median: 33 seconds), thereby hampering the adoption of benchmarks for continuous performance assessment. 
On the other hand, underestimation often leads to performance measurements that significantly deviate from those collected in the steady state (median 7\%), thus leading to poor results quality and potentially wrong judgements.
The latter side effect can be partially mitigated by running an adequate number of forks (\eg 5). Indeed, as we have shown in RQ$_2$, forks play a significant role in reducing performance deviations of non-steady measurements. Unfortunately, they also largely increase benchmark execution time, and this may be impractical in real-word contexts.
Another option is to leverage automated techniques that can effectively estimate the end of the warmup time at run-time.
Prior work tried to address this challenge through dynamic reconfiguration \citep{Laaber2020}.

Based on the results for RQ$_5$, dynamic reconfiguration techniques significantly improve the effectiveness of the state-of-practice. The achieved results show that dynamic reconfiguration techniques outperform developer static configurations in terms of \emph{warmup estimation error} with statical significance (p $\leq$ 0.05) and large/medium effect sizes (see \secref{sec:dyn_vs_dev_wee}). Nevertheless, this improvement may come at the expense of an increased microbenchmark execution time. For example, $RCIW$ produces higher estimates of the warmup time with non-negligible effect size in 20 out of the \systems systems (see \secref{sec:dyn_vs_dev_wt}). 
On the other hand, $CV$ and $KLD$ have more heterogeneous behaviors depending on the system, but they still report higher warmup estimates than those of developers in 10 systems.
Further empirical studies are needed to assess whether such time increase is acceptable for practitioners.

The results for RQ$_4$ also highlight a substantial diversity among different dynamic reconfiguration techniques. One peculiar example is $RCIW$ that, on one hand, induces the highest increase in microbenchmark execution time, but on the other hand, it provides the most reliable set of performance measurements.
Microbenchmark practitioners that do not have specific concerns on time (\eg small benchmarks suites) should adopt $RCIW$ for a reliable steady state performance assessment.
In the other cases, $KLD$ and $CV$ represent the best alternatives.

Despite the promising results highlighted in RQ$_5$, our findings suggest room for improvement for dynamic reconfiguration.
As shown for RQ$_4$, all dynamic reconfiguration techniques lead to a substantial estimation error, with $RCIW$ providing by far the largest error (median: 48 seconds), and, $CV$ and $KLD$ producing smaller, but still relevant, estimation errors (median of 19 and 17 seconds, respectively).
These errors induce significant, though diverse, side effects depending on the technique.
For example, $RCIW$ induces substantial side effects in terms of time waste (median: 46 seconds), while $KLD$ and $CV$ induce more frequent and impactful side effects on the reliability of performance measurements (median performance deviation of 10\% and 9\%, respectively).
Nonetheless, half of the warmup estimates of dynamic reconfiguration techniques can be reduced by at least 96\%, when only considering overestimated forks.
These results highlight a large space for improvement in dynamic reconfiguration techniques, and call for further research on designing and developing more effective dynamic reconfiguration techniques.
For example, future research may explore the use of other stability metrics (\eg autocorrelation metrics, other confidence interval metrics~\citep{Fieller1954}), or combinations of them to more effectively determine the end of the warmup phase.
Another suggestion is to focus more on improving warmup estimation accuracy, rather than finding stability criteria that are suitable to both stop forks and warmup iterations. 
Indeed, based on our results, dynamically stopping forks does not produce any tangible improvement over developer static configurations. In this regard, our suggestion is to allow practitioners manually configuring forks based on their own needs and time constraints. Nonetheless, we always recommend to run at least 5 forks (\ie the default in JMH) to mitigate the impact of non-steady measurements. Practitioners may decide to run less forks when time represents a major concern, however they should be aware of potential implications on results quality.

\section{Threats to validity} \label{sec:threats}

Our study may be affected by different threats that span from how we collected performance data to the subject project domains, and we describe them in the following.

\paragraph{Construct validity}

We assume that the benchmarks that reach a steady state will be able to do it within the execution time we defined.
There may be benchmarks that need more time to show some stability.
However, we chose the execution time to be considerably longer than the time we found in developer configurations (171 times longer on average).
Also, as done in other studies~\citep{Laaber2020}, measurement time, the number of iterations, and the number of forks are fixed for every benchmark.
The consequence is that the number of invocations varies from one benchmark to another. 
Nonetheless, no benchmark is invoked less than 3000 times per fork, that is 1000 times more than in \citep{Barrett2017}.
The number of forks is fixed at 10, as recommended in \citep{Barrett2017}.

Our experiment was performed in an environment where we tried our best to reduce the measurement noise and external influencing factors~\citep{DBLP:journals/tse/PapadopoulosVBH21,DBLP:conf/asplos/MytkowiczDHS09}.
Such settings are effective in improving the accuracy of results, but may not represent the more common environment in which developers execute the benchmarks.
However, general reference environments can hardly exist for benchmarks.
In fact, different developers can potentially execute the same microbenchmark on a wide range of different machines/environments, given the inherently distributed nature of open-source software development.
On top of that, there is an increasing interest in promoting the adoption of microbenchmarks in CI \citep{Laaber2021, Laaber2020,Laaber2019, Laaber2018}, which are most likely uncontrolled/noisy environments. All these aspects make it impractical to identify a reference environment for each benchmark/system.
For this reason, we have deliberately chosen to execute benchmarks on the same bare-metal server, using precautionary measures to mitigate measurement noise and external influencing factors (see \secref{sec:microbench} for details).
In this respect, we preferred to control the confounding variables rather than losing accuracy in more noisy and unreproducible settings.
This consideration was especially motivated by the first goal of our study (RQ$_1$), \ie checking whether benchmarks reach a steady state of performance.
Indeed, we prefer to run benchmarks on an environment that is considerably less subject to noise than those of developers, rather than possibly causing non-steady executions due to the noise that is specific to our environment.
Nonetheless, we do believe that future research should further strengthen our investigation beyond controlled environments, \ie by studying steady state performance even in uncontrolled and noisy environments. For example, future studies may replicate our experiments in cloud contexts to assess how these virtualized environments affect steady state performance. We envision that this latter step is crucial to foster the adoption of Java microbenchmarking in CI.

Developer configurations for warmup iterations $wi$ and measurement iterations $i$ are derived from the JSON reports generated by JMH (details in \secref{sec:bench_conf}).
We do not consider other ways in which benchmarks might be executed by developers, by using, for example, different command line arguments at launch time or by configuring additional parameters in build automation pipelines.
Nonetheless, to the best of our knowledge, no study so far showed evidence of microbenchmarks being used as part of build automation pipelines~\citep{Rausch2017, Vassallo2017, Beller2017}.

Following the methodology of \cite{Laaber2020}, we performed post-hoc analysis to derive measurements for developers and dynamic configurations.
Indeed, a comprehensive execution of all benchmarks across all developer static configurations and dynamic reconfiguration alternatives would have made our experimental evaluation impractical due to extremely long  execution times (the execution of all \benchmarks benchmarks using our single setup took about 93 days).
In order to mitigate the impact of post-hoc analysis on the validity of our study, we employed a carefully defined process to derive measurements.
We first estimated the time spent in each warmup/measurement iteration using the average execution time of each iteration as observed in our microbenchmarking setup. Then, we derived the \emph{estimated warmup time} $wt$ and the set of measurements $M^{conf}$ based on the JMH configurations provided by software developers or dynamic reconfiguration techniques.
The detailed process to estimate $wt$ and derive $M^{conf}$ can be found in \secref{sec:bench_conf}.
Nevertheless, our results may be marginally affected by approximations in converting measurement samples from our configuration to the others.

\paragraph{Internal validity}

The steady state is detected on the basis of execution time measurements.
We do not consider other event-based stability criteria, such as JIT activity, as these are not part of any JMH report and, therefore, are not something we can compare against when examining developer configurations.
However, considering such additional criteria may lead to a different steady state classification.

Before performing the change point detection, we filtered the outliers using Tukey's fences on a sliding window, as described in \secref{sec:exp_steadystate}. The parametrization of this procedure might affect the detection of the steady state. However, we tried to select parameters that would result in a very conservative outliers filtering. In fact, we only filter $0.27\%$ of the datapoints. Nonetheless, the outliers filtering is a necessary procedure when employing change point detection algorithms, as the vast majority of such algorithms cannot distinguish between actual changes and outliers~\citep{fearnhead2019outliers}, thus leading to an overestimation of changes.

\paragraph{External validity}

We only focused on GitHub repositories.
Therefore, it is unclear if the findings are valid for other open-source hosting platforms or industrial software.
Nonetheless, we conducted our experiment on 30 systems, a number larger than most recent empirical studies on performance (\eg see \citeauthor{Laaber2018,Laaber2020,DBLP:conf/icse/DingCS20,DBLP:conf/kbse/ReicheltKH19}).

We chose to limit the scope of the experiment to JMH microbenchmarks, because JMH is a mature and widely adopted Java microbenchmark harness.
We ran the benchmarks on JVM 8 (JDK 1.8.0 update 241) or 11 (JDK 11.0.6), depending on the requirements of the specific system.
Using other JVM versions or JVMs from other vendors may change the results.
All the benchmarks were executed on the same bare-metal server running Linux.
Nothing can be said about other hardware characteristics or other operating systems.

\paragraph{Conclusion validity}
The changepoint detection method we used assumes independence in the time series.
Even when the data contain some dependence, changepoint methods can still be used, provided that larger penalty values are used~\citep{ANTOCH1997291,Barrett2017}.
The penalty values we dynamically compute for each fork, as explained in \secref{sec:exp_steadystate}, tend to be larger (the average penalty is 504.54) than the value used, for example, in \cite{Barrett2017}.
Also, we manually inspected some of the time series to ensure that the segmentation was reasonable given the goal of the experiment.

Wherever possible, we used appropriate statistical procedures with p-value and effect size measures to test the significance of the differences and their magnitude.

\section{Related work} \label{sec:related}

There are different perspectives of tackling performance analysis of software systems, through models at runtime \citep{GieseL020,CORTELLESSA2022111084}, or by means of benchmarking.
Recently, benchmarking technique has played a key role to discover potential performance flaws \citep{DBLP:conf/wosp/StefanHBT17}. 

One of the main problems with benchmarking results is the reliability of the data. 
Recently, different approaches have defined rigorous processes to interpret those data. 
For example, some approaches rely on statistical inference for identifying and measuring the reliability of benchmarking results~\citep{Kalibera2013, kalibera2020}. 
Other approaches, instead, have presented performance analysis methodologies to extract data in a more reliable way \citep{Georges2007}. 
\cite{Barrett2017} introduced a fully automated statistical approach based on changepoint analysis.
In their work, \cite{Barrett2017} studied a set of small and deterministic VM benchmarks across different types of VMs, including the JVM.
They found that VM microbenchmarks may not always reach a steady state of performance.

On the other hand, performance benchmarking is a time-demanding process. 
Recently, some approaches investigated solutions to reduce the time for performance analysis while preserving reliable results~\citep{DBLP:conf/issta/MostafaWX17, He2019, alghamdi_2020}. 
\cite{He2019} have studied the reduction of performance testing in the cloud. They introduced a statistical tool, namely PT4Cloud, that provides stop conditions in order to obtain reliable performance indices. Another way to reduce testing time is, for example, by reusing the ``functional'' unit tests, which are likely available and maintained. For example, \cite{DBLP:journals/ase/BulejBHKMTT17} extended ``functional'' unit tests with performance knowledge by equipping them with stochastic performance logic.

Java Microbenchmark Harness (JMH) is a popular benchmarking framework for Java software. JMH allows defining performance testing to reduce variability in the measurements as well as external factor contributions during the microbenchmark testing phase. Hence, different studies spanned over different JMH aspects \citep{Costa2019, Samoaa2021, Laaber2019, Laaber2018}. 
A JMH microbenchmark might be affected by bad practices that could degrade performance results. 
\citep{Costa2019} studied those bad practices by analyzing a corpus of 123 OSS, and they extracted those bad practices that more likely lead to bad performance indicators.
\cite{Laaber2020} focused their study on reducing the required execution time of microbenchmarking tests through dynamic reconfiguration. They have defined three stability criteria to dynamically estimate the end of the warmup phase and halt warmup iterations accordingly. \cite{Samoaa2021} studied, instead, the impact of benchmark parameters and how they affect performance results.

In this work, we studied the effectiveness of modern Java microbenchmarking for steady state performance assessment.
Similarly to \cite{Barrett2017}, we investigated whether microbenchmarks reach a steady state of performance. However, unlike them, we studied this aspect in the more defined scope of ``testing-oriented'' Java benchmarks, \ie JMH benchmarks specifically designed to assess performance of a particular software.
Our results are consistent with those gathered by \cite{Barrett2017}, thus confirming that, even in a  different context, Java benchmarks may not always reach a steady state of performance.

\cite{Costa2019} broadly studied bad practices in JMH benchmarks, instead we specifically investigated the effectiveness of developer configurations for steady state performance assessment. \cite{Laaber2020} presented dynamic reconfiguration as a viable alternative to developer static configurations. In their study, they compared dynamic reconfiguration to JMH default configurations, and they observed a significant reduction in execution time with a negligible loss of result quality. In our study, instead, we evaluated the effectiveness of dynamic reconfiguration for steady state performance assessment. Furthermore, we showed, through a rigorous comparison, that dynamic reconfiguration is significantly more effective than developer configurations and, as such, it produces less pronounced side effects.

\section{Conclusion} \label{sec:conclusion}
This paper presents a comprehensive investigation on Java steady state performance assessment.
Through a rigorous assessment, we showed that Java microbenchmarks do not always reach a steady state of performance, thus confirming the finding of \cite{Barrett2017} in the more defined scope of ``testing-oriented'' Java microbenchmarks.

Even when microbenchmarks consistently reach a steady state of performance, a reliable assessment remains far from trivial.
According to our results, the current state-of-practice, which mostly relies on developer static configurations, show poor effectiveness for steady state performance assessment.
Developers often fail to accurately estimate the end of the warmup phase, thereby causing either large time wastes or poor results quality.
Dynamic reconfiguration provides a significant leap forward over developer static configurations by providing more accurate warmup estimates and less pronounced side effects. Still, the achieved results highlight non-trivial estimation errors, large time wastes, and distorted performance measurements.

The findings of our work have implications for both practitioners and researchers.

For the former, it is important to be aware that benchmark forks may not always reach a steady state of performance. The recommendation here is to perform an adequate number of forks (\eg 10) to mitigate the noise introduced by ``non-steady'' forks.
Another important lesson for practitioners is to favor dynamic reconfiguration over static configuration when possible.
Indeed, when compared to developer configurations, dynamic reconfiguration techniques provide more accurate estimates of warmup time, though this improvement may (sometimes) come at the expense of a more time-consuming performance assessment process.
Further empirical studies are needed to assess whether this ``cost'' is acceptable for practitioners.
Nonetheless, the achieved results are also helpful for suggesting which technique to use depending on the practitioner's need.

On the researchers' side, given the promising results of dynamic reconfiguration and the large room for improvement suggested by our investigation, we envision research aimed at designing novel and more effective dynamic reconfiguration techniques to (i) reduce the time effort devoted to performance assessment and (ii) strengthen the reliability of performance measurements.
This is a direction we aim to investigate in future work.

We have made the code and the data used in our study publicly available to encourage further research on this topic.

\section*{Data Availability} \label{sec:replication}
To aid reproducibility we provide the data and scripts needed to replicate our findings. The complete replication package is available at 
\center{\href{https://doi.org/10.5281/zenodo.7058361}{\texttt{DOI:10.5281/zenodo.7058361}}}

\begin{acknowledgements}
We thank the anonymous reviewers for their constructive comments, which guided us in improving the paper.
Luca Traini is grateful for the financial support  by ``Fondo Territori Lavoro e Conoscenza CGIL, CSIL and UIL'' through the project ``Territori Aperti''.
Daniele Di Pompeo is supported by the EMERGE project at Centre of EXcellence on Connected, Geo-Localized and Cybersecure Vehicle (EX-Emerge), funded by Italian Government under CIPE resolution n. 70/2017.
Michele Tucci is supported by the OP RDE project No. CZ.02.2.69/\-0.0/\-0.0/\-18\_053/\-0016976 ``International mobility of research, technical and administrative staff at Charles University''.
\end{acknowledgements}

\bibliographystyle{spbasic}
\bibliography{main.bib}

\end{document}